\documentclass[12pt]{iopart}
\usepackage{bm}
\usepackage{iopams}

\usepackage{graphicx,epsfig}

\newcommand{\text}[1]{\mbox{\scriptsize{#1}}}

\begin{document}

\title{Nonequilibrium dynamics of a fast oscillator coupled to Glauber spins}
\author{L L Bonilla$^1$, A Prados$^2$ and A Carpio$^3$}
\address{$^1$G. Mill\'an Institute for Fluid Dynamics, Nanoscience and Industrial Mathematics, Universidad Carlos III de Madrid, 28911 Legan\'es, Spain}
\address{$^2$ F\'{\i}sica Te\'{o}rica, Universidad de Sevilla,
Apartado de Correos 1065, E-41080, Sevilla, Spain}
\address{$^3$ Departamento de Matem\'atica Aplicada, Universidad Complutense de Madrid, 28040 Madrid, Spain}
\eads{\mailto{bonilla@ing.uc3m.es},\mailto{prados@us.es},\mailto{carpio@mat.ucm.es}}

\begin{abstract}
A fast harmonic oscillator is linearly coupled with a system of Ising spins that are in contact with a thermal bath, and evolve under a slow Glauber dynamics at dimensionless temperature $\theta$. The spins have a coupling constant proportional to the oscillator position. The oscillator-spin interaction produces a second order phase transition at $\theta=1$ with the oscillator position as its order parameter: the equilibrium position is zero for $\theta>1$ and non-zero for $\theta< 1$. For $\theta<1$, the dynamics of this system is quite different from relaxation to equilibrium. For most initial conditions, the oscillator position performs modulated oscillations about one of the stable equilibrium positions with a long relaxation time. For random initial conditions and a sufficiently large spin system, the unstable zero position of the oscillator is stabilized after a relaxation time proportional to $\theta$. If the spin system is smaller, the situation is the same until the oscillator position is close to zero, then it crosses over to a neighborhood of a stable equilibrium position about which keeps oscillating for an exponentially long relaxation time. These results of stochastic simulations are predicted by modulation equations obtained from a multiple scale analysis of macroscopic equations.
\end{abstract}


\maketitle

\section{Introduction}
\label{s1}
Many physical processes are modeled by means of an oscillator coupled
to a thermal bath or to spin systems. Among numerous examples, we can mention the classical version of the spin-phonon system describing the collective Jahn-Teller effect \cite{fed73,rik78}, decoherence of a spin representing a two-level system due to coupling to a boson bath (the spin-boson system) \cite{leg87}, a classical oscillator coupled to a spin causes wave function collapse thereof \cite{bon92}, single molecule magnets or nuclear spins modeled as large spins coupled to a boson bath \cite{hic08}, mass spectrometry through a nanoelectromechanical oscillator whose resonant frequency decreases as single molecules are added thereto \cite{boi09},  a 1/2-spin representing a nonlinear Josephson phase quantum bit is coupled to an oscillator (superconducting resonator) and to  a classical signal \cite{hof09,oco10},  etc.

Recently, we have studied a mechanical system (a classical harmonic oscillator) coupled to a chain of Ising spins in contact with a thermal bath at temperature $T$, so that
 \begin{equation}
    \frac{d^2x}{dt^2}+\omega_0^2 x=\frac{\mu_0}{m\sqrt{N}} \sum_{i=1}^N \sigma_i \sigma_{i+1} \, .
\end{equation}
The spins have an energy $-\mu_0 x \sum_{i=1}^N \sigma_i \sigma_{i+1}/\sqrt{N}$, and they flip stochastically according to Glauber dynamics at a rate $W_i(\bm{\sigma}|x,p)=(\alpha/2)\, [1-\gamma\sigma_{i} (\sigma_{i-1}+\sigma_{i+1})/2]$, with $\gamma=\tanh(2\mu_0x/(k_BT\sqrt{N}))$ \cite{Gl63,PBC10}. There is a second order phase transition at a critical temperature $T_c=\mu^2_0/(m\omega_0^2 k_B)$ and the equilibrium position of the oscillator is its order parameter. Above the critical temperature, the oscillator equilibrium position is the same as that of the uncoupled oscillator. Below the critical temperature, two symmetric nonzero equilibrium positions issue forth from zero as in the diagram of a pitchfork bifurcation. In the limit of fast relaxation of the spins compared to the natural period of the oscillator, $\omega_0/\alpha\ll 1$, and ignoring fluctuations, the oscillator position satisfies an effective equation having a nonlinear friction and a nonlinear force term \cite{PBC10}. The stationary solutions of the reduced dynamics coincide with the oscillator equilibrium positions for each temperature. Numerical simulations confirm the theory except for very low temperatures (sufficiently lower than $T_c$). In this temperature range, the numerical simulations show underdamped oscillations towards a nonzero equilibrium position of the oscillator while the theory predicts monotonic evolution towards the same value \cite{PBC10}.  Breakdown of the theoretical predictions is expected for sufficiently low temperatures because the relaxation time of the spins increases as the temperature decreases \cite{Re80,ByP94,ByP93b,ByP96} and the assumption that the spins relax in a time scale much shorter than the oscillator natural period is no longer true.

In this paper, we consider the same model of an oscillator coupled to Ising spins that evolve according to Glauber dynamics but in the opposite limit of a short
oscillator period compared to the spin relaxation time, i.e. $\omega_0\ll\alpha$. The main characteristics of the model are reviewed in section \ref{s2}. We use nondimensional variables in which the natural frequency of the oscillator and the transition temperature are both equal to one. In a mesoscopic description, the dynamical equations of the system are stochastic due to the coupling of the spins with the thermal bath at temperature $T$. These equations are equivalent to a master equation for the joint probability density ${\cal P}(x,p,\bm{\sigma},t)$ of finding the oscillator with position $x$ and momentum $p$, and the spins in configuration $\bm{\sigma}$ at time $t$. Its stationary solution is the canonical distribution, which describes the equilibrium behavior of the system, including the second order phase transition described in the previous paragraph. The proof showing that it is the \textit{only} stationary solution and that any solution of the master equation for finitely many spins tends to it in the long time limit is outlined in \ref{appH}.

The rest of the paper is as follows. In section \ref{s2}, we recall the model introduced in \cite{PBC10} and derive the nondimensional macroscopic equations valid in the limit of infinitely many spins and in the absence of fluctuations. In section \ref{s3}, we analyze these equations in the limit of large oscillator frequency as compared with the reciprocal relaxation time of the spin system, $\delta=\alpha/\omega_0\ll 1$. We show that there are two well separated time scales governing the behavior of our system: the fast time scale corresponding to the natural period of the oscillator and the slow time scale $\tau=\delta t$ over which the oscillator and the spin correlations relax to equilibrium. The oscillator position $x^{(0)}$ is approximately given by the order one spin correlation, $\widetilde{C_1}=N^{-1}\sum_{i=1}^{N} \langle\sigma_i \sigma_{i+1}\rangle$ (as $N\to\infty$), plus a modulated oscillatory term. The  equations for the spin correlations are the typical ones for Glauber dynamics, but with their parameters averaged over an oscillator period.

The main results in this paper are summarized in section \ref{s3}. The averaged equations for the slowly varying spin correlations and the envelope of the fast oscillations in $x^{(0)}$ constitute our first result. These averaged equations are approximated in two limits corresponding to our results 2 and 3: near the critical temperature $\theta=1$ and at low temperatures $\theta\to 0$. For most initial configurations producing a nonzero $\widetilde{C_1}$, the correlations evolve towards their equilibrium values. Near the critical temperature there is critical slowing down (on a time scale $|1-\theta|\tau$) in the already slow evolution of $\widetilde{C_1}$ toward its equilibrium value while the oscillation envelope decays to zero over the time scale $\tau$. The behavior for low temperatures is more surprising. The correlations decay algebraically (long time tails) to equilibrium, while the amplitude of the oscillations in the oscillator position does not vanish (except on an exponentially long time scale which is outside the scope of standard numerical calculations).  On the other hand, if the initial configuration of the spin system is random, the system approaches the unstable equilibrium solution $\widetilde{C_1}=0$, $\widetilde{x}=0$ for $\theta\leq 1$ (we have checked this for temperatures as low as $\theta=0.37$) instead of going to the stable equilibrium solution with nonvanishing $\widetilde{C_1}$ and $\widetilde{x}$. For initial conditions arbitrarily close to the random ones, the system first approaches the unstable equilibrium solution until it is quite close to it. Then it crosses over towards the stable solution having nonzero $\widetilde{C_1}$ while the oscillator position undergoes fast oscillations about $\widetilde{C_1}$ with an exponentially small damping.  This departure of the equilibrium behavior for exceedingly long time intervals is reminiscent of dynamical glassy behavior in spin glasses and other slow relaxing systems \cite{Bo92,BCKyM98}.

The comparison between our results and numerical simulations is excellent, as discussed in section \ref{s4}. Section \ref{s5} contains detailed derivations of the results given in section \ref{s3}. Section \ref{s6} contains our conclusions and the appendices are devoted to proving the H theorem for our system (\ref{appH}), and to technical details on the approximations used near the critical temperature (\ref{app_bif}) and for low temperatures (\ref{app_lowtemp}).

\section{Dynamics}
\label{s2}

Let us summarize the main aspects of the model introduced in \cite{PBC10}, where more details can be found. The system consists of a one dimensional harmonic oscillator (mass $m$, frequency $\omega_0$, position $x$ and momentum $p$) and $N\gg 1$ internal degrees of freedom modeled by Ising spins ($\sigma_i=\pm1$, $i=1,\ldots,N$) in contact with a heat bath at temperature $T$. The system has an energy
\numparts
\begin{eqnarray}
{\cal H}(x,p,\bm{\sigma}) & = & \frac{p^2}{2m}+\frac{1}{2}m\omega_0^2 x^2-\frac{\mu_0}{\sqrt{N}} x \sum_{i=1}^N \sigma_i \sigma_{i+1}  \, . \label{2.2a}
\end{eqnarray}
\endnumparts
The first two terms on the right hand side (rhs) of (\ref{2.2a}) correspond to the energy of the uncoupled oscillator, while the last term stands for the interaction energy between the oscillator and the spins. The latter can also be understood as a nearest neighbour interaction between the spins with a coupling constant $J_{\text{eff}}=\mu_0 x/\sqrt{N}$ which is proportional to the oscillator position $x$. The parameter $\mu_0$ measures the strength of the coupling between the oscillator and the Ising system.

The dynamics of the oscillator is governed by Hamilton's equations of motion, i.e.
\begin{equation}\label{2.3c}
    \ddot{x}+\omega_0^2 x=\frac{\mu_0}{m\sqrt{N}} \sum_{i=1}^N \sigma_i \sigma_{i+1} \, ,
\end{equation}
and the Ising spins evolve according to Glauber spin flip dynamics at temperature $T$. Thus at any time $t$, the system may experience a transition from $(x,p,\bm{\sigma})$ to $(x,p,R_i\bm{\sigma})$ at a rate given by \cite{Gl63}
\begin{equation}\label{2.4}
W_i(\bm{\sigma}|x,p)=\frac{\alpha}{2} \left[
1-\frac{\gamma}{2} \sigma_{i} (\sigma_{i-1}+\sigma_{i+1}) \right] \,, \qquad \gamma=\tanh\left( \frac{2 \mu_0 x}{k_{B}T\sqrt{N}}\right) \,,
\end{equation}
where $R_i\bm{\sigma}$ is the configuration obtained from $\bm{\sigma}$ by flipping the $i$-th spin, $k_{B}$ is the Boltzmann constant and $T$ is the temperature of the system. The parameter $\alpha$ gives the characteristic attempt rate for the transitions in the Ising system.

\begin{table}[ht]
\begin{center}\begin{tabular}{ccccc}
 \hline
 $x$ & $p$ &${\cal P}$ & $W_i$ & $t$  \\
$\frac{\mu_0 \sqrt{N}}{m\omega_0^2}$ & $\frac{\mu_0 \sqrt{N}}{\omega_0}$ & $\frac{m\omega_0^3}{\mu_0^2 N}$ &$\alpha$ & $\frac{1}{\omega_0}$ \\
 \hline
\end{tabular}
\end{center}
\caption{Nondimensional units and parameters.}
\label{t1}
\end{table}

It is convenient to introduce nondimensional variables according to $x^*=x/[x]$, $t^*=t/[t]$, \dots, where the units $[x]$, $[t]$, etc are as defined in Table \ref{t1}. Dropping the asterisks so as not to clutter our formulas, we obtain the nondimensional equations:
\begin{eqnarray}\label{2.16}
   && \frac{d^2 x}{dt^2}+x=\frac{1}{N} \sum_{i=1}^N \sigma_i \sigma_{i+1} \, ,\\
&&  \left[ \partial_t +p\, \partial_x + \left(\frac{1}{N}
 \sum_{i=1}^n \sigma_i \sigma_{i+1}-x\right) \partial_p \right]
 {\cal P}(x,p,\bm{\sigma},t)\nonumber \\
 && \quad \quad = \delta\, \sum_{i=1}^N \left[ W_i(R_i\bm{\sigma}|x,p) {\cal P}(x,p,R_i\bm{\sigma},t)-
W_i(\bm{\sigma}|x,p){\cal P}(x,p,\bm{\sigma},t) \right],   \label{2.17}\\
&& W_i(\bm{\sigma}|x,p) = \frac{1}{2}-\frac{\gamma(x)}{4}\,
\sigma_i (\sigma_{i-1}+\sigma_{i+1}), \label{2.18}\\
&&\gamma(x)=\tanh\left(\frac{2x}{\theta}\right) .\label{2.18bis}
\end{eqnarray}
Here $\delta$ and $\theta$ are dimensionless parameters given by
\begin{equation}\label{2.15}
 \delta=\frac{\alpha}{\omega_0}\,,\quad\theta=\frac{T}{T_c}\,,\quad   T_c=\frac{\mu_0^2}{m\omega_0^2 k_B}\,,
\end{equation}
$\delta$ is the ratio of the characteristic spin rate $\alpha$ to the oscillator natural frequency $\omega_0$ and $\theta$ is the dimensionless temperature. At $\theta=1$ ($T=T_c$), there is a second order phase transition whose order parameter is the equilibrium position of the oscillator \cite{PBC10}. Equation (\ref{2.16}) is equivalent to the oscillator equation (\ref{2.3c}), while (\ref{2.17}) is the master equation for the joint probability density ${\cal P}(x,p,\bm{\sigma},t)$ of finding at time $t$ the oscillator with nondimensional position $x$ and momentum $p$, and the spins in a configuration $\bm{\sigma}=\{\sigma_1,\sigma_2,\ldots,\sigma_N\}$. The equilibrium canonical density is the stationary solution
\begin{eqnarray}\label{2.19}
    {\cal P}_{\text{eq}}(x,p,\bm{\sigma})= \frac{1}{Z} \exp\left[-\frac{N}{\theta}\,
    {\cal H}(x,p,\bm{\sigma}) \right] ,\\
    {\cal H}(x,p,\bm{\sigma})=\frac{p^2+x^2}{2}-\frac{x}{N}\sum_{i=1}^N
    \sigma_i\sigma_{i+1} \,, \label{2.20}
\end{eqnarray}
where the partition function $Z$ guarantees that ${\cal P}_{\text{eq}}(x,p,\bm{\sigma})$ is normalized to unity. Summing over all the spin configurations $\bm{\sigma}$, we derive the marginal probability ${\cal P}_{\text{eq}}(x,p)$
\begin{eqnarray}
    {\cal P}_{\text{eq}}(x,p)=\frac{1}{Z} \exp \left[-\frac{N}{\theta}\left(\frac{p^2}{2}
   +\mathcal{V}_{\text{eff}}(x)\right) \right],   \label{2.21}\\
    \mathcal{V}_{\text{eff}}(x)= \frac{x^2}{2}- \theta\left[\ln\cosh\left(\frac{x}{\theta}\right)+\ln 2\right] \, .\label{2.22}
\end{eqnarray}
For finite $N$, the H-theorem of \ref{appH} proves that (\ref{2.19}) is globally stable and that it is the only stationary solution of the master equation (\ref{2.17}). The maxima of ${\cal P}_{\text{eq}}(x,p)$, $(\widetilde{x}_{\text{eq}},\widetilde{p}_{\text{eq}})$, coincide with the equilibrium mean values of $x$ and $p$ in the limit as $N\to\infty$. They are given by the extrema of $\mathcal{V}_{\text{eff}}(x)$ in (\ref{2.22}), i.e. by the solutions of the equation
\begin{equation}\label{2.24}
    \left.\frac{\rmd {\cal V}_{\text{eff}}(x)}{\rmd x}\right|_{x=\widetilde{x}_{\text{eq}}}=0 \, , \qquad
  \widetilde{x}_{\text{eq}}- \tanh \left(\frac{\widetilde{x}_{\text{eq}}}{\theta}\right)=0 \, .
\end{equation}
Clearly $\widetilde{x}_{\text{eq}}=0$ is always a solution for any $\theta$. For $\theta>1$, it is the only solution, it corresponds to a maximum of ${\cal P}_{ \text{eq}}$ and is therefore stable. At $\theta=1$ two new stable equilibria bifurcate from $\widetilde{x}_{\text{eq}}=0$ and exist for $\theta<1$. As $\theta\rightarrow 1^-$, we have
\begin{equation}\label{2.25}
    \widetilde{x}_{\text{eq}}\sim\pm \sqrt{3 \left(1-\theta\right)} \, .
\end{equation}

In the limit as $N\to\infty$, we split the variables $x=\widetilde{x}+\Delta x$, $\sigma_i\sigma_{i+n}=\widetilde{C_n}+\Delta C_{i,n}$, where $\widetilde{x}= \langle x\rangle$ and $\widetilde{C_n}=\langle\sigma_i\sigma_{i+n}\rangle$ (independent of $i$, provided we only consider homogeneous initial correlations $\langle\sigma_i\sigma_{i+n}\rangle (t=0)$ that depend on $n$ but not on $i$). Then we insert the result in the following equations obtained from  (\ref{2.16})-(\ref{2.18bis}),
\begin{equation}\label{2.26}
   \frac{\rmd^2 \langle x\rangle}{\rmd t^2}+ \langle x\rangle =\frac{1}{N}\sum_{i=1}^\infty \langle C_{i,1}\rangle \,,
\end{equation}
\begin{equation}\label{2.27}
\fl \qquad \frac{\rmd}{\rmd t}\langle C_{i,n} \rangle= \frac{\delta}{2} \langle \gamma(x) \left( C_{i,n-1}+C_{i,n+1}+ C_{i-1,n+1}+C_{i+1,n-1} \right) \rangle-2 \delta \langle C_{i,n} \rangle
\end{equation}
(for $n\geq 1$ and $i=1,\ldots, N$), and ignore the fluctuations $\Delta x$ and $\Delta C_{i,n}$. We obtain the following system of equations for the macroscopic variables $\widetilde{x}$ and $\widetilde{C_n}$ \cite{PBC10}:
\begin{eqnarray}\label{2.28}
  \frac{\rmd^2\widetilde{x}}{\rmd t^2}+ \widetilde{x} =\widetilde{C_1} \, ,  \\
  \label{2.29}
\frac{\rmd}{\rmd t}\widetilde{C_n}=  \delta \gamma(\widetilde{x}) \left( \widetilde{C_{n-1}}+\widetilde{C_{n+1}} \right)-2 \delta\,\widetilde{C_n}, \, n\geq 1, \qquad \widetilde{C_0}=1,\\
\gamma(x)=\tanh\left(\frac{2x}{\theta}\right).\label{2.30}
\end{eqnarray}
The mean-field or macroscopic dynamical behavior of the oscillator-spin system is found by solving the equations (\ref{2.28}) and (\ref{2.29}) with the boundary condition $\widetilde{C_0}=1$ and appropriate initial conditions.

\section{Fast oscillator dynamics. General results}
\label{s3}
The limit of a fast oscillator compared to spin relaxation describes the region of low temperatures of our system (no matter what the oscillator natural frequency is) because the spin relaxation time becomes arbitrarily large as the temperature decreases towards zero. We have obtained the following results:

\textit{Result 1. In the limit as $\delta\to 0$, the system of nondimensional equations (\ref{2.28}) - (\ref{2.29}) can be approximated by the multiscale solution:}
\numparts
\begin{eqnarray}
&&\widetilde{x}(t;\delta)= x^{(0)}(\chi,\tau)+\Or(\delta), \quad \widetilde{C_n}(t;\delta)= C_n^{(0)}(\tau)+\Or(\delta),  \label{b4a}\\
&& x^{(0)}(\chi,\tau)= C_1^{(0)}(\tau) + R(\tau)\,\sin\chi,\quad \tau=\delta t,\, \chi=t+\phi(\tau), \label{b4b}
\end{eqnarray}
\endnumparts
\numparts
in which $C_n^{(0)}(\tau)$, $R(\tau)$ and $\phi(\tau)$ solve
\begin{eqnarray}
&&\frac{\rmd C_n^{(0)}}{\rmd \tau}= \overline{\gamma(x^{(0)}(\chi,\tau))}\,
(C_{n-1}^{(0)} + C_{n+1}^{(0)})-2 C_n^{(0)}, \quad C_0^{(0)}=1, \label{b5a}\\
&&\frac{\rmd R}{\rmd\tau} = -(1+C_{2}^{(0)})\,\overline{\gamma(x^{(0)}(\chi,\tau))\sin \chi},  \quad \frac{\rmd \phi}{\rmd\tau} =0. \label{b5b}
\end{eqnarray}
\endnumparts
\textit{Here we have defined the time averages over the fast periodic variable $\chi$ (keeping the slow variable $\tau$ fixed) as}
\begin{eqnarray}
\overline{f(\chi)}=\frac{1}{2\pi}\int_{-\pi}^\pi f(\chi)\, \rmd \chi\, .  \label{b5}
\end{eqnarray}

This result will be derived later in section \ref{s4}. Note that, according to (\ref{b4a}) and (\ref{b4b}), the oscillator velocity is
\begin{eqnarray}
\frac{\rmd}{\rmd t}\widetilde{x}(t;\delta)=\frac{\partial x^{(0)}}{\partial\chi}(t+\phi,\delta t) + \Or(\delta).  \label{b6}
\end{eqnarray}
Then the initial conditions
\begin{equation}\label{b6b}
\widetilde{x}(0;\delta)\equiv x_0\,, \quad \dot{\widetilde{x}}(0;\delta)\equiv v_0\,, \quad  \widetilde{C_n}(0;\delta)\equiv\widetilde{C_n}(0)
\end{equation}
yield the following initial conditions for the approximate quantities:
\begin{eqnarray}
R(0)=\sqrt{[x_0-\widetilde{C_1}(0)]^2+v_0^2}, \quad \sin\phi=\frac{x_0-\widetilde{C_1}(0)}{R(0)},  \label{b7}
\end{eqnarray}
for $R(0)\neq 0$. If $x_0=\widetilde{C_1}(0)$ and $v_0=0$, $R(0)=0$ and (\ref{b5b}) produces $R(\tau)=0$ so that there are no modulated oscillations for this choice of initial data. $0=v_0=d\widetilde{x}/dt=0$ yields $dC_1^{(0)}/d\tau=0$ at $t=0$.

The stationary solutions of (\ref{b5a}) and (\ref{b5b}), $C_{n,\text{eq}}^{(0)}$ and $x_{\text{eq}}^{(0)}$, are constants, independent on $\chi$ and $\tau$. Therefore $R=0$, $x_{\text{eq}}^{(0)}=C_{1,\text{eq}}^{(0)}$, and we have
\begin{eqnarray}
 -2C_{n,\text{eq}}^{(0)}+\gamma(C_{1,\text{eq}}^{(0)}) \left[ C_{n-1,\text{eq}}^{(0)}+C_{n+1,\text{eq}}^{(0)} \right]=0 \,, \qquad C_{0,\text{eq}}^{(0)}=1 \,. \label{3.20b}
\end{eqnarray}
This system has the unique solution
\begin{equation}\label{3.21}
    C_{n,\text{eq}}^{(0)}=\left(C_{1,\text{eq}}^{(0)}\right)^{n}  ,
\end{equation}
with
\begin{equation}\label{3.22}
    C_{1,\text{eq}}^{(0)} =\frac{1}{\gamma(C_{1,\text{eq}}^{(0)})}-\sqrt{\frac{1}{[\gamma(C_{1,\text{eq}}^{(0)})]^2}-1}=
    \tanh\left(\frac{C_{1,\text{eq}}^{(0)}}{\theta}\right)\,,
\end{equation}
which determines the equilibrium value of $C_1^{(0)}$ or $x^{(0)}$, since both of them are equal. Equation (\ref{3.22}) is equivalent to (\ref{2.24}), the bifurcation equation. Then the stationary solutions of the approximate dynamical equations (\ref{b5a})-(\ref{b5b}) coincide with the exact equilibrium solutions obtained in the previous section. There is only one stationary solution $R=C_n^{(0)}=0$ ($n>0$) for $\theta>1$ (which is stable) and three stationary solutions $R=0$, $C_{n,\text{eq}}^{(0)}=\left(C_{1,\text{eq}}^{(0)}\right)^{n}$, with $C_{1,\text{eq}}^{(0)}$ solving (\ref{3.22}) for $\theta<1$. In the latter case, $C_{1,\text{eq}}^{(0)}=0$ is unstable whereas the solutions with nonzero $C_{1,\text{eq}}^{(0)}$ are stable and bifurcate from the zero solution at $\theta=1$. This was to be expected, since the equilibrium distribution does not depend on $\delta$, and it shows the consistency of the multiple scales approximation.

Equations (\ref{b5a})-(\ref{b5b}) have solutions $C_n^{(0)}\equiv 0$ ($n\geq 1$), with $R(\tau)$ decreasing from $R(0)>0$ to 0 as $\tau\to\infty$ ($dR/d\tau\leq 0$ because $\gamma(R(\tau)\sin \chi)\sin\chi\geq 0$ in $\chi\in [-\pi,\pi]$. Then its average appearing in (\ref{b5b}) is positive for $R>0$ and it vanishes for $R=0$). For $\theta<1$, this solution lies on the stable invariant manifold of the saddle point $R=0$, $C_n^{(0)}\equiv 0$ ($n\geq 1$): linearizing (\ref{b5a})-(\ref{b5b}) for $n=1$ about the saddle point, we find
\numparts
\begin{eqnarray}
&&\frac{\rmd C_1^{(0)}}{\rmd \tau}\sim 2 \left(\frac{1}{\theta}-1\right) C_1^{(0)},  \label{b5c}\\
&&\frac{\rmd R}{\rmd\tau} \sim -\frac{R}{\theta}. \label{b5d}
\end{eqnarray}
\endnumparts
Setting $\widetilde{C_n}\equiv 0$ ($n\geq 1$) in (\ref{2.28})-(\ref{2.30}) we no longer obtain an invariant manifold as it is the case with $C_n^{(0)}\equiv 0$ for the averaged equations (\ref{b5a})-(\ref{b5b}). In fact, (\ref{2.29}) becomes $d\widetilde{C_1}/dt= \gamma(\widetilde{x})\neq 0$ for $n=1$ if $\widetilde{C_n}\equiv 0$ ($n\geq 1$) and $\widetilde{C_0}\equiv 1$.

There are two limits in which we can find approximate solutions of the averaged equations (\ref{b5a}) and (\ref{b5b}), namely $\theta\to 1$ (the critical temperature) and $\theta\to 0$.

\subsection{The limit $\theta\to 1$}

\textit{Result 2. In the limit as $\theta\to 1$, the solutions $C_n^{(0)}(\tau)$ and $R(\tau)$ of (\ref{b5a}) and (\ref{b5b}) are given by
\numparts
\begin{eqnarray}
&&C_1^{(0)}(\tau)= [\mbox{sign}\,C_1^{(0)}(0)]\,\sqrt{\frac{3\, (1-\theta)}{1+e^{-4 (1-\theta)\tau}\left(\frac{3\, (1-\theta)}{C_1^{(0)}(0)^2}-1\right)}}+\Or(|1-\theta|),    \label{b1a}\\
&& C_n^{(0)}(\tau)= [C_1^{(0)}(\tau)]^n [1+\Or(|1-\theta|^{1/2})]\,, n\geq 2 \label{b1b}\\
&&R(\tau)=R(0)\, e^{-\tau} +\Or(|1-\theta|),  \label{b1c}
\end{eqnarray}
\endnumparts
provided the initial conditions $R(0)$ and $C_1^{(0)}(0)$ are both of order $\epsilon=\sqrt{|1-\theta|}$. }

These formulas will be derived later in section \ref{s4} by using bifurcation theory (we understand sign$(0) = 0$ in the previous formula, so that $C_1^{(0)}(\tau)=0$ if $C_1^{(0)}(0)=0$). In \ref{app_bif}, we show that (\ref{b1a}) and (\ref{b1b}) with $\tau=\delta\, t$ also hold for $\widetilde{C_1}=\widetilde{x}$ and $\widetilde{C_n}$, respectively, except for exponentially decreasing terms.

\subsection{The limit $\theta\to 0$}

\textit{Result 3. In the limit as $\theta\to 0^+$, Equations (\ref{b5a}) and (\ref{b5b}) can be approximated by
\numparts
\begin{eqnarray}
&& \frac{\rmd C_n^{(0)}}{\rmd \tau}= S \left(C_{n-1}^{(0)} + C_{n+1}^{(0)}\right) - 2 C_n^{(0)}, \quad C_0^{(0)}=1, \label{b2a}\\
&&\frac{\rmd R}{\rmd\tau}=-2\left(1+C_2^{(0)}\right) e^{-4|C_1^{(0)}|/\theta} I_1\left(\frac{4R}{\theta}\right), \label{b2b}
\end{eqnarray}
\endnumparts
with $S=$sign$\, C_1^{(0)}(\tau)$, whenever $R<|C_1^{(0)}|$, and for $R>|C_1^{(0)}|$ by}
\numparts
\begin{eqnarray}
&& \frac{\rmd C_n^{(0)}}{\rmd \tau}= \frac{2}{\pi}\, \left(C_{n-1}^{(0)} + C_{n+1}^{(0)}\right)\, \arcsin\left(\frac{C_1^{(0)}}{R}\right)-2 C_n^{(0)}, \quad C_0^{(0)}=1, \label{b3a}\\
&&\frac{\rmd R}{\rmd\tau} =-\frac{2}{\pi}\, \left(1+C_2^{(0)}\right)\sqrt{1- \left(\frac{C_1^{(0)}}{R}\right)^2}. \label{b3b}
\end{eqnarray}
\endnumparts

Note that rescaling $C_n^{(0)}\to S^n C_n^{(0)}$, we can reduce the case $S=-1$ in (\ref{b2a}) to the case $S=1$. For $C_1^{(0)}<0$, $R$ and $C_n^{(0)}$ for even $n$ remain the same as in the case $C_1^{(0)}>0$ whereas the correlations $C_n^{(0)}$ for odd $n$ change their sign. For $C_1^{(0)}>0$,
\begin{equation}\label{5.5}
C_n^{(0)}(\tau)= 1-e^{-2\tau} \sum_{j=1}^\infty [1-C_j^{(0)}(0)]  [I_{|n-j|}(2\tau) - I_{n+j}(2\tau)] \,,
\end{equation}
where $I_n(z)$ is the modified Bessel function of the first kind with index $n$, solves (\ref{b2a}). Then the asymptotic properties of $I_n(z)$ \cite{Be99} yield as $\tau\to\infty$ the long time tail result
\begin{equation}\label{5.5b}
    C_n^{(0)}(\tau)\sim 1-n (4\pi\tau^3)^{-1/2}\sum_{j=1}^{\infty} j [1 -C_j^{(0)}(0)],
\end{equation}
provided that the sum is finite. As $\tau\to\infty$, (\ref{b2b}) becomes
\begin{equation}\label{5.6}
    \frac{\rmd R}{\rmd\tau} \sim -4 e^{-4/\theta} I_1\left(\frac{4R}{\theta}\right) \, ,
\end{equation}
$R\to 0$ and, using the relation $I_1(x)\sim x/2$ \cite{Be99} in (\ref{5.6}), this equation becomes
\begin{equation}\label{5.7}
    \frac{\rmd R}{\rmd\tau} \sim -\frac{8}{\theta} e^{-4/\theta} R \quad (\tau\to\infty).
\end{equation}
$R$ decays exponentially with a characteristic time which diverges exponentially as $\theta\to 0$. The amplitude of the oscillations around the steady value $x_{\text{eq}}=1$ is damped only after an exponentially large time $\theta e^{4/\theta}/8$.

Let us consider now equilibrium-like initial conditions for the Ising system, $C_n(0)=r^n$, with $0\leq r \leq 1$. In this case, the solution (\ref{5.5}) becomes
\begin{equation}\label{5.8}
 \fl   C_n^{(0)}(\tau)=\frac{2}{\pi} \int_0^\pi \rmd q\, \frac{r \sin q\sin nq}{1+r^2-2r\cos q} e^{-2\tau(1-\cos q)}+ \int_0^\tau e^{-2t} \left[I_{n-1}(2t)-I_{n+1}(2t)\right] \, dt.
\end{equation}
For $n=1$, (\ref{5.8}) increases monotonically from $C_1^{(0)}(0)=r$ to its asymptotic value $1$. Since $R$ decreases with time, $C_1^{(0)}(\tau)>R(\tau)$ if this condition holds at $\tau=0$. Using  the asymptotic expression of $I_{n\pm 1}(2t)$ for large $t$ \cite{Be99}, we find
\begin{equation}\label{5.8b}
    C_1^{(0)}(\tau)\sim 1-\frac{1}{\sqrt{\pi\tau}}+\Or(\tau^{-3/2})  \,.
\end{equation}
For this subset of initial conditions, the long time behavior of the $C_1$ is independent of the initial conditions. It must be noted that this is consistent with (\ref{5.5b}), since for these initial conditions the sum in (\ref{5.5b}) diverges.

Let us now consider equilibrium-like initial conditions with $0<r\ll 1$, so that $R(0) \gg C_1^{(0)}(0)>0$. Then (\ref{b3a})-(\ref{b3b}) become
\numparts
\begin{eqnarray}
&& \frac{\rmd C_n^{(0)}}{\rmd \tau}= \frac{2C_1^{(0)}}{\pi R}\, C_{n-1}^{(0)} -2 C_n^{(0)}, \quad C_0^{(0)}=1, \label{b3aa}\\
&&\frac{\rmd R}{\rmd\tau} =-\frac{2}{\pi}. \label{b3bb}
\end{eqnarray}
\endnumparts
We have assumed that $C_n^{(0)}=\Or(r^n)$ and $R(0)=\Or(1)$ as $r\to 0$. Then (\ref{b3aa}) yields $C_1^{(0)}=C_1^{(0)}(0)\, e^{-(1-2/\pi)\tau}$ and (\ref{b3bb}) gives $R(\tau)=R(0)-2\tau/\pi$. The correlations remain small whereas $R(\tau)$ becomes zero at a time
\begin{eqnarray}
\tau_0=\frac{\pi R(0)}{2}= \frac{\pi}{2}\,\sqrt{[x_0-\widetilde{C_1}(0)]^2+v_0^2}.
\label{b3c}
\end{eqnarray}
At a slightly smaller time than this, $R(\tau)\sim C_1^{(0)}(\tau)$ and the approximations (\ref{b3aa}) and (\ref{b3bb}) break down after some time proportional to $\ln (1/r)$. If $r=0$, then we are on the stable invariant manifold of the stationary saddle point solution mentioned above, all correlations remain zero and $R$ will decay to zero at $\tau=\tau_0$ and remain there indefinitely. However the stationary state $R(\tau)= C_1^{(0)}(\tau)=0$ is unstable as indicated by the linearized equation (\ref{b5c}) and any numerical error could give rise to a ``virtual'' small $r$ and lead to  the neighborhood of one of the stable stationary (equilibrium) solutions after the break down time before mentioned. The solutions of the macroscopic system (\ref{2.28})-(\ref{2.30}) will exhibit a different behavior because this system does not have the same invariant manifold as the averaged system (\ref{b5a})-(\ref{b5b}). These conclusions will be checked with numerical solutions in section \ref{s4}.

\section{Comparison with numerical simulations}
\label{s4}
To check the validity of our approximations, we shall first compare the solutions of the macroscopic equations (\ref{2.28})-(\ref{2.29}) to the averaged system (\ref{b5a})-(\ref{b5b}) (Result 1 in section \ref{s3}) and to one of the approximations thereof: either bifurcation theory for $\theta$ close to 1 (Result 2, equations (\ref{b1a})-(\ref{b1c}) in section \ref{s3}) or the set of two equations (\ref{b2a})-(\ref{b3b}) for low temperature $\theta$ (Result 3 in section \ref{s3}). Later we will compare our theory with direct numerical simulations of the oscillator-spin system.

\subsection{Comparison with numerical solutions of the macroscopic equations}

\begin{figure}[htbp]
\begin{center}
\includegraphics[width=6cm]{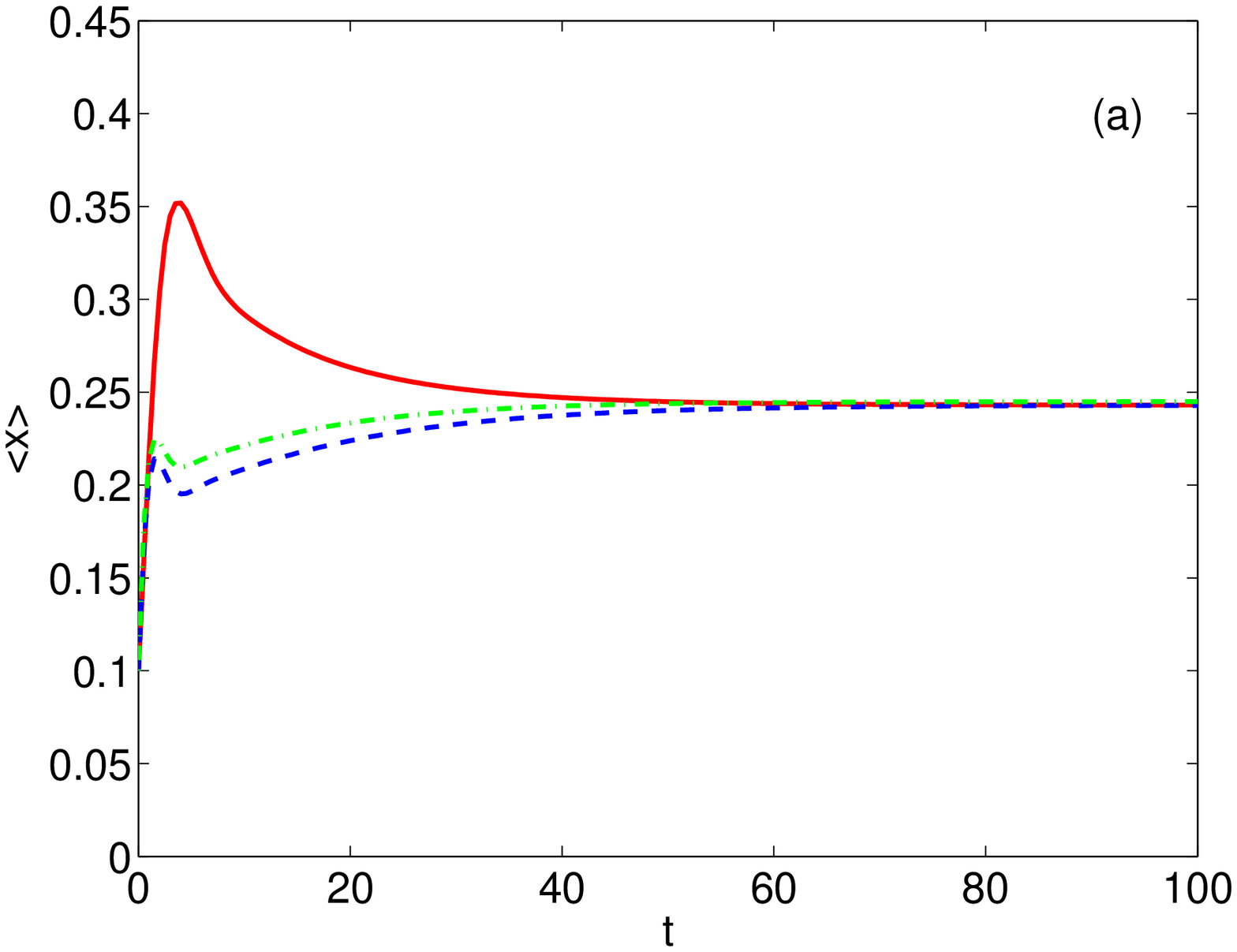}
\includegraphics[width=6cm]{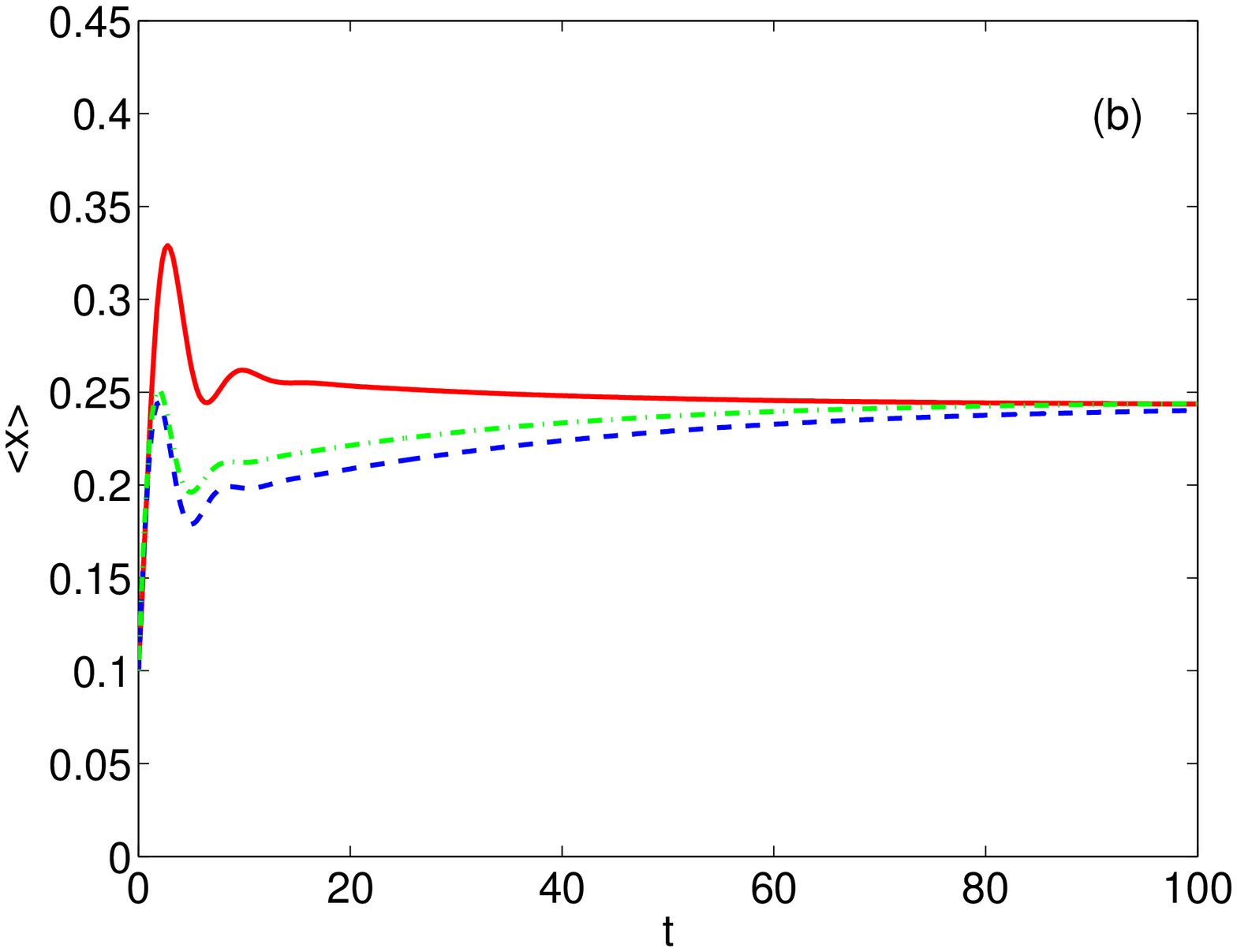}
\includegraphics[width=6cm]{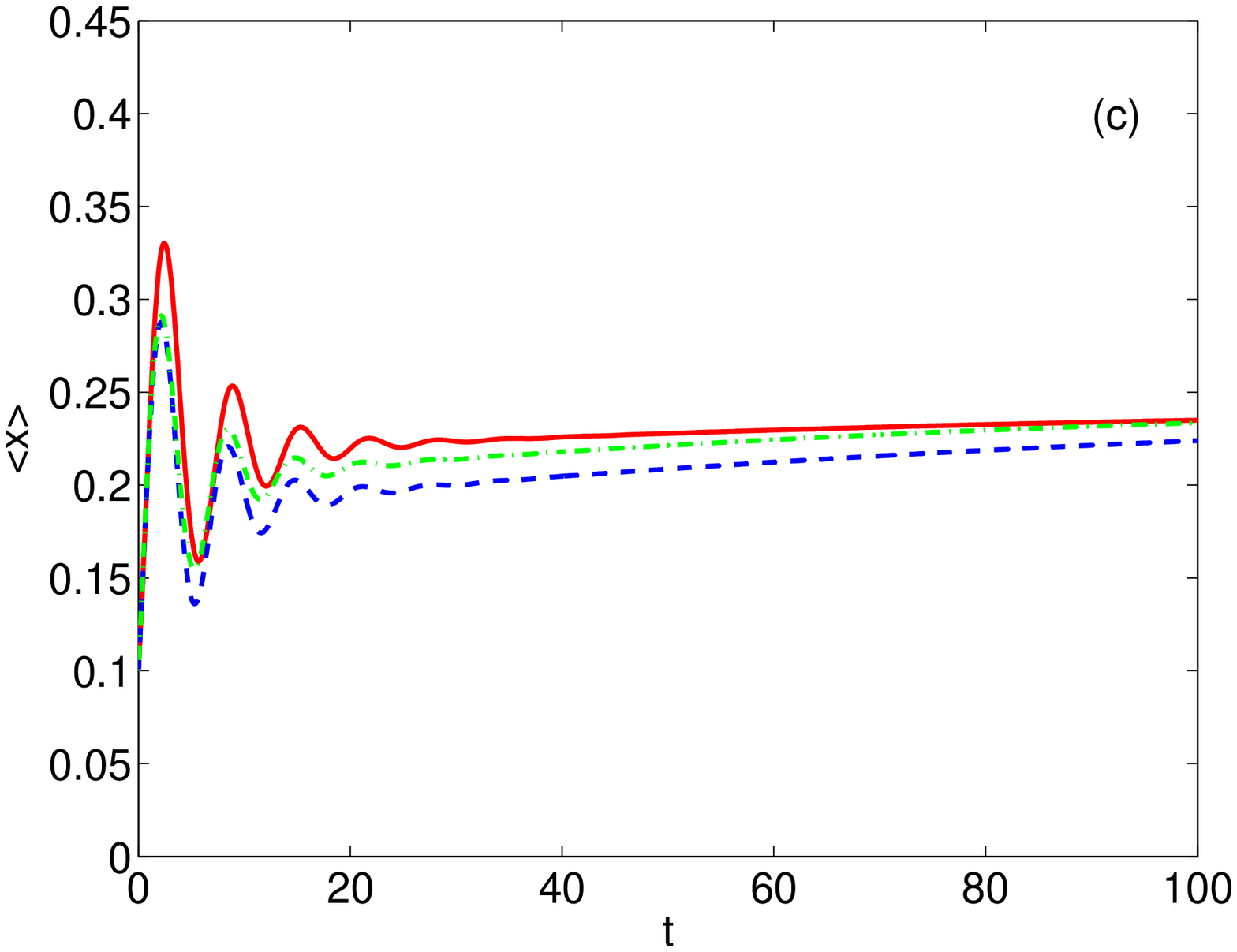}
\includegraphics[width=6cm]{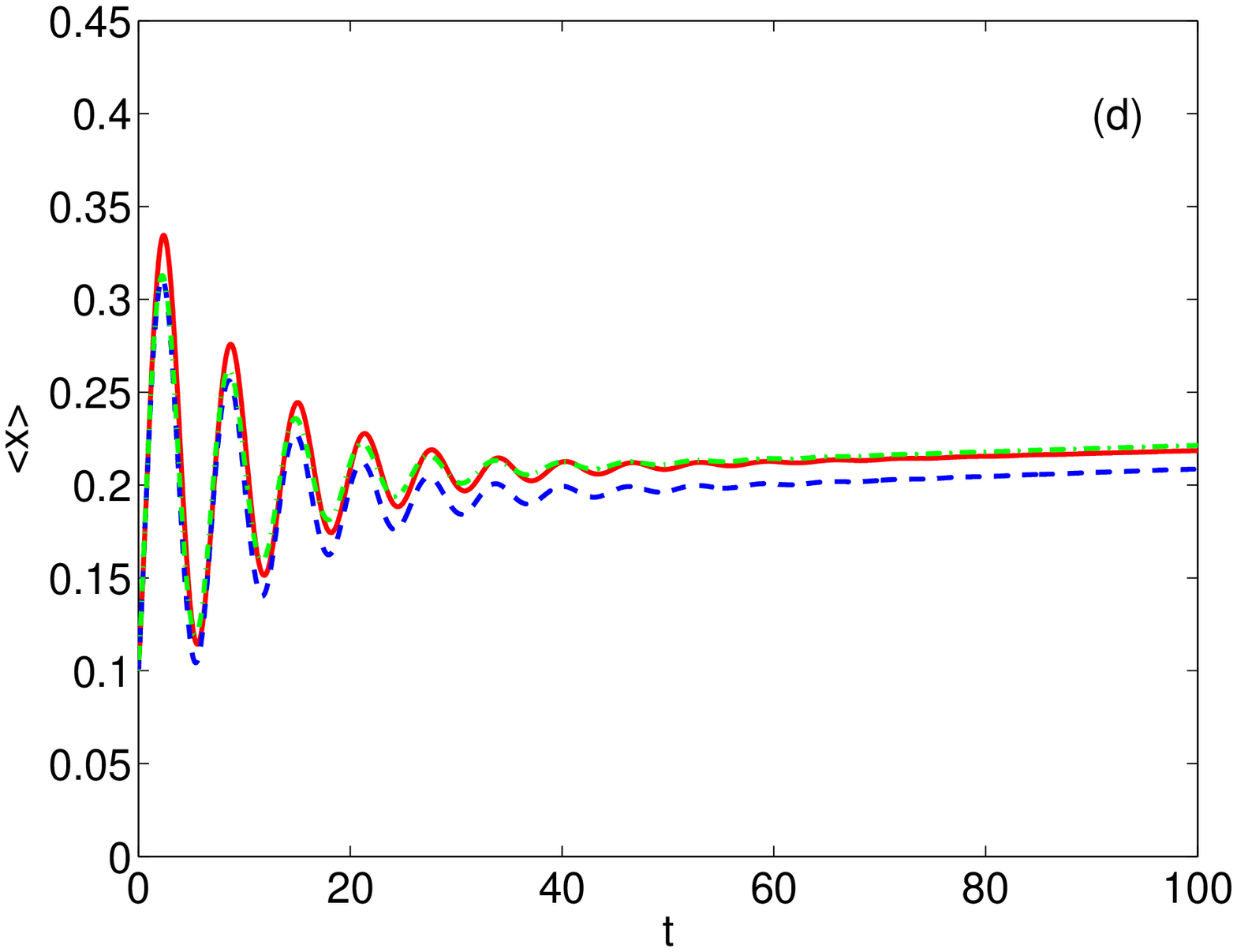}
\includegraphics[width=6cm]{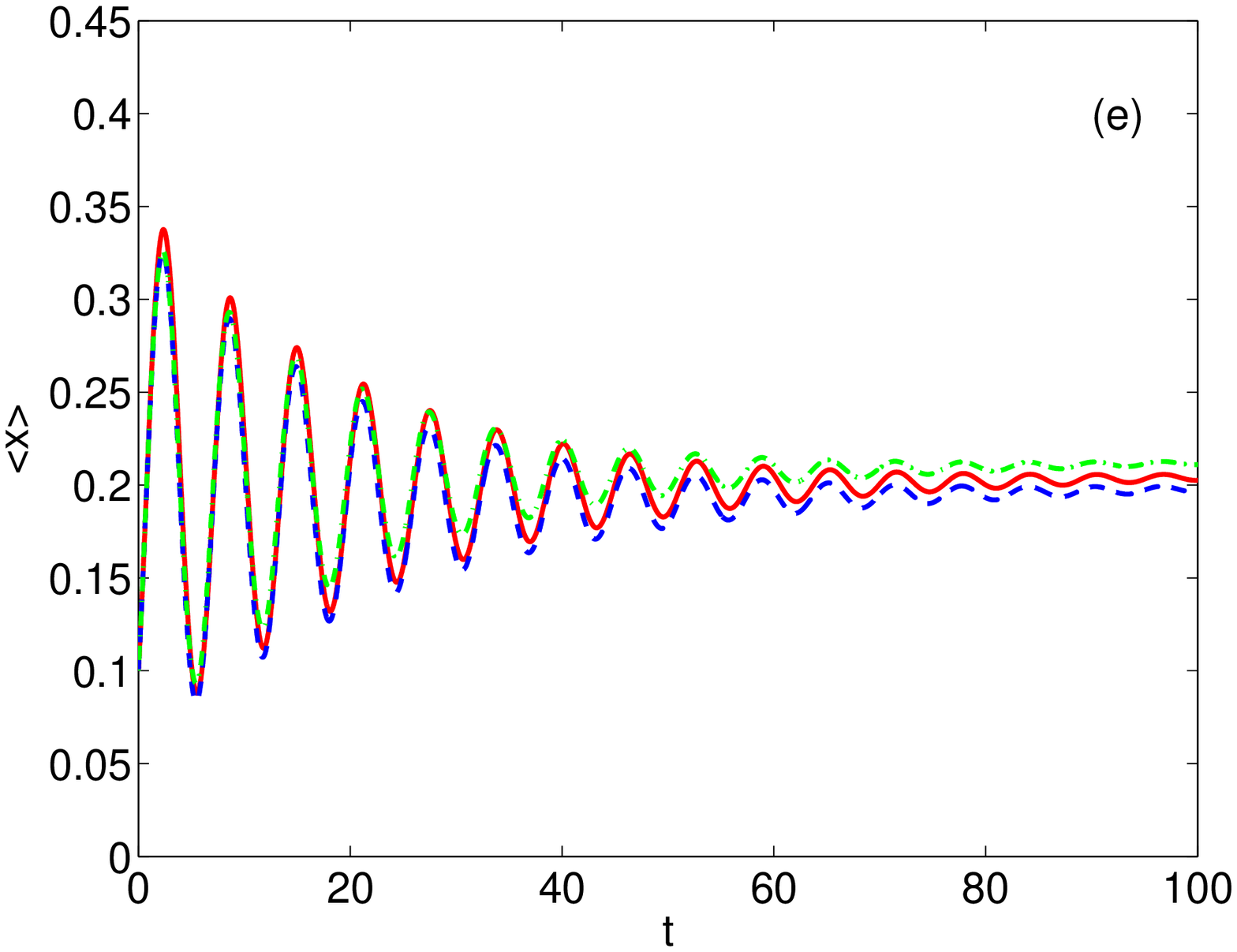}
\includegraphics[width=6cm]{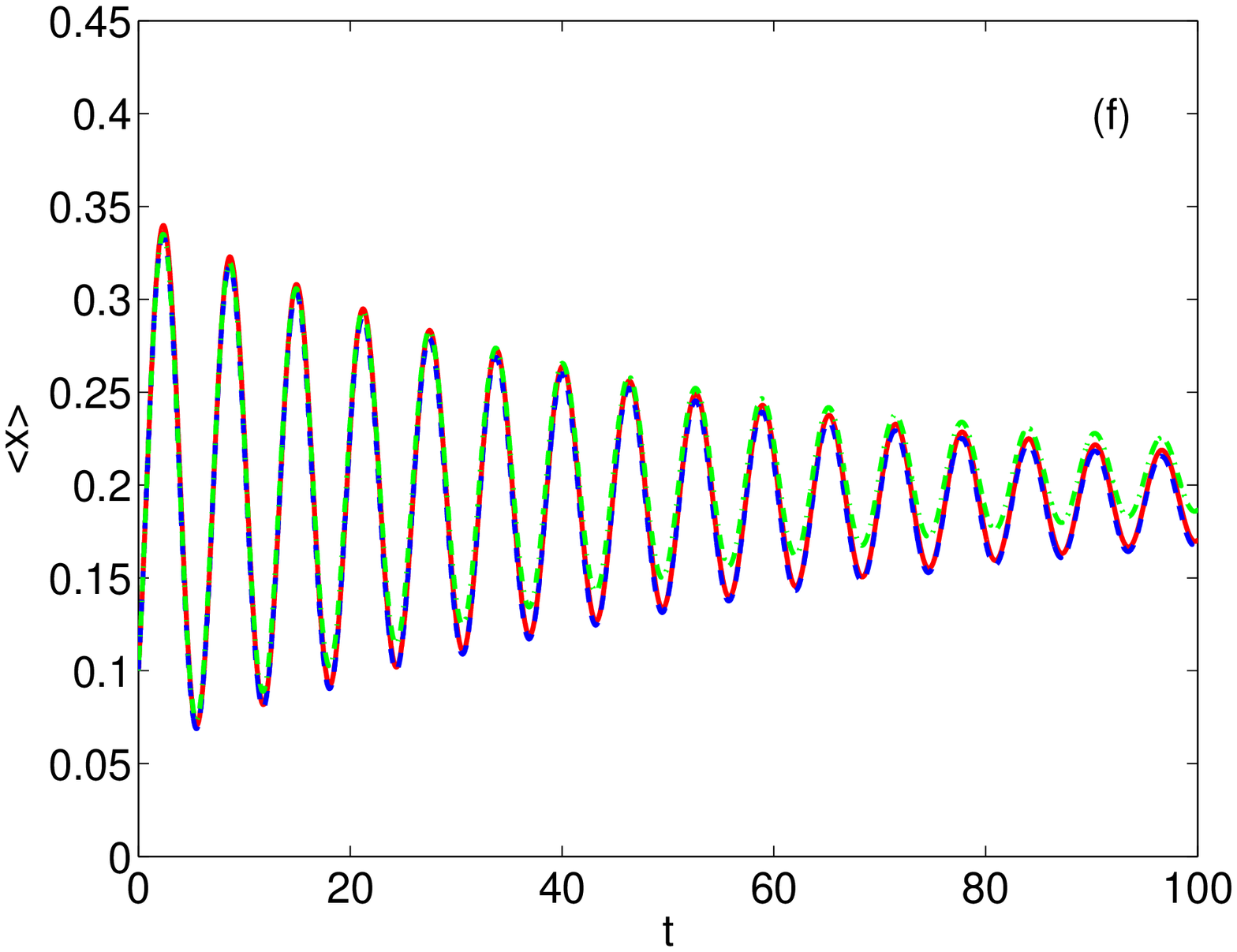}
\includegraphics[width=6cm]{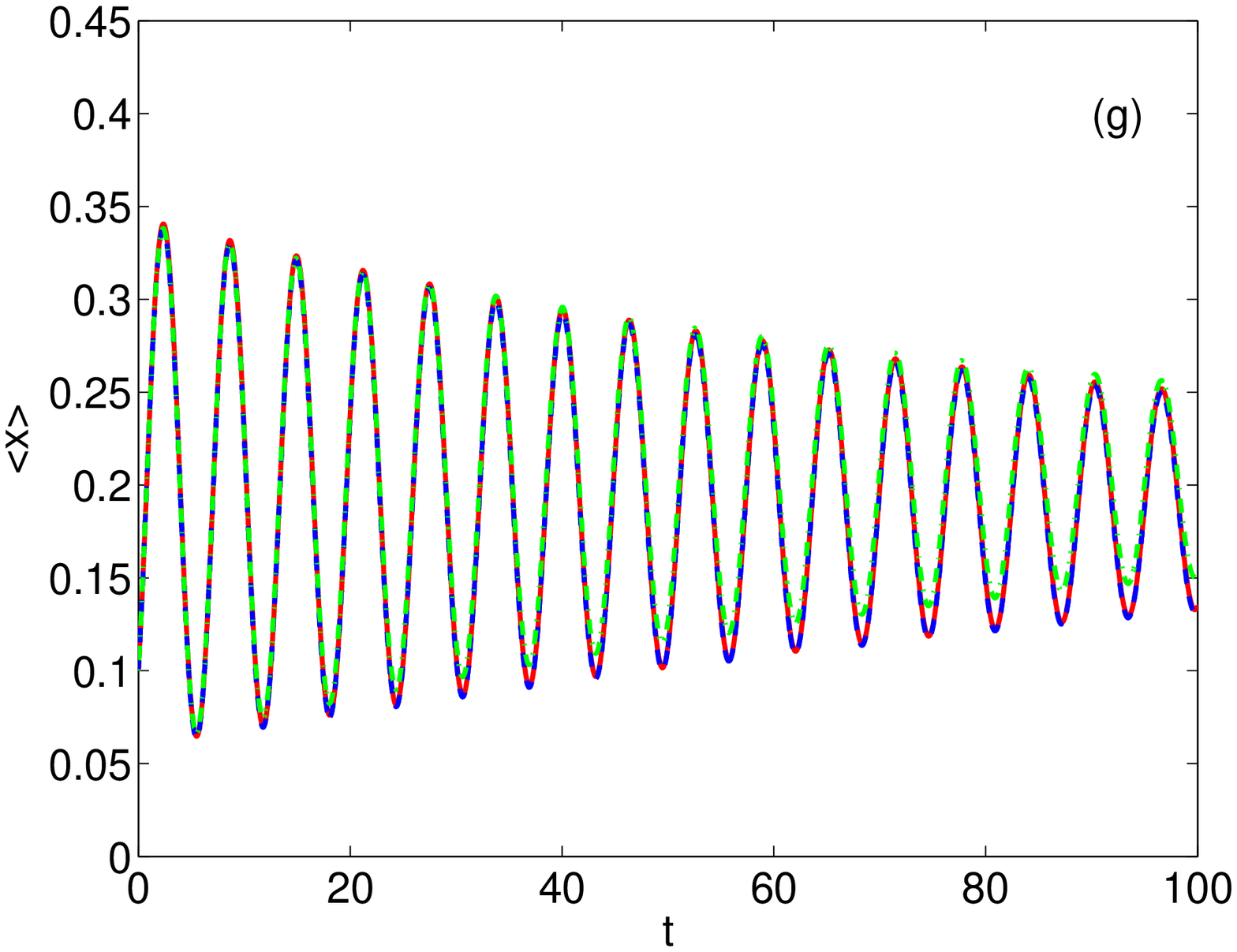}
\caption{Numerical solution of the macroscopic system (solid red line), Result 1 (dashed blue line) and Result 2 (dot-dashed green line) for the following values of $\delta$: (a) 1, (b) 0.5, (c) 0.2, (d) 0.1, (e) 0.05, (f) 0.02, (g) 0.01. In all cases, $\theta=0.98$, and initial data are $x_0=0.1$, $v_0=0.1$, $\widetilde{C_1}(0)=x_0+0.1$, $\widetilde{C_n}(0)=0$, $n\geq 2$. The macroscopic system has been truncated by imposing $\widetilde{C_{1000}}\equiv 0$. }
\label{f1}
\end{center}
\end{figure}

\begin{figure}[htbp]
\begin{center}
\includegraphics[width=6cm]{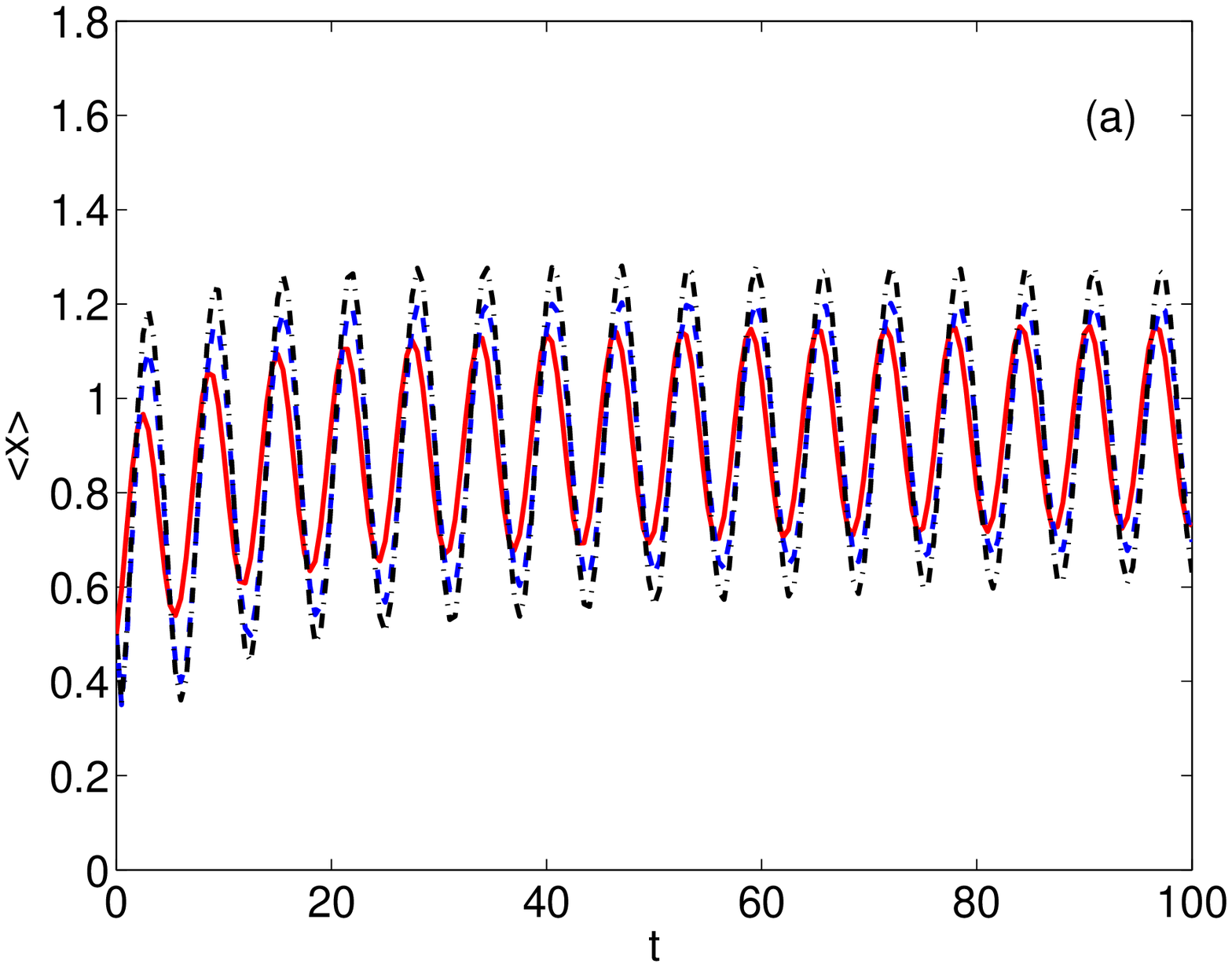}
\includegraphics[width=6cm]{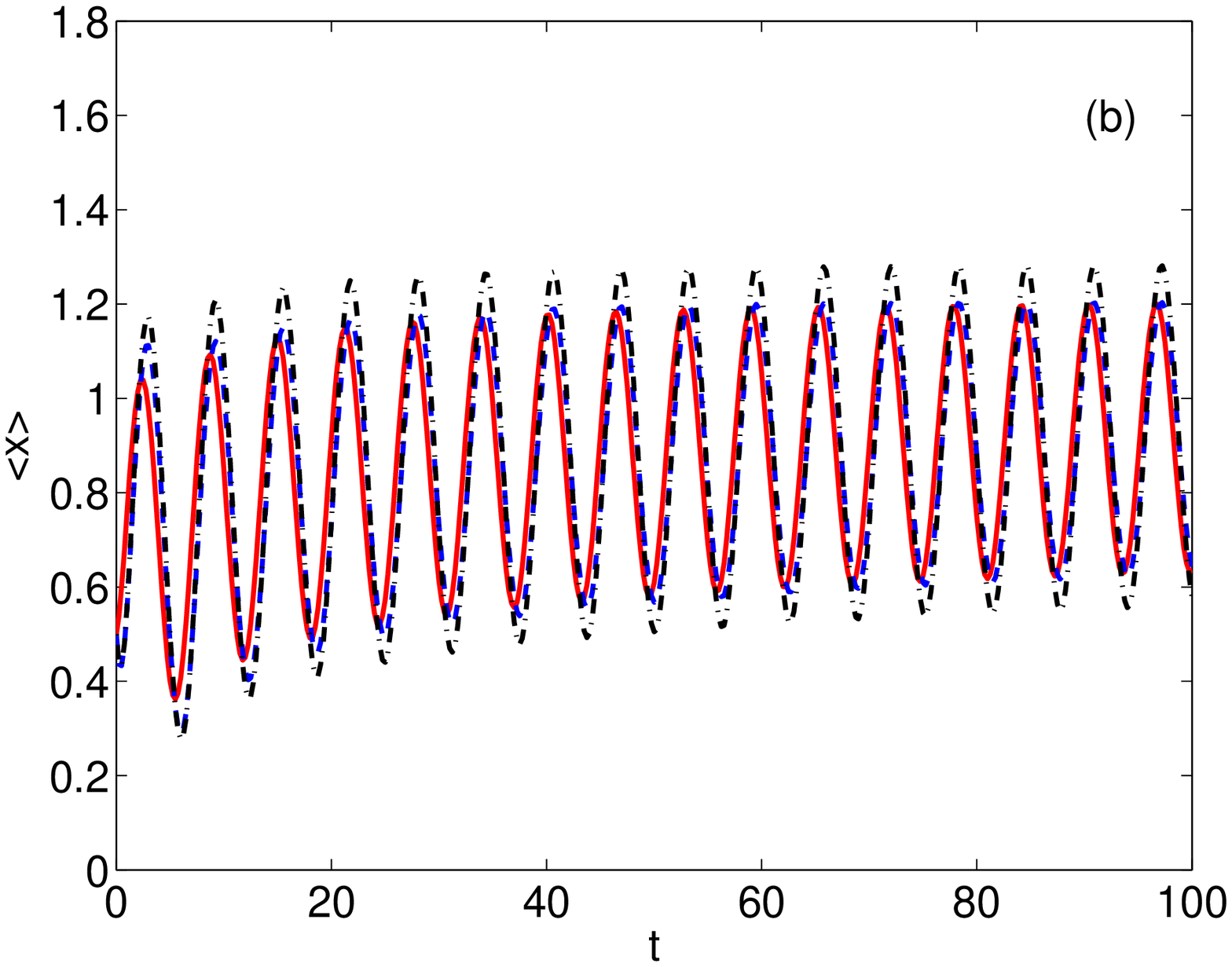}
\includegraphics[width=6cm]{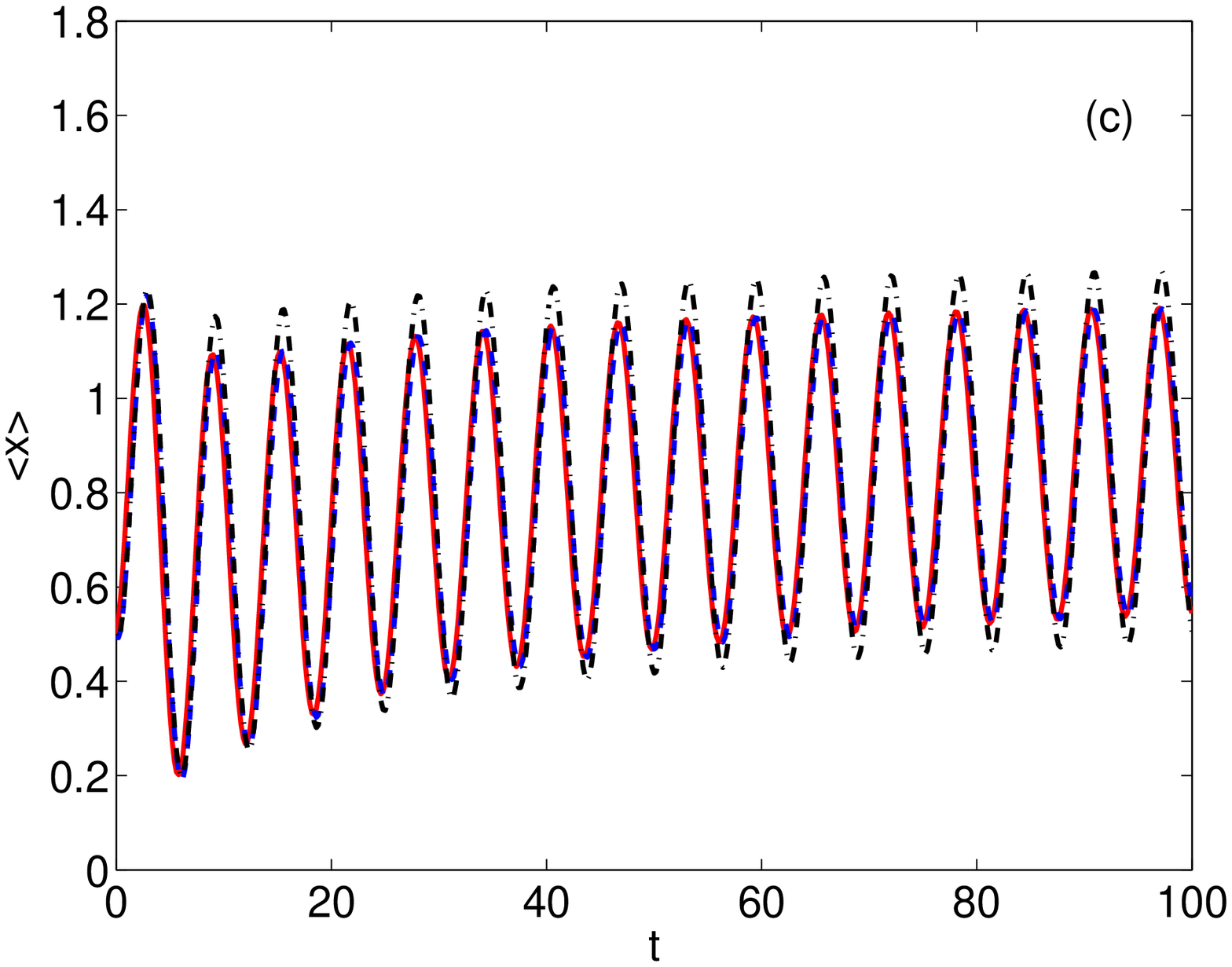}
\includegraphics[width=6cm]{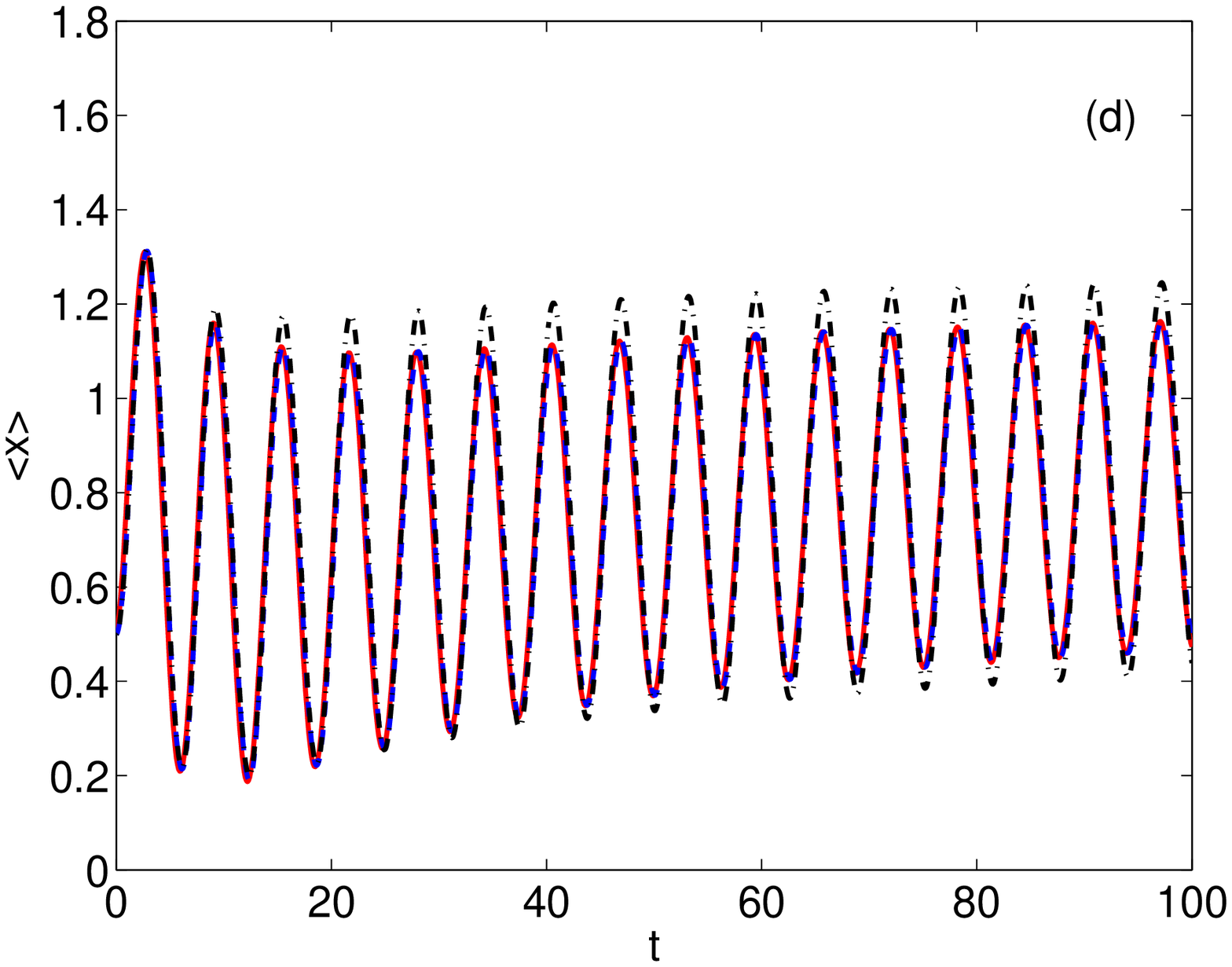}
\includegraphics[width=6cm]{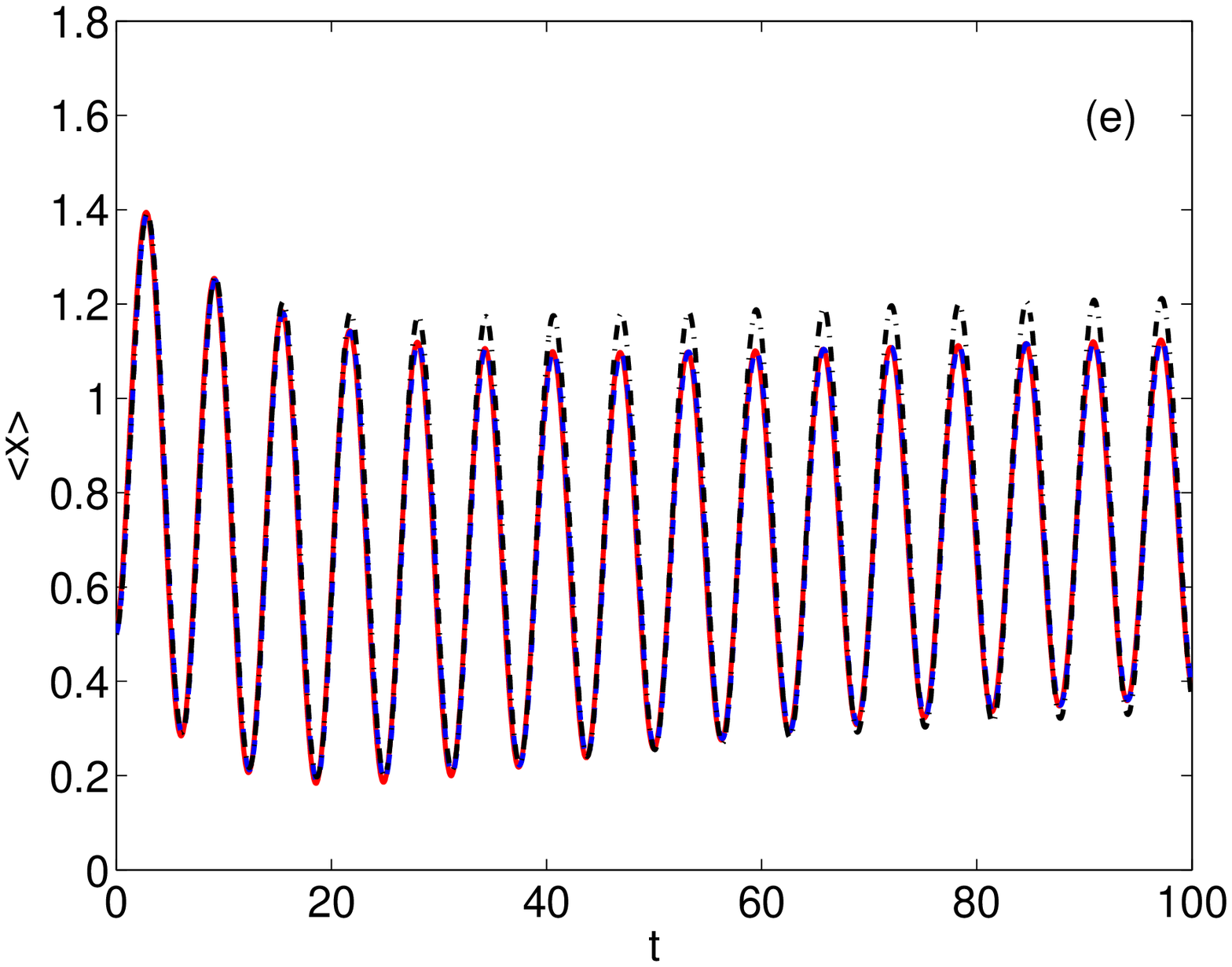}
\includegraphics[width=6cm]{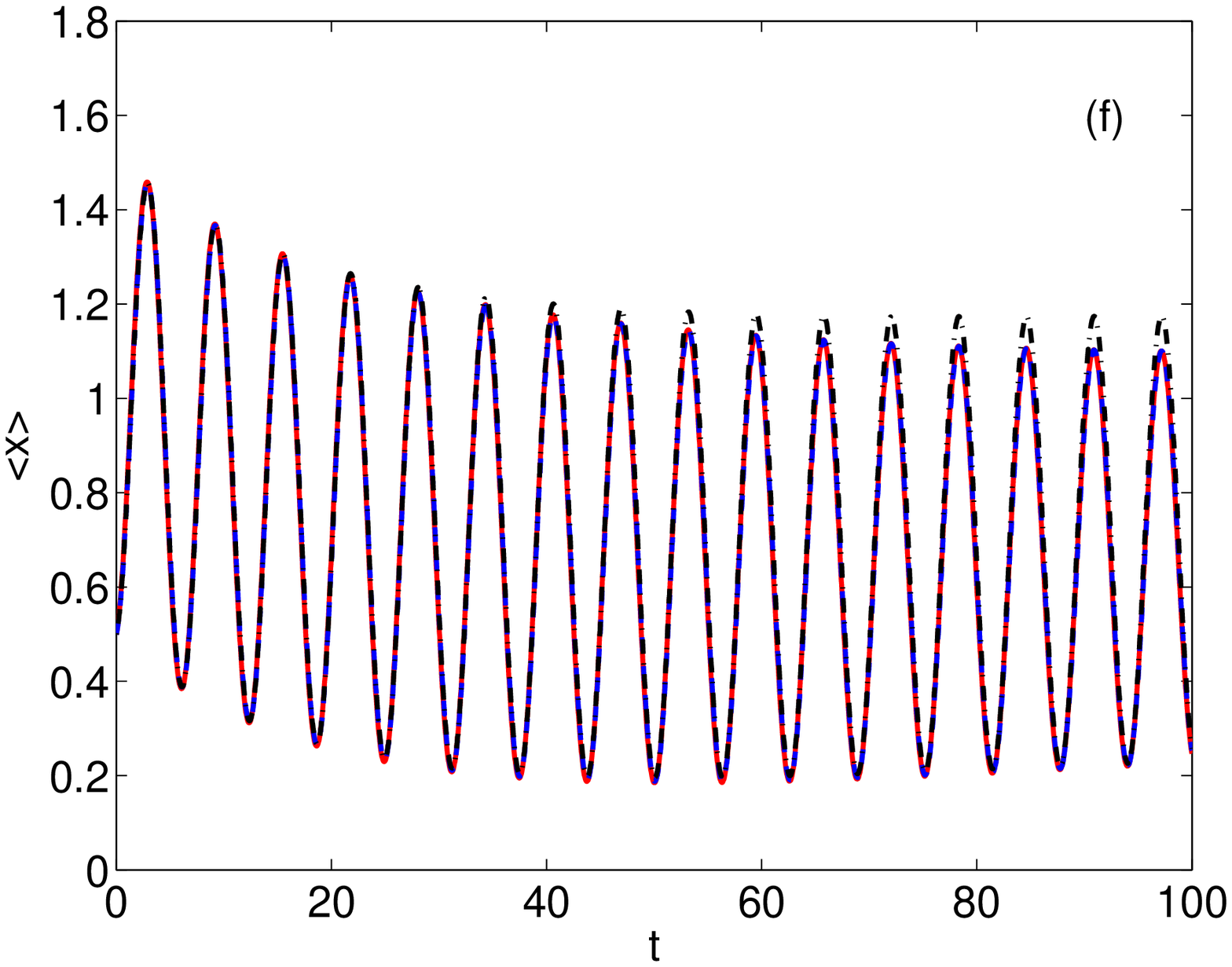}
\includegraphics[width=6cm]{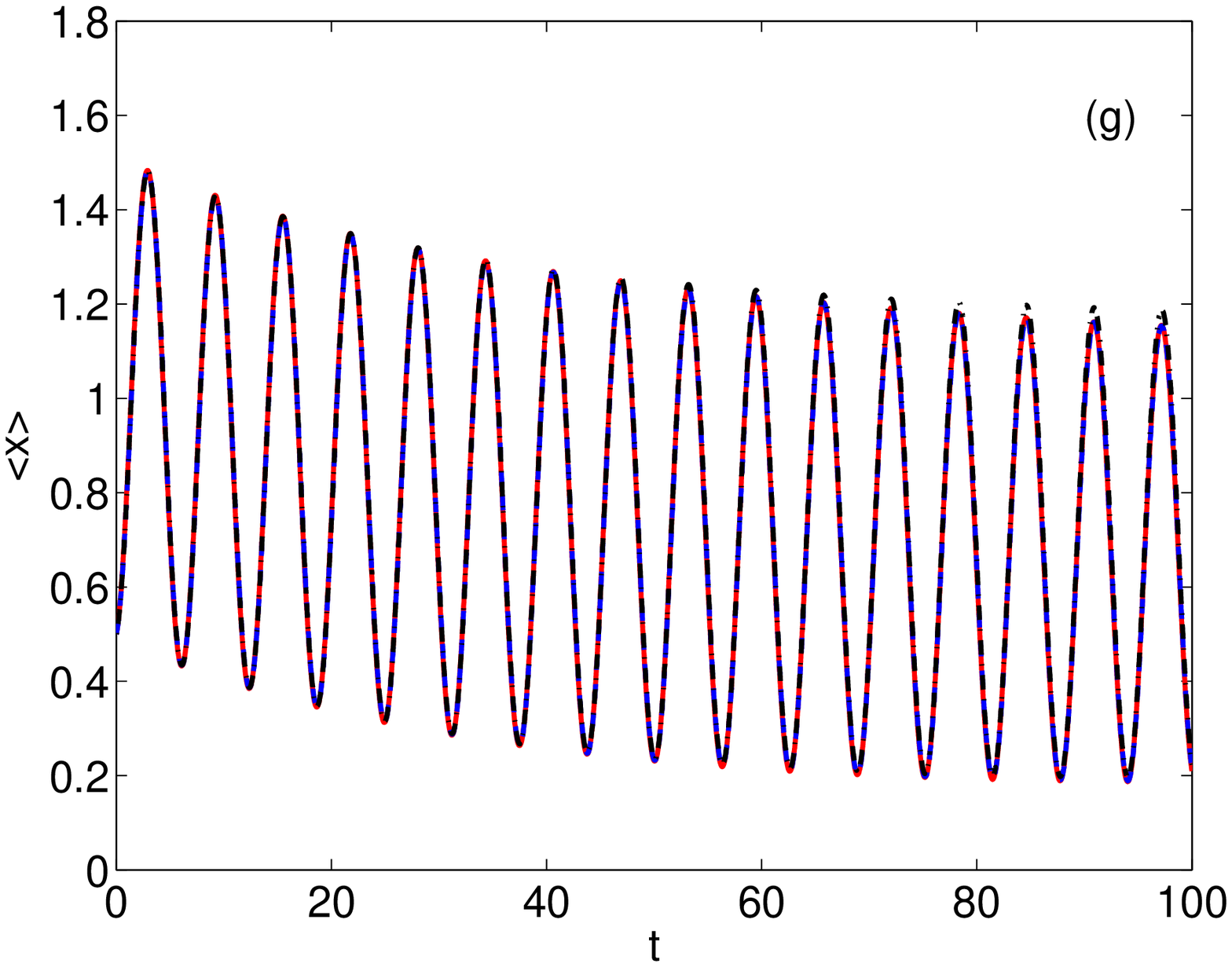}
\caption{Numerical solution of the macroscopic system (solid red line), Result 1 (dashed blue line) and Result 3 (dot-dashed black line) for the following values of $\delta$: (a) 1, (b) 0.5, (c) 0.2, (d) 0.1, (e) 0.05, (f) 0.02, (g) 0.01. Here $\theta=0.37$, initial data are $x_0=0.5$, $v_0=0.1$, $\widetilde{C_1}(0)=x_0+0.5$, $\widetilde{C_n}(0)=0$, $n\geq 2$, and the other parameter values as in figure \ref{f1}. }
\label{f2}
\end{center}
\end{figure}

Figures \ref{f1} and \ref{f2} compare the direct numerical solutions of the macroscopic equations to the approximations given in Results 1, 2 and 3 of section \ref{s3}. Near the critical temperature, figure \ref{f1} shows that the approximations given by Results 1 and 2 (averaged equations and bifurcation approximation) are excellent for sufficiently small $\delta$ (0.1 and smaller values), whereas there is a small transient that is not captured by the approximations for larger values of $\delta$. For those larger values, the approximation of the averaged equations given by bifurcation theory (Result 2) is somewhat better than the averaged equations themselves (Result 1). This is not surprising as the bifurcation equation for the averaged equations (\ref{b5a})-(\ref{b5b}) is the same as that for the full macroscopic equations as shown in \ref{app_bif}. An even better approximation can be found if the initial time layer is built by directly using the transients for the macroscopic equations. For low temperature, $\theta=0.37$ ($|\widetilde{x}_{\text{eq}}|=|\widetilde{C_{1,\text{eq}}}|\simeq 0.99$), figure \ref{f2} compares the solution of the macroscopic equations to the approximations given by Results 1 and 3 in section \ref{s3}. We observe that even the rougher approximation given by Result 3 ranks from good to excellent as $\delta$ decreases.

\begin{figure}[htbp]
\begin{center}
\includegraphics[width=6cm]{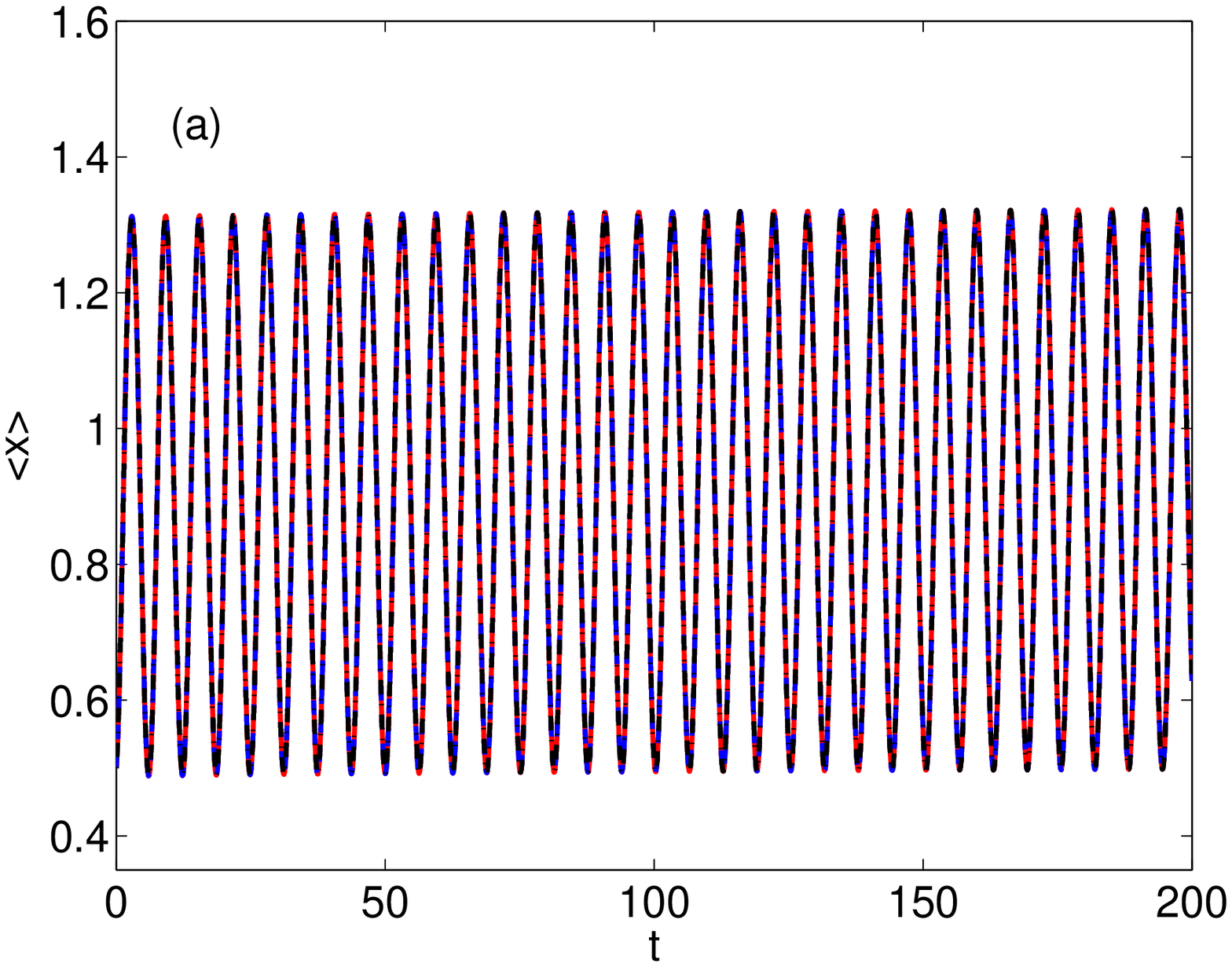}
\includegraphics[width=6cm]{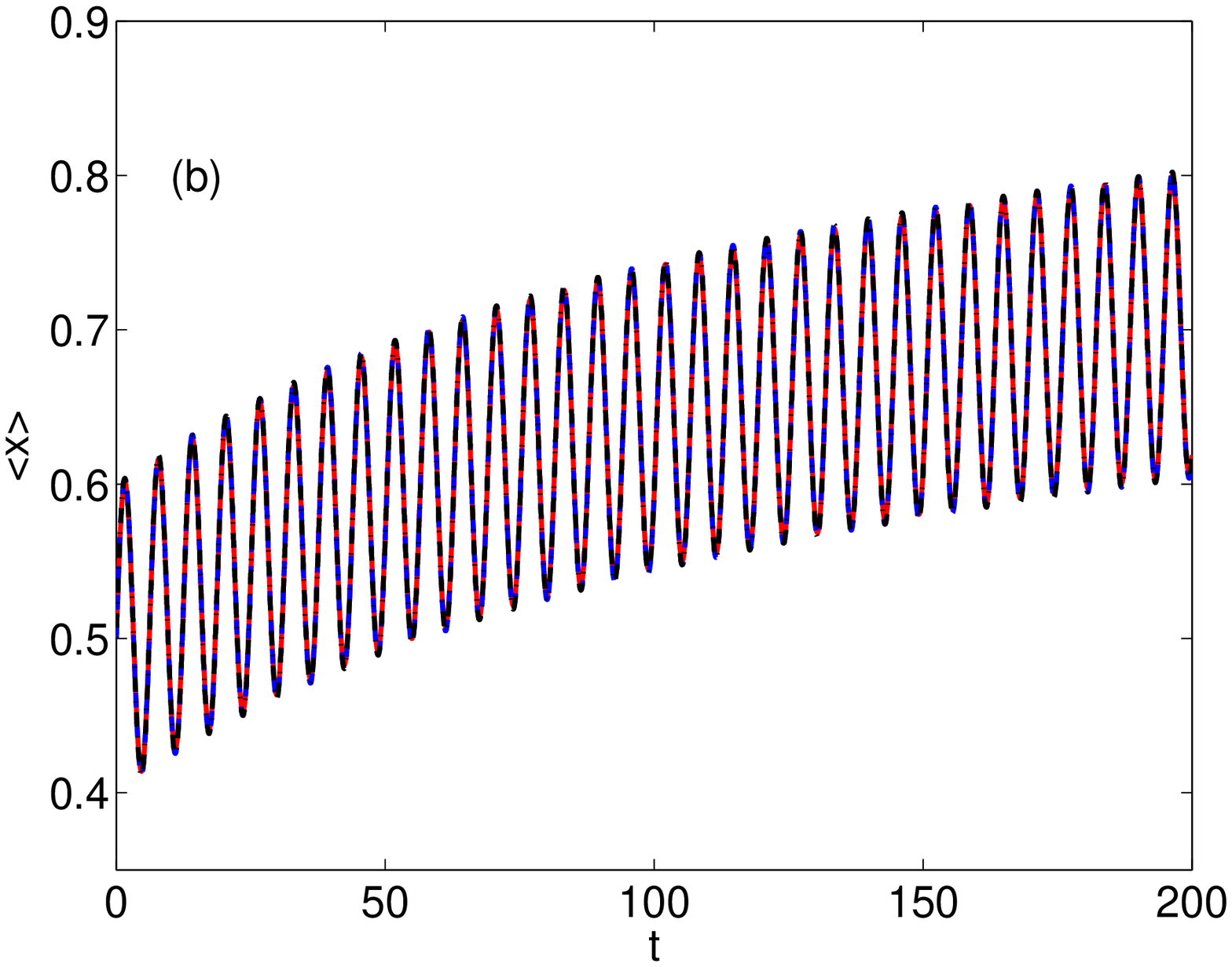}
\includegraphics[width=6cm]{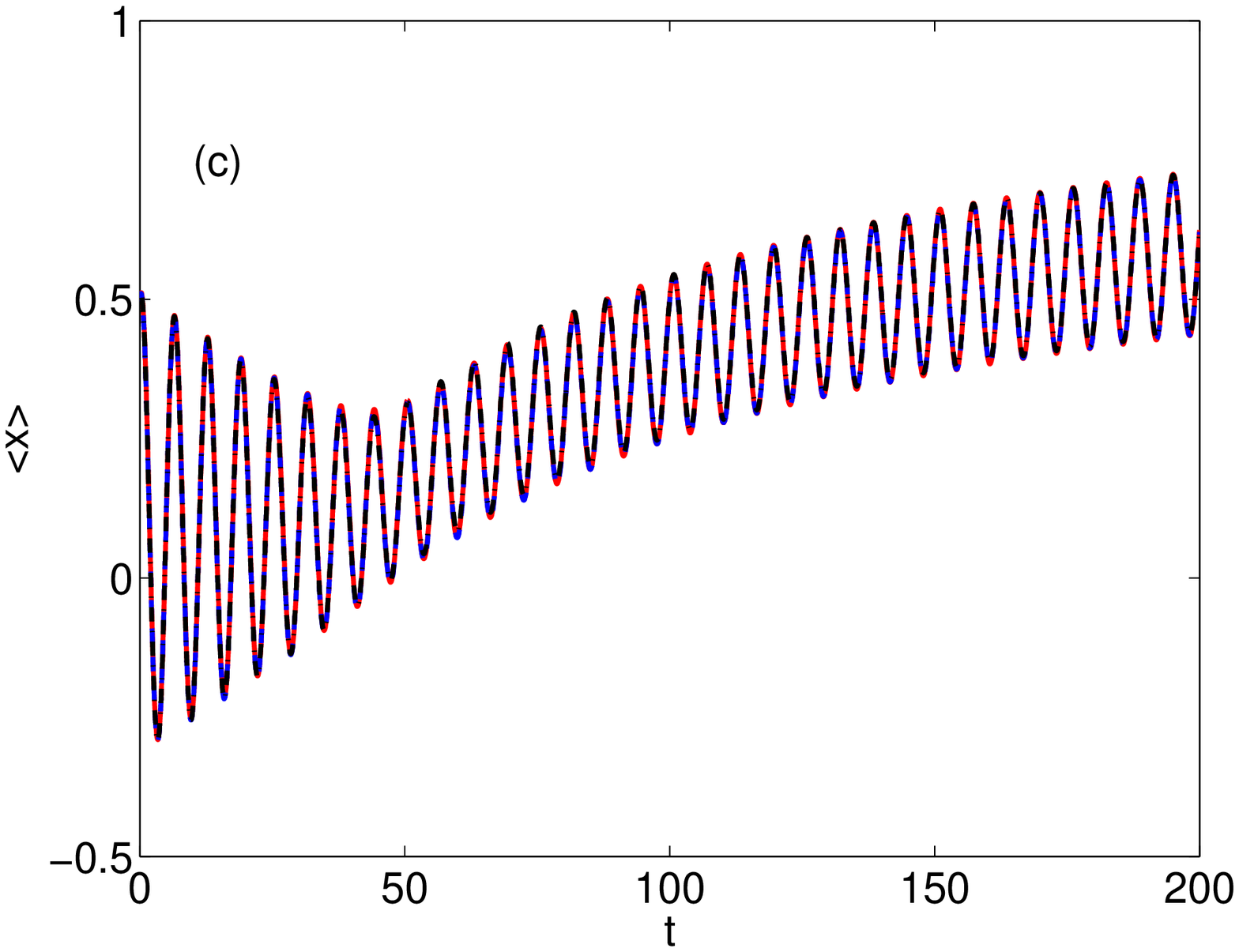}
\includegraphics[width=6cm]{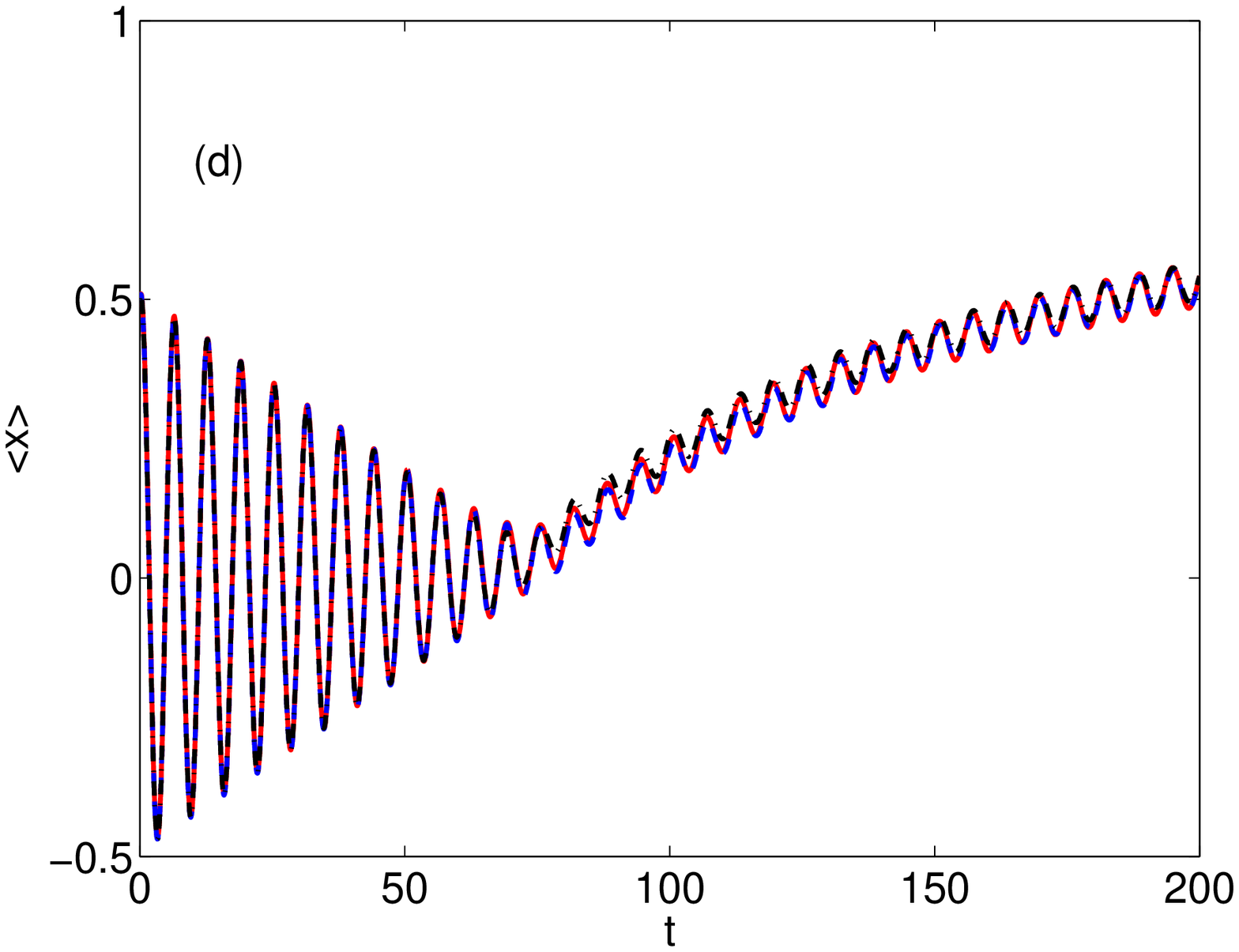}
\includegraphics[width=6cm]{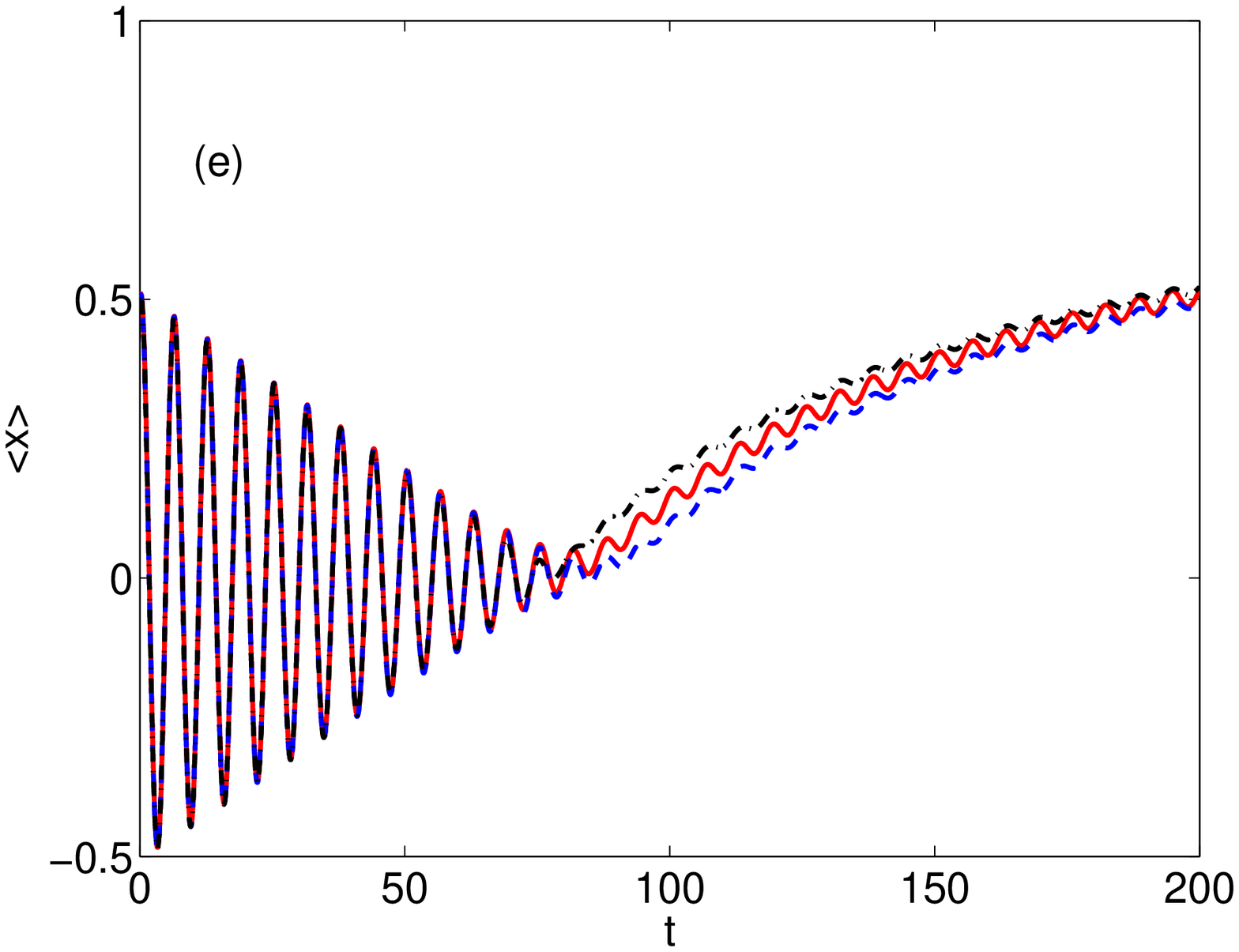}
\includegraphics[width=6cm]{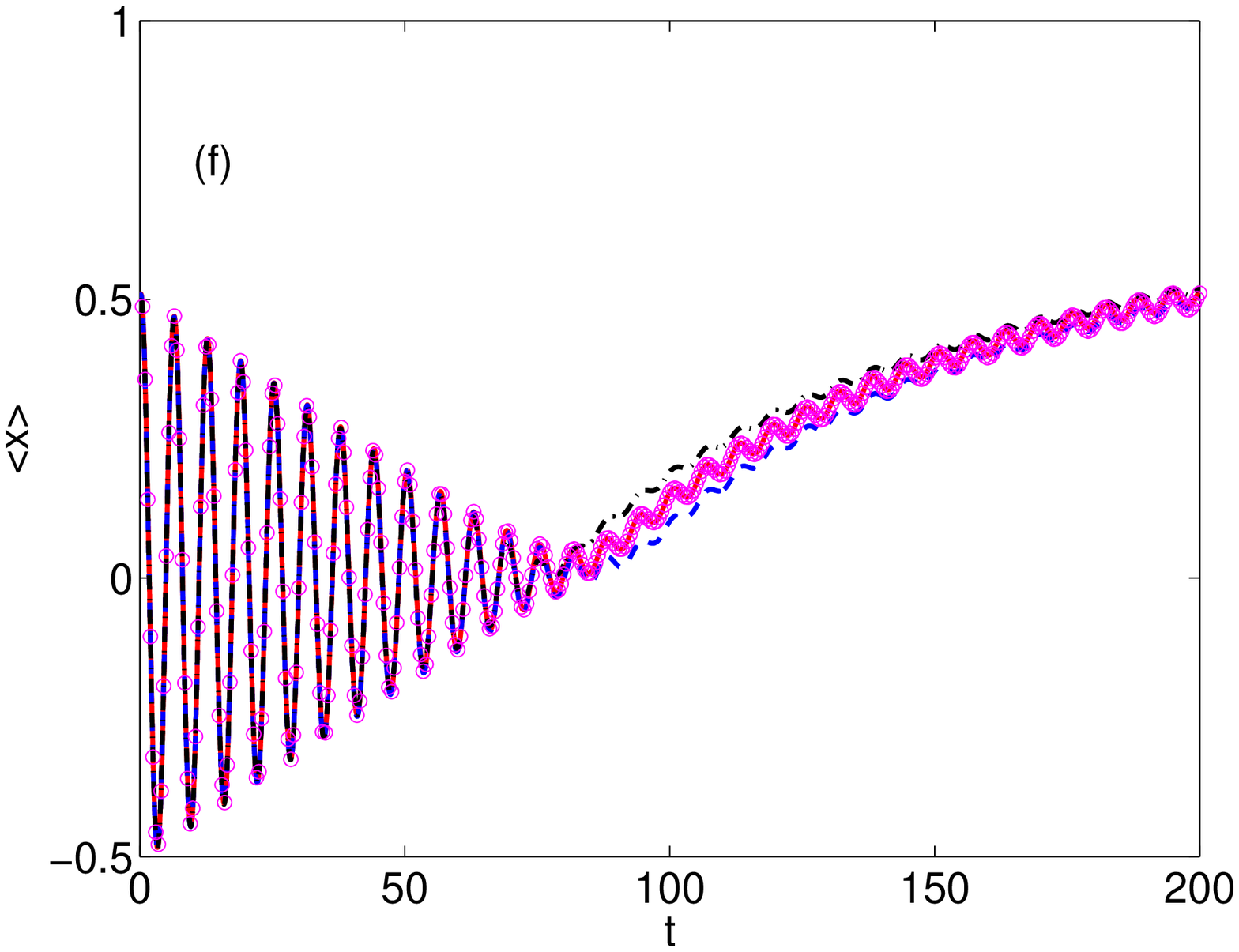}
\includegraphics[width=6cm]{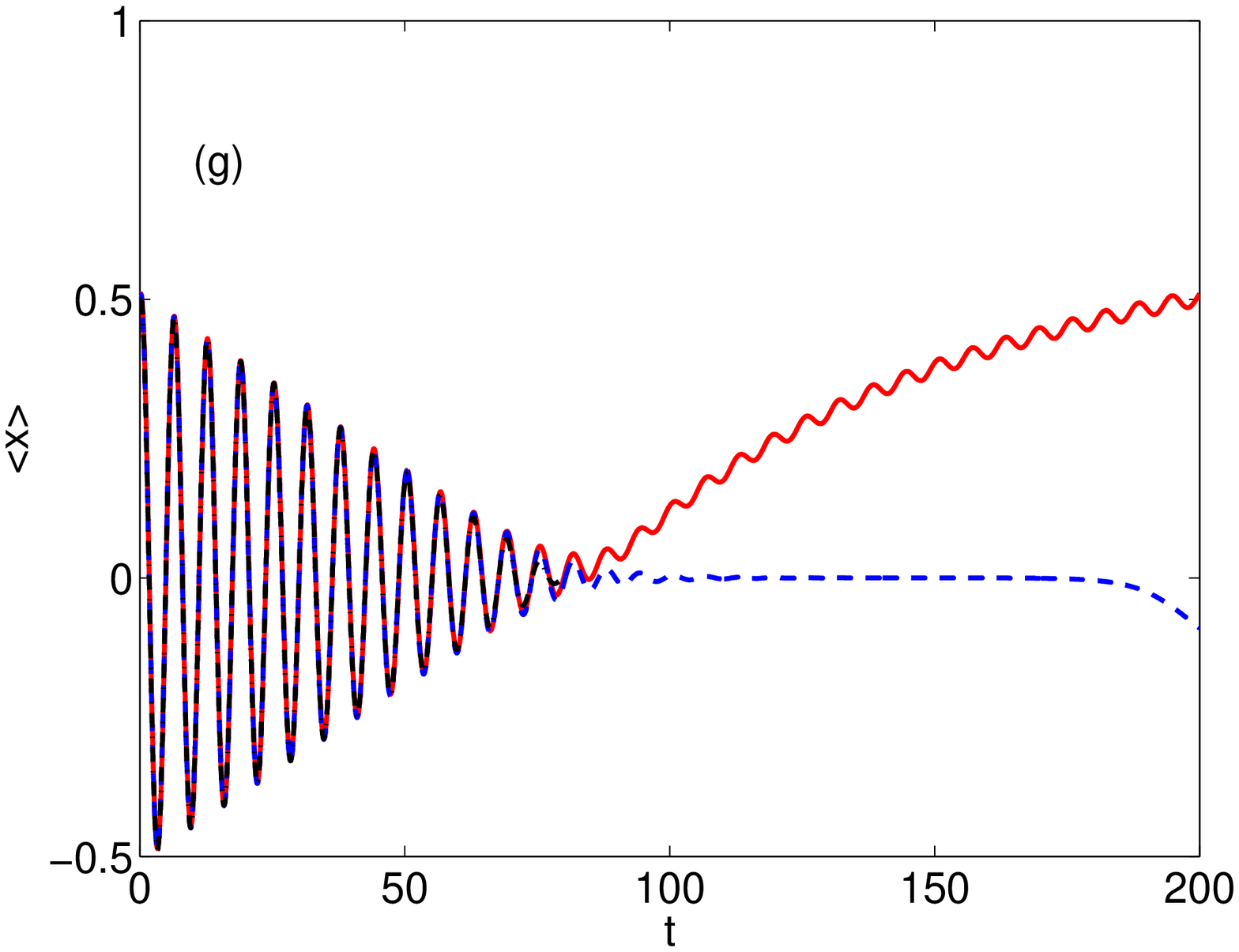}
\includegraphics[width=6cm]{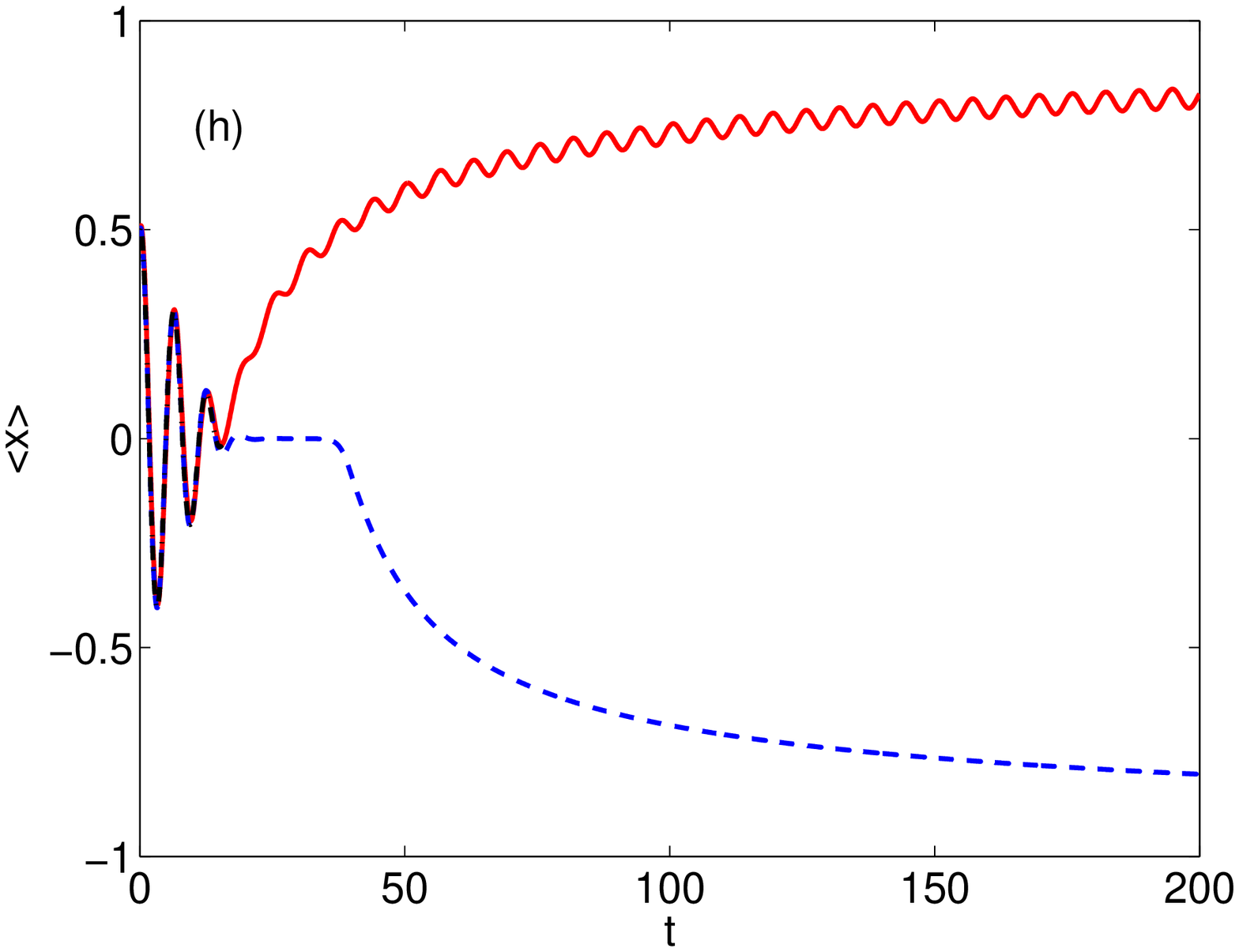}
\caption{Numerical solution of the macroscopic system (solid red line), Result 1 (dashed blue line) and Result 3 (dot-dashed black line) for $\theta=0.1$, $\delta=0.01$, $x_0=0.5$, $v_0=0.1$, $\widetilde{C_1}(0)=r^n$ with $r=$: (a) 0.9, (b) 0.5, (c) 0.1, (d) 0.01, (e) 0.001, (f) same as (e) but including results of stochastic simulations (magenta circles), (g) 0. (h) corresponds to $\delta=0.05$ and $r=0$. Note that the approximation given by (\ref{b3a})-(\ref{b3b}) (Result 3) breaks down when $R=0$ and the corresponding line is interrupted in (g)-(h) after the breakdown time. }
\label{f3}
\end{center}
\end{figure}

Next, we probe the dependence of our approximation on initial conditions. In figure \ref{f3}, we fix $\theta=0.1$ and $\delta=0.01$, and use equilibrium-like initial correlations $\widetilde{C_1}(0)=r^n$ for different values of $r$ that range from very small to near 1. We find that Results 1 and 3 approximate very well the solution of the macroscopic equations except for very small $r$. As $r$ decreases, the system first tends towards the unstable solution $\widetilde{x}=0$ (attempting to stabilize it) and then it crosses over to the basin of attraction of the stable equilibrium with $\widetilde{x}$ close to 1. For $r=0$ (or $r\ll 1$), the approximations given by Results 1 and 3 indicate stabilization of $\widetilde{x}=0$ at the time (\ref{b3c}), i.e.\ $t_0=\tau_0/\delta= 80.09$. This is confirmed by figure \ref{f3}(g) and it agrees with the fact that $C_n^{(0)}\equiv 0$ ($n\geq 1$), $R=R(\tau)$ is an invariant stable manifold of the averaged equations (\ref{b5a})-(\ref{b5b}). For the case $r=0$, $R$ and $C_n^{(0)}$ should remain zero. However, figures \ref{f3}(g) and (h) show that numerical errors build up and send $R(\tau)$ given by the solution of the averaged equations (\ref{b5a})-(\ref{b5b}) to one of the stable equilibrium values (in the case of figures \ref{f3}(g) and (h) to the wrong one, near $\widetilde{x}=-1$) after some time $\tau_1>\tau_0$.  Since zero correlations do not correspond to an invariant stable manifold of the macroscopic equations (\ref{2.28})-(\ref{2.30}), the solutions thereof depart in the vicinity of zero and after some time from the solutions of the averaged system (\ref{b5a})-(\ref{b5b}), cf figures \ref{f3}(g)-(h).

\begin{figure}[htbp]
\begin{center}
\includegraphics[width=6cm]{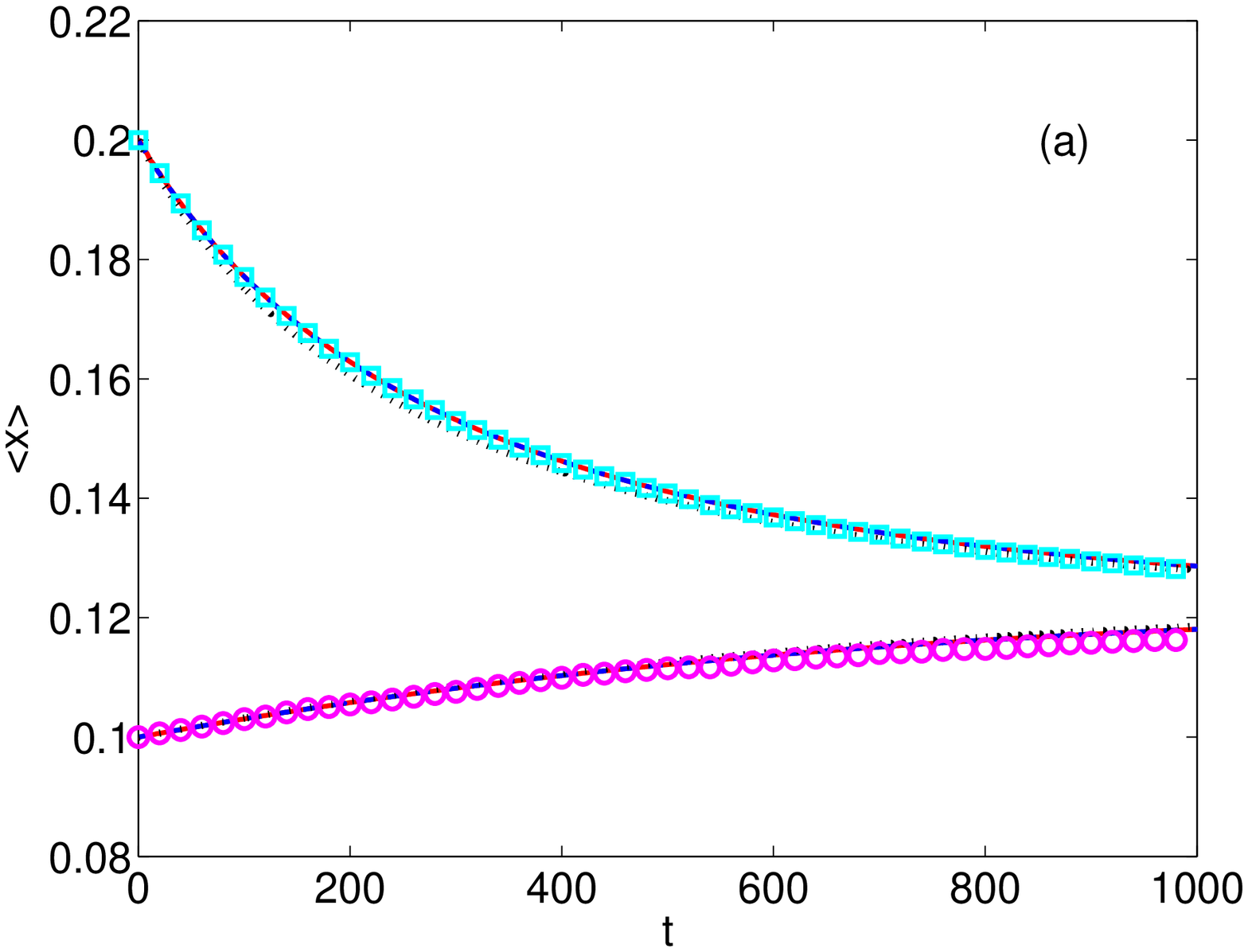}
\includegraphics[width=6cm]{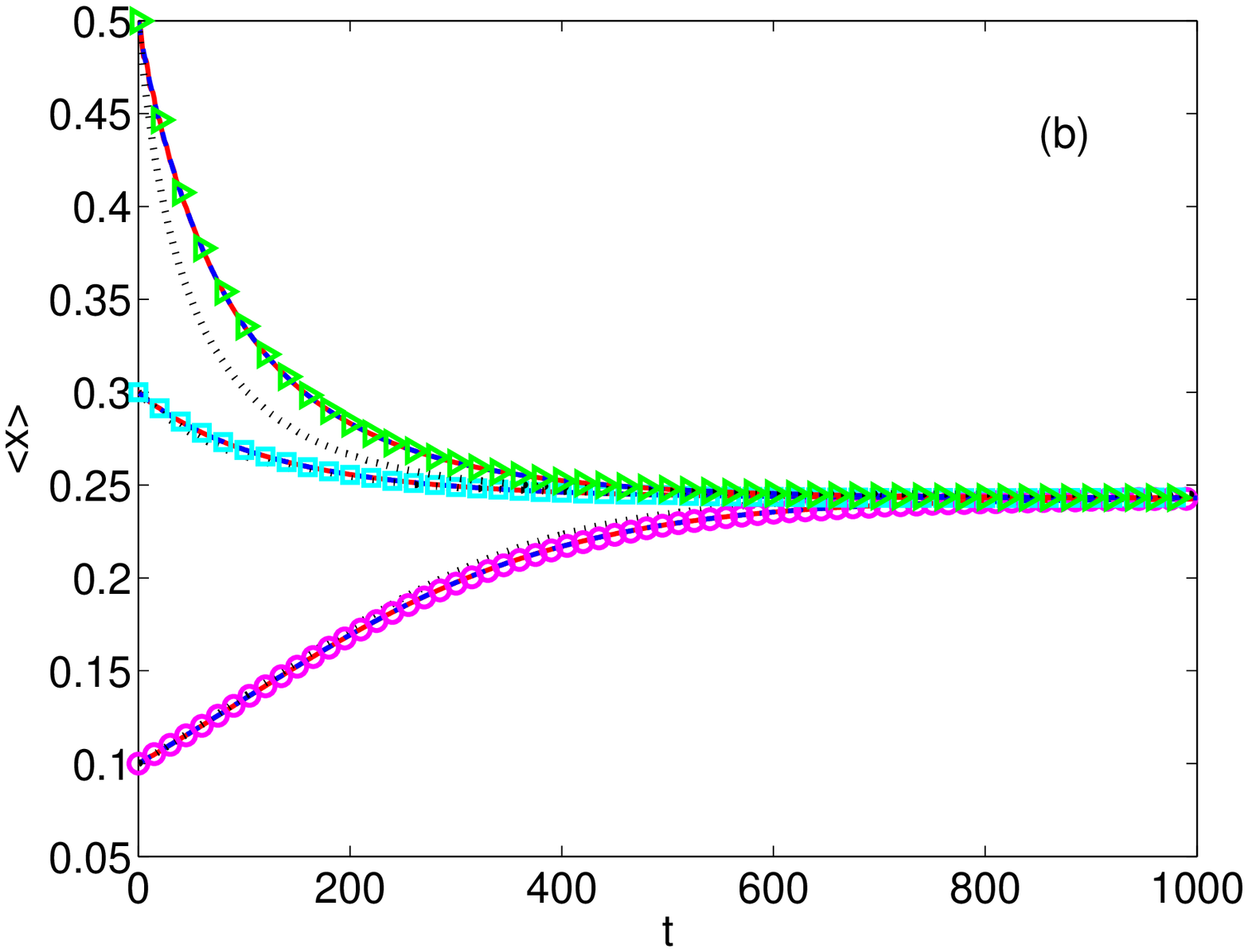}
\caption{Averaged trajectories $\langle x(t)\rangle=\widetilde{x}(t)$ for: (a) $\theta=0.995$, $v_0=0$, $N_T=10^3$, $\widetilde{C_1}(0)=x_0=0.2$ (upper curve, cyan squares) and $\widetilde{C_1}(0)=x_0=0.1$ (lower curve, magenta circles); (b) $\theta=0.98$, $N_T=10^3$ and from top to bottom, $\widetilde{C_1}(0)=x_0=0.5$ (green triangles), $0.3$ (cyan squares), $0.1$ (magenta circles). Other parameter values are $\delta=0.1$, $N=10^6$. Solid red line: macroscopic equations (\ref{2.28})-(\ref{2.29}), dashed blue line: averaged equations (\ref{b5a})-(\ref{b5b}), dotted black line: bifurcation theory (\ref{b1a})-(\ref{b1c}). Most of these theoretical predictions are indistinguishable in the scale of the figure.}
\label{fig1}
\end{center}
\end{figure}

\begin{figure}[htbp]
\begin{center}
\includegraphics[width=9cm]{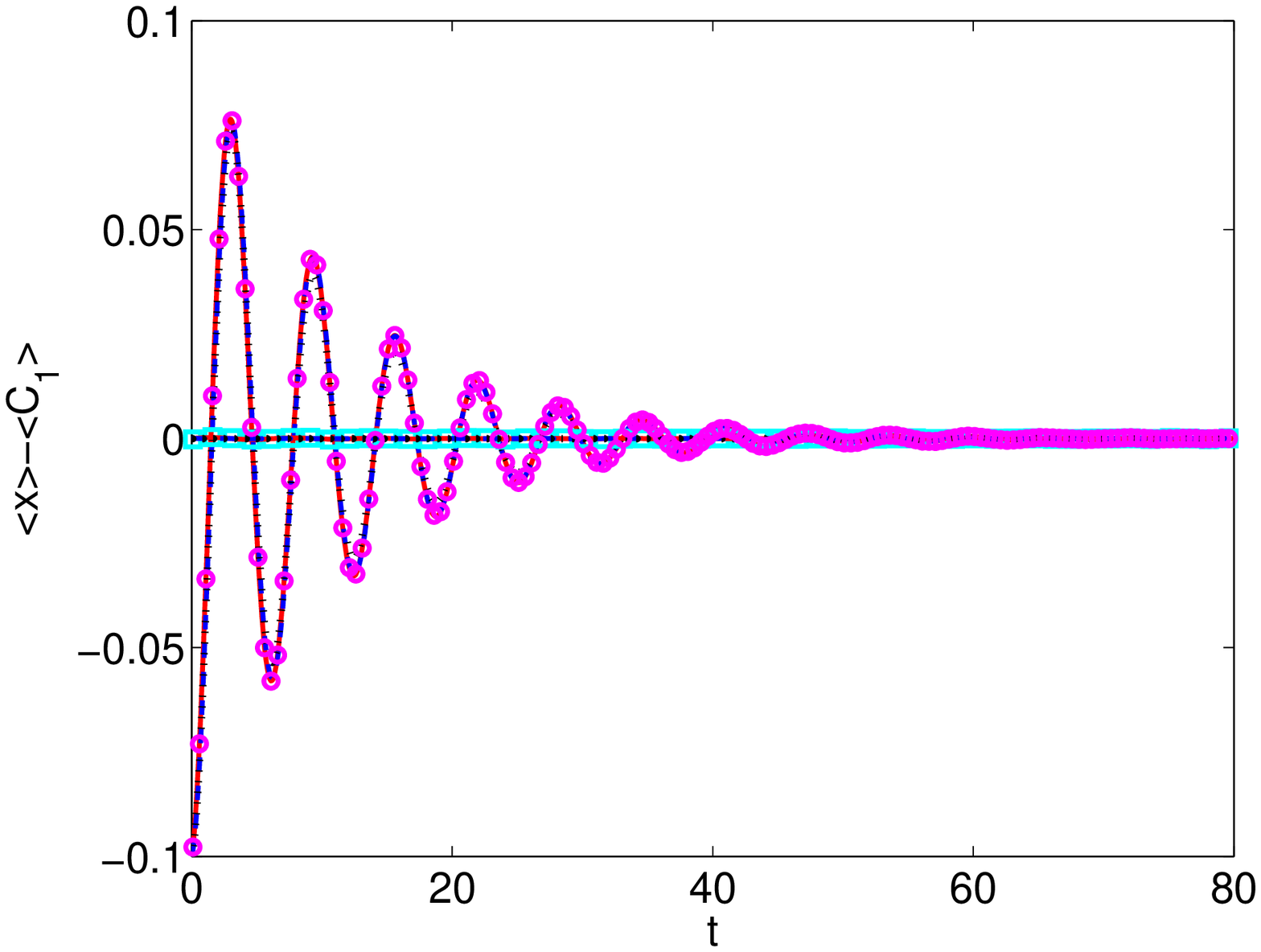}
\caption{Difference $\langle x(t)\rangle-\langle C_1(t)\rangle=\widetilde{x}(t)-\widetilde{C_1}(t)$ for $\widetilde{C_1}(0)=0.2$, $x_0=0.1$, $v_0=0$, $\theta=0.995$, $\delta=0.1$, $N=10^6$, $N_T=100$. Stochastic equations (magenta circles and cyan squares), macroscopic system (solid red line), Result 1 (dashed blue line) and Result 2 (dot-dashed black line). Also plotted $\widetilde{x}(t)-\widetilde{C_1}(t)=0$ for $x_0=0.2$. All predictions are indistinguishable in the scale of the figure.}
\label{fig2}
\end{center}
\end{figure}

\subsection{Comparison with numerical simulations of the oscillator-spin system}

We now compare our approximate theories to direct numerical simulations of the oscillator-spin system with Glauber dynamics. We choose $\omega=10$ and $\alpha=1$, so that $\delta=0.1$. We have carried out simulations for a temperature $\theta=0.995$, very close to the critical temperature (figure \ref{fig1}(a)), which corresponds to $\epsilon=\sqrt{0.005}=0.07$. A large number of particles must be considered in the simulations, because of the divergence of the fluctuations at the critical temperature. Namely, the number of spins is $N=10^6$ and we have averaged over $N_T=1000$ trajectories. The initial conditions have been chosen such that $x_0=\widetilde{C_1}(0)$ and $v_0=0$. Therefore, we expect that $R(\tau)=0$ for all times, and $x(t)=C_1(t)$, both of them relaxing to their equilibrium value $\widetilde{x}_{\text{eq}}=\widetilde{C_{1,\text{eq}}}\simeq 0.122$. Two different initial values of $x(0)$ have been considered, one above the equilibrium value $x_0=0.2$, and one below it, $x_0=0.1$. In both cases, the simulation curves agree with the theoretical prediction given by (\ref{b1a}). For a lower temperature, $\theta=0.98$ the theory still gives a good description of the simulation results while the initial conditions are not too big (figure \ref{fig1}(b)). This value of $\theta$ corresponds to $\epsilon=\sqrt{0.02}=0.14$, which is not so small (in fact, the steady value for the oscillator position is $0.24$, and its maximum value is unity).
A further test of the theory is done in figure \ref{fig2}, for the same values of the parameters as in figure \ref{fig1}(a), but $x_0\neq \widetilde{C_1}(0)$.
The initial conditions are $\widetilde{C_1}(0)=0.2$ but $x_0=0.1$ with vanishing initial velocity, so that we obtain from (\ref{b7})
\begin{equation}\label{b30}
    R(\tau)=0.1 e^{-\tau} \, , \qquad \phi=-\frac{\pi}{2}   \,\Rightarrow \widetilde{x}(t)-\widetilde{C_1}(t)=-0.1 e^{-\tau} \cos t \, .
\end{equation}
In figure \ref{fig2}, we observe that the simulations of the macroscopic equations, the averaged equations (Result 1) and bifurcation theory (Result 2) all follow exactly (\ref{b30}) and agree very well with direct stochastic simulations.

\begin{figure}[htbp]
\begin{center}
\includegraphics[width=9cm]{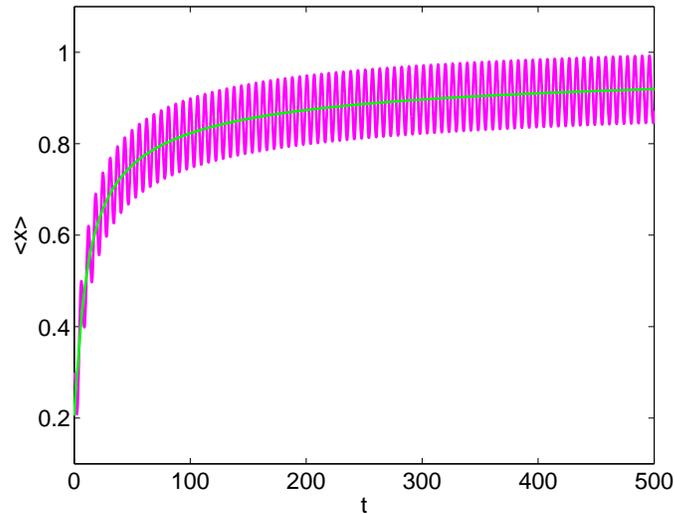}
\caption{Averaged trajectories $\langle x(t)\rangle=\widetilde{x}(t)$ (magenta) and $\langle C_1(t)\rangle=\widetilde{C_1}(t)$ (green) for $r=0.2$, $x_0=0.25$, $v_0=0$, $\theta=0.37$, $\delta=0.1$, $N=10^5$, $N_T=10^2$. }
\label{fig3}
\end{center}
\end{figure}

\begin{figure}[htbp]
\begin{center}
\includegraphics[width=6cm]{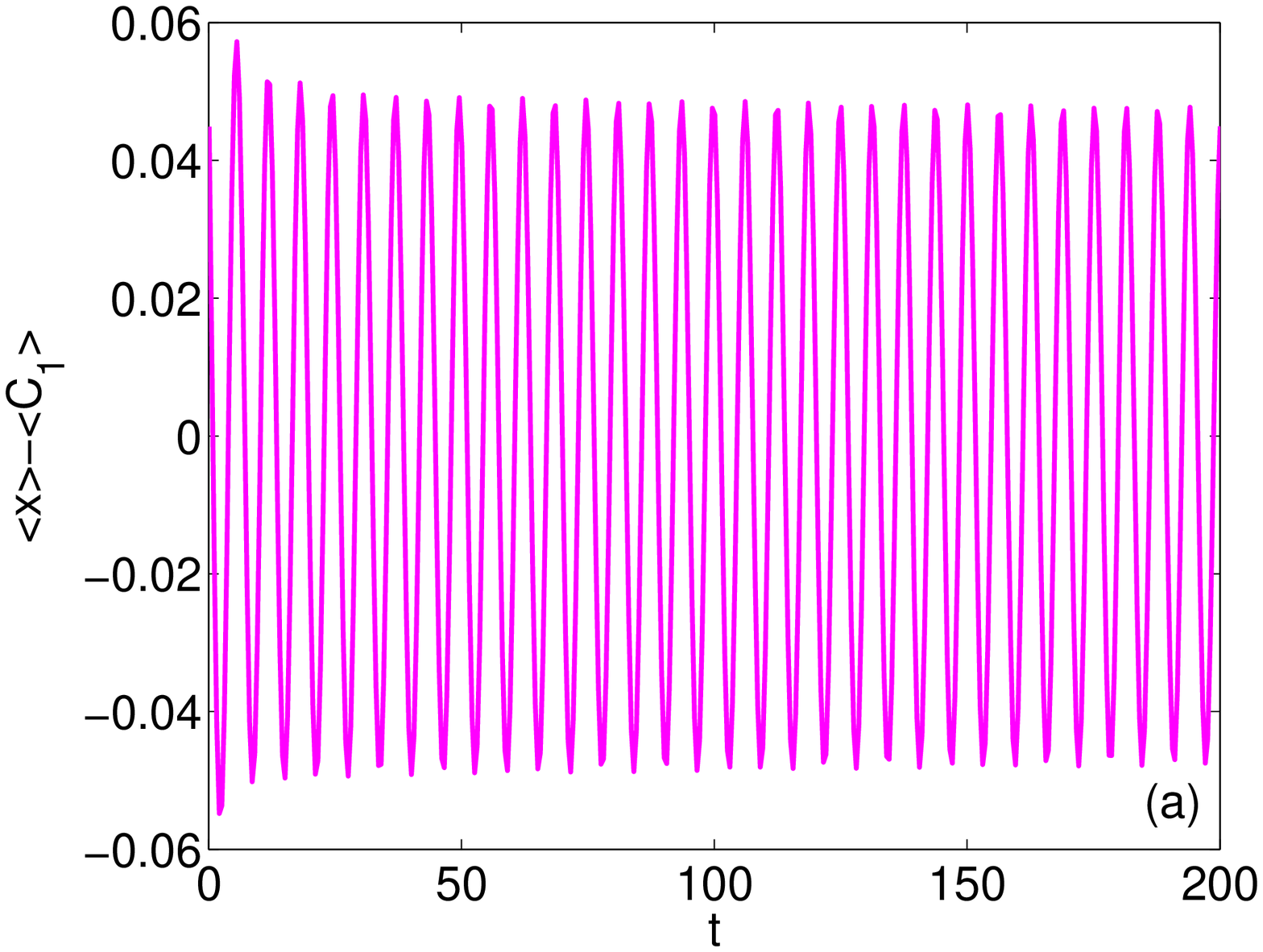}
\includegraphics[width=6cm]{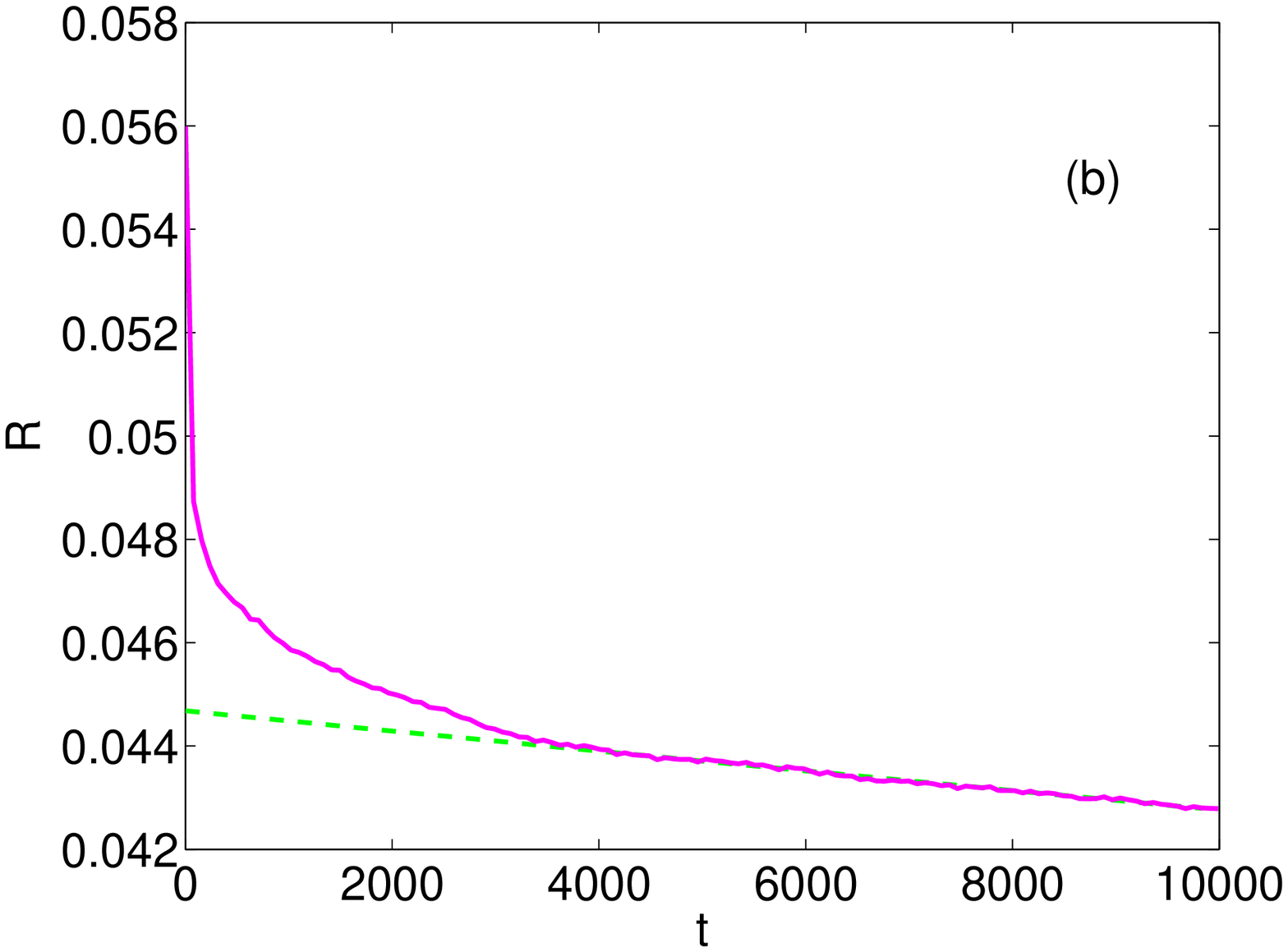}
\caption{(a) Difference $\langle x(t)\rangle-\langle C_1(t)\rangle=\widetilde{x}(t)-\widetilde{C_1}(t)$ from stochastic simulations with $r=0.2$, $x_0=0.25$, $v_0=0$, $\theta=0.37$, $\delta=0.1$, $N=10^5$, $N_T=10^2$. (b) Amplitude $R$ (log-scale) vs time $t$. The dashed line is the best long time ($t\geq 5000$) exponential decay fit, $R=0.045 \exp(-4.39\times 10^{-6} t)$. This is consistent with $R\propto \exp(-4.36\times 10^{-6} t)$ obtained from (\ref{5.7}) for very long times ($\tau=\delta t\geq 500$).}
\label{fig4}
\end{center}
\end{figure}

Considering now equilibrium-like initial conditions for the Ising system, i.e.\ $\widetilde{C_n}(0)=r^n$ with $0\leq r \leq 1$, the low temperature correlations are given by (\ref{5.8}). In figure \ref{fig3}, we have chosen parameter values $\theta=0.37$, $\delta=0.1$ and initial conditions $x_0=0.25$, $v_0=0$, $r=0.2$. In this way, as $x_0-\widetilde{C_1}(0)$ is nonzero, we expect to observe the damped oscillations of $\widetilde{x}(t)$ around $\widetilde{C_1}(t)$. The theoretical curve for $\widetilde{C_1}(t)$ is not plotted, because it is indistinguishable from the numerical one, and the oscillations of $\widetilde{x}(t)$ around $\widetilde{C_1}(t)$ are clearly observed. In figure \ref{fig4}(a), the difference $\widetilde{x}(t)-\widetilde{C_1}(t)$ is shown, for the same values of the parameters as in figure \ref{fig3}. We observe clear modulated oscillations with an envelope $R$ that decreases over a very long time scale, as shown in figure \ref{fig4}(b) (note that the time scale thereof is much longer than that in figure \ref{fig4}(a)). The long time behavior of $R$ is very well described by the exponential decay in equation (\ref{5.7}).

\begin{figure}[htbp]
\begin{center}
\includegraphics[width=6cm]{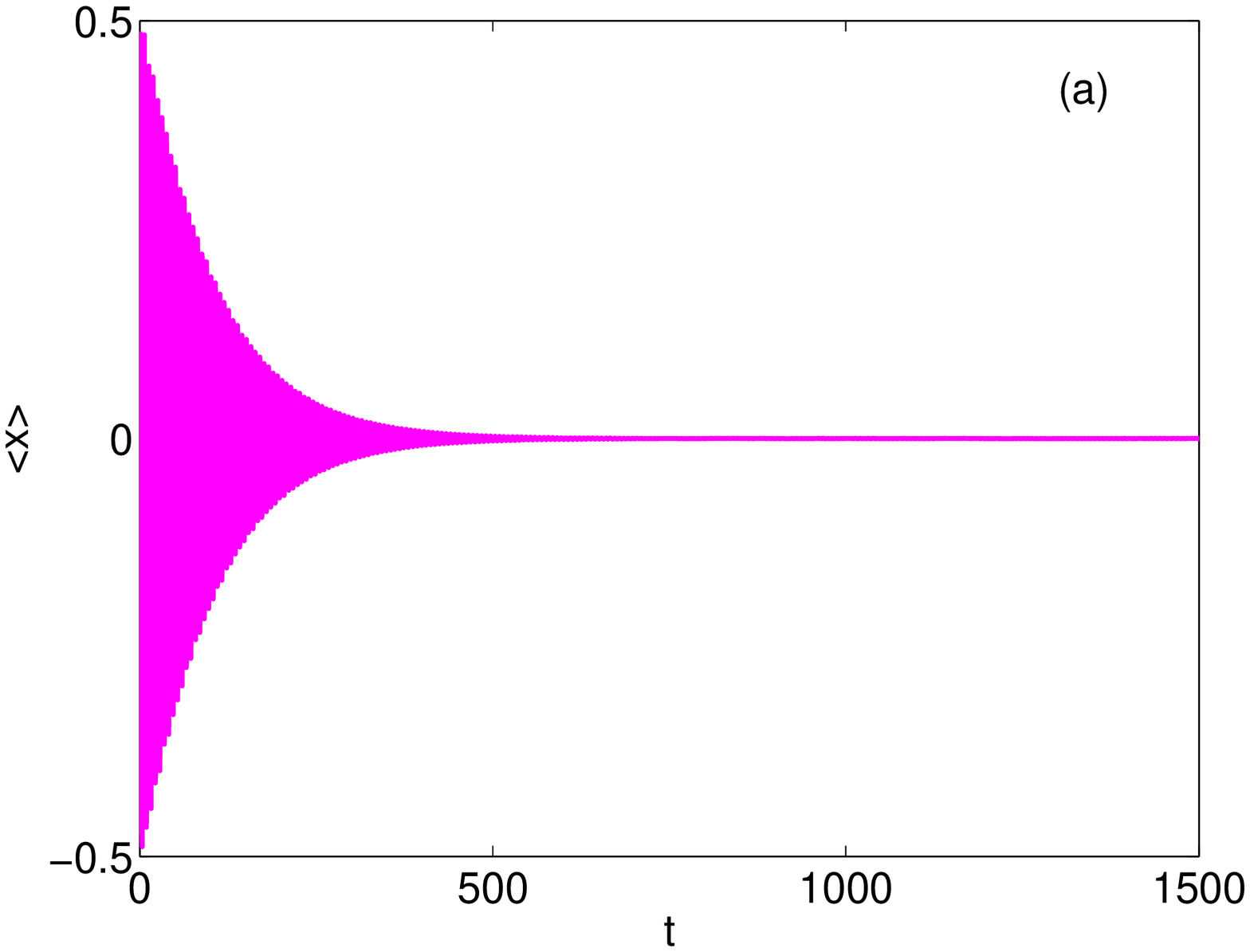}
\includegraphics[width=6cm]{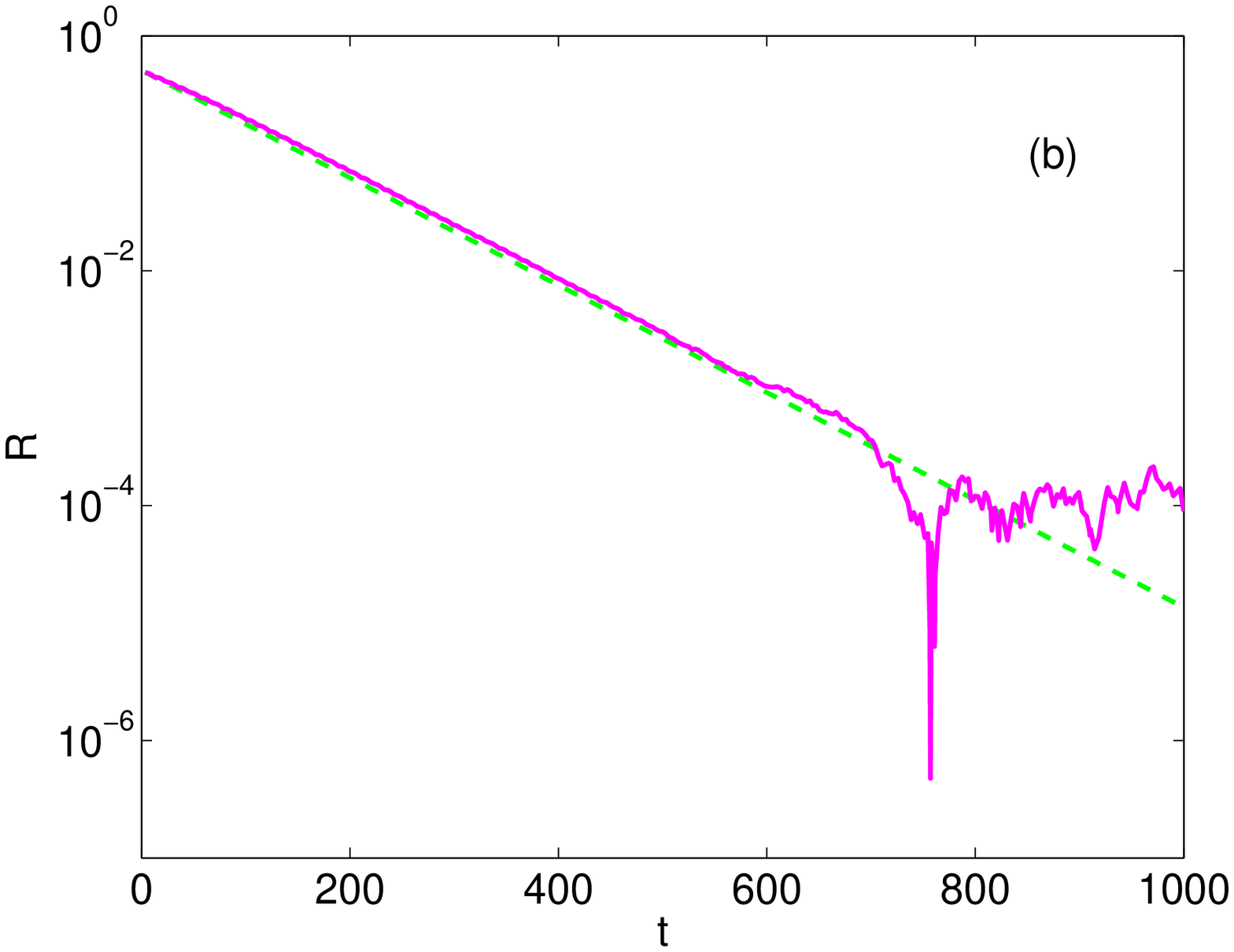}
\includegraphics[width=6cm]{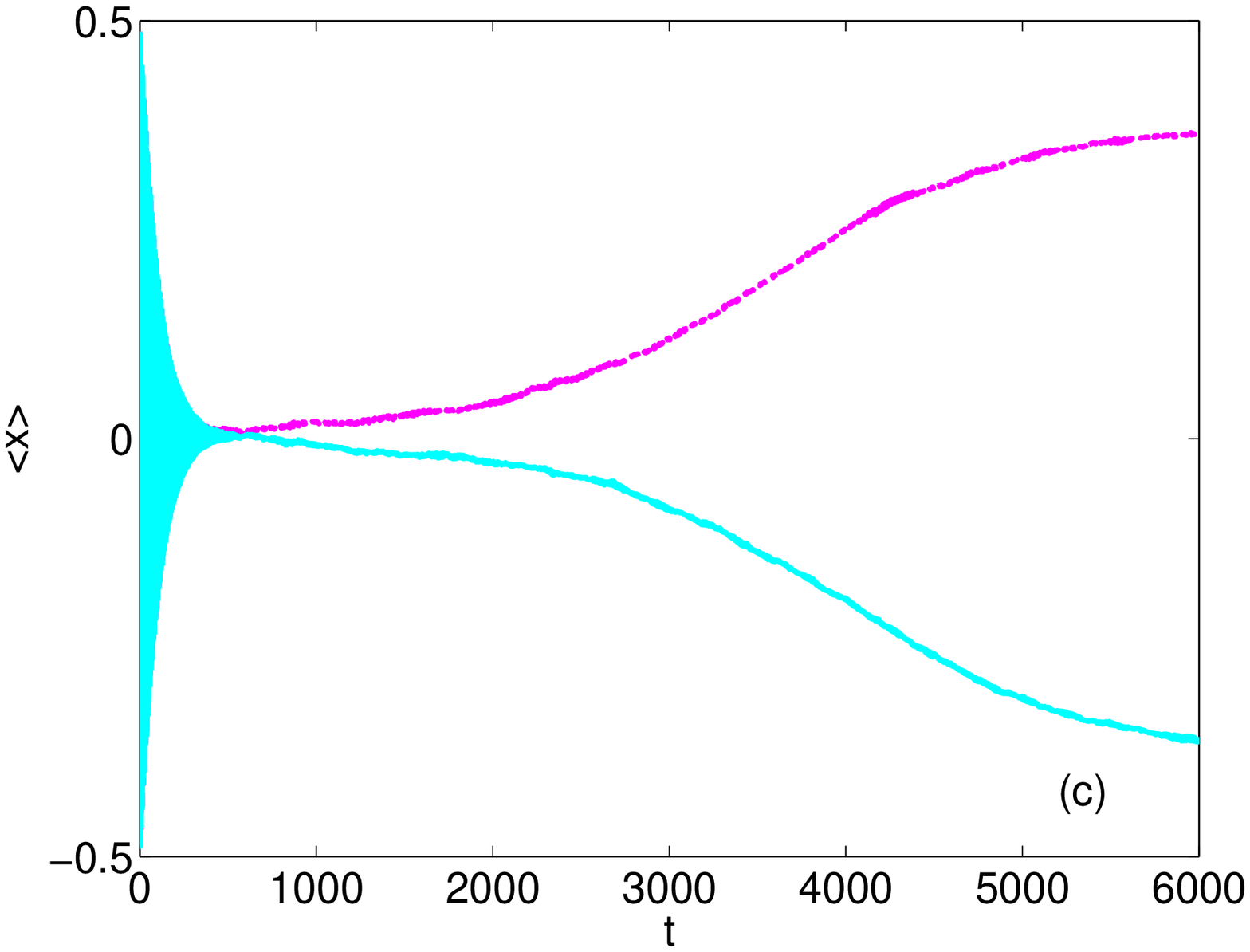}
\caption{(a) Oscillator position $\langle x(t)\rangle$ for $r=0$, $x_0=0.5$, $v_0=0$, $\theta=0.95$, $\delta=0.01$, $N=10^8$, $N_T=1$. This individual trajectory shows  the stabilization of the unstable state $\widetilde{x}_{\text{eq}}=0$. (b) Amplitude $R$ (log-scale) vs time. (c) Oscillator position $\langle x(t)\rangle$ for a smaller system, $N=10^6$. We have depicted two typical trajectories, corresponding to different choices of the initial spin configuration.}
\label{fig5}
\end{center}
\end{figure}

\begin{figure}[htbp]
\begin{center}
\includegraphics[width=6cm]{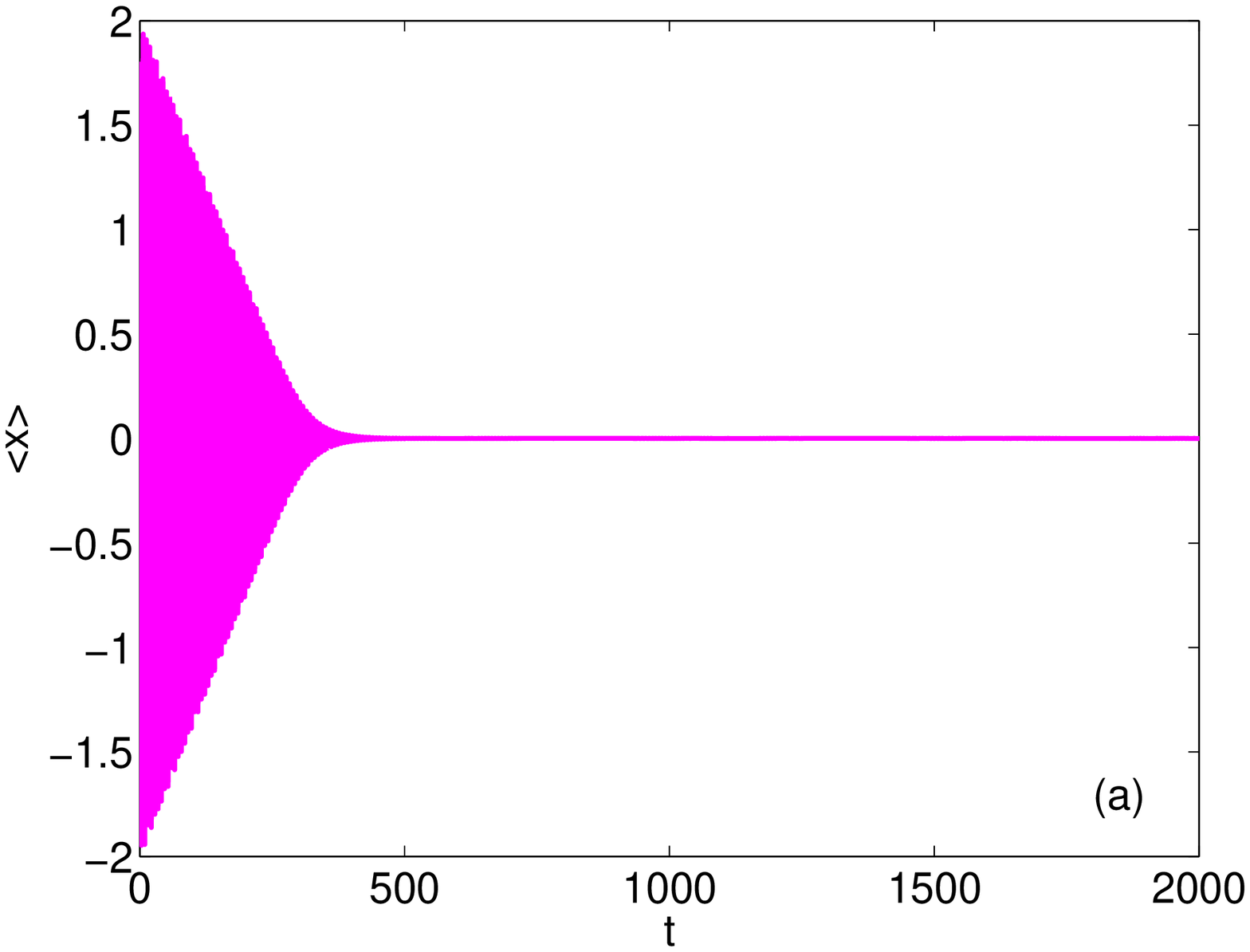}
\includegraphics[width=6cm]{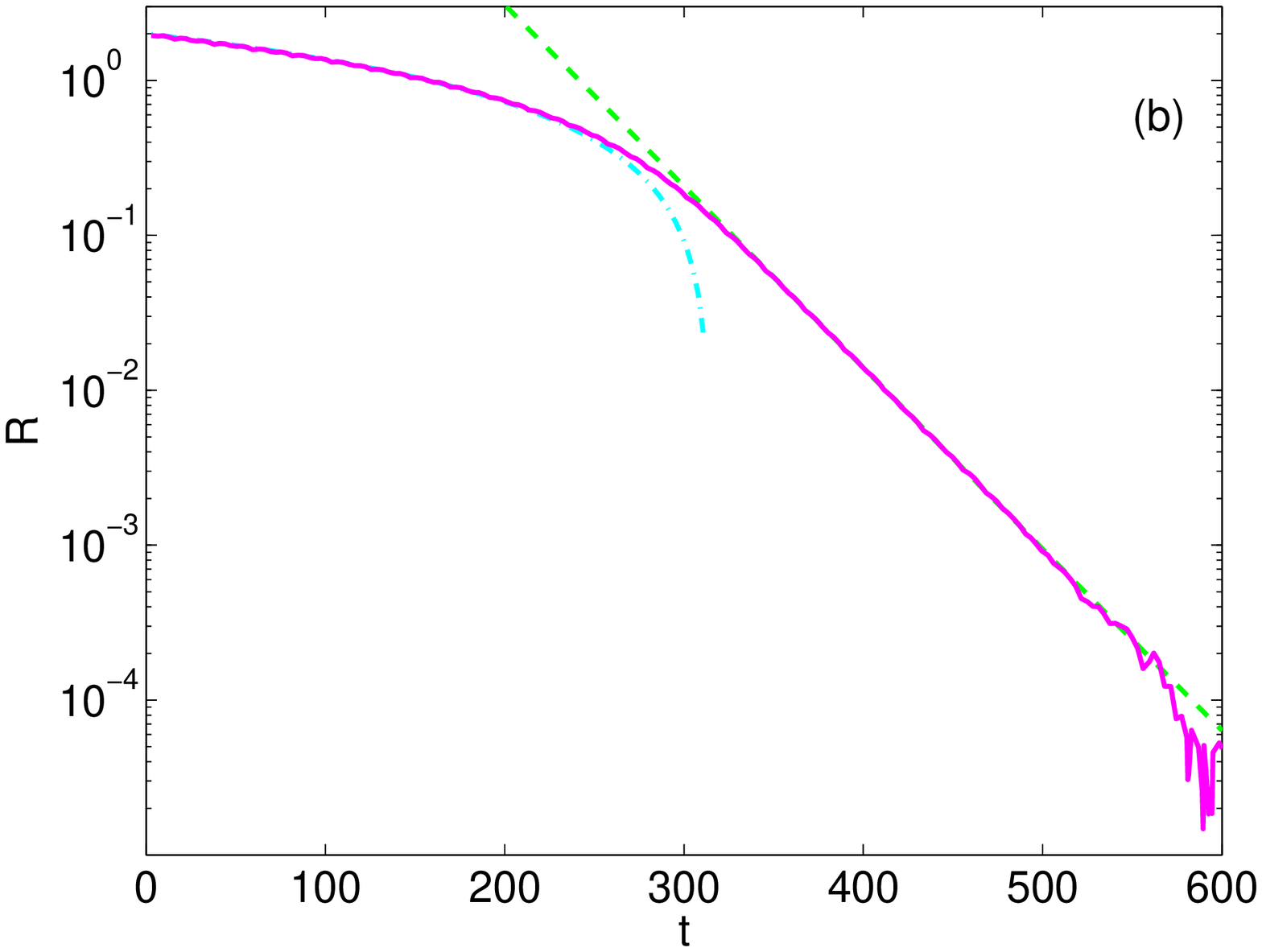}
\includegraphics[width=6cm]{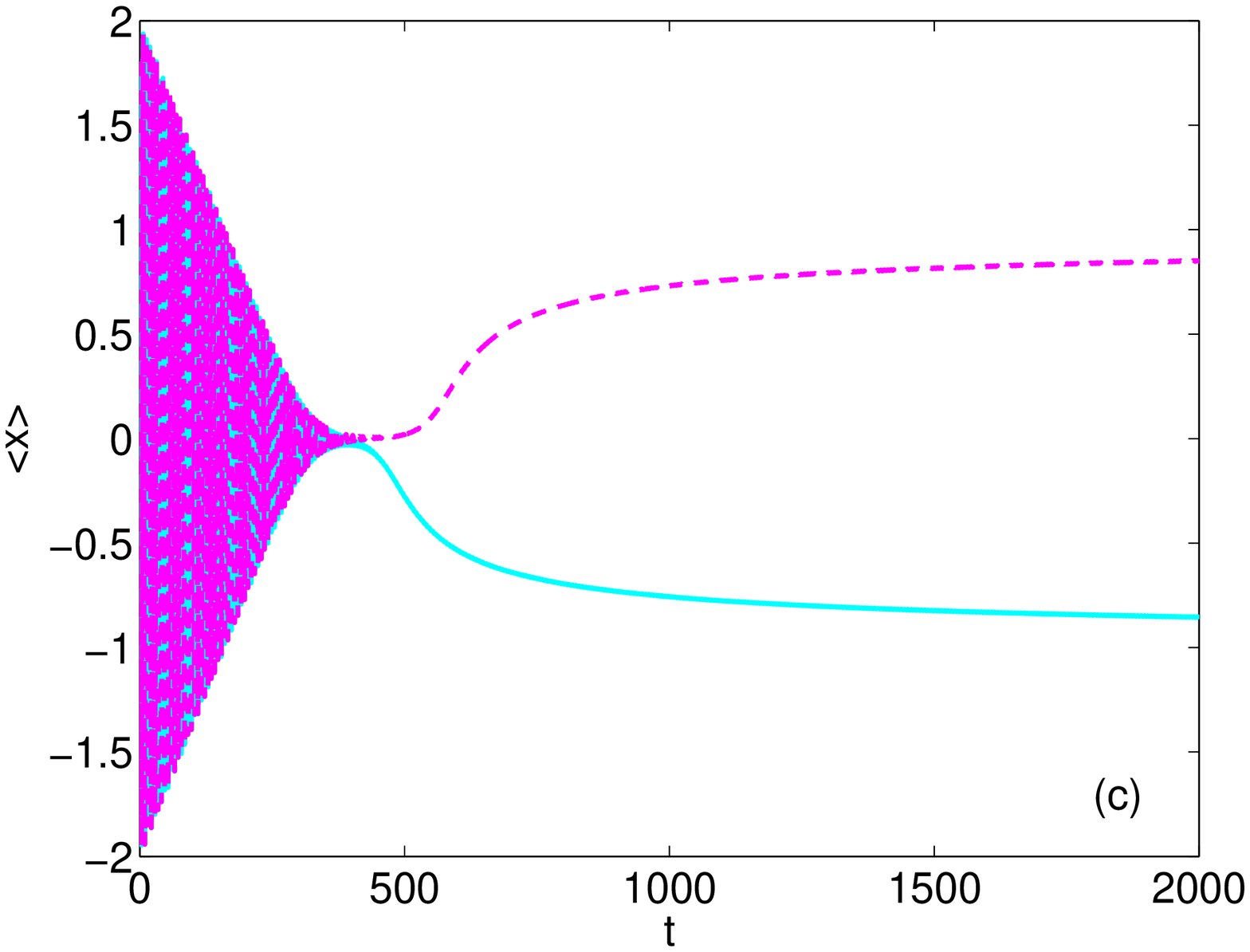}
\caption{Same as in figure \ref{fig5} for $\theta=0.37$, $x_0=2$. In (b), $R$ vs time shows the initial linear regime, (\ref{b3bb}), and the long time exponential decay, (\ref{b5d}). }
\label{fig6}
\end{center}
\end{figure}

Next we consider the case of random initial conditions, $\widetilde{C_n}(0)=0$ for $n\geq 1$. For $\theta<1$, (\ref{b5a})-(\ref{b5b}) have the solutions $\widetilde{C_n}^{(0)}(\tau)=0$, $n\geq 1$, with $R(\tau)$ decreasing monotonically to zero, according to (\ref{b5c})-(\ref{b5d}). We have investigated this stabilization of the unstable equilibrium solution $\widetilde{x}_{\text{eq}}=\widetilde{C_{1,\text{eq}}}=0$ by stochastic simulations of the oscillator-spin system, initially prepared in a completely random configuration. In figures \ref{fig5} and \ref{fig6}, we have chosen $x_0=0.5$ and $v_0=0.1$, respectively, but we have observed the same behavior for all the other values of $x_0$ and $v_0$ we have tested. For  $\theta= 0.95$, figure \ref{fig5}(a) shows modulated oscillations of the oscillator position with an envelope that decreases monotonically to zero and remains there, in fact stabilizing the unstable solution $\widetilde{x}_{\text{eq}}=0$. We depict the oscillation envelope in \ref{fig5}(b), together with the theoretically predicted exponential decay given by (\ref{b5d}). The agreement is excellent (note the logarithmic scale in the vertical axis). It must be stressed that, in both figures, we are plotting only one trajectory for a very large system, $N=10^8$. All the individual trajectories (for fixed $x_0$ and $v_0$) coincide. For a smaller system, the fluctuations are larger and the system finishes in one of the two stable stationary states, $\widetilde{x}_{\text{eq}}>0$ or $\widetilde{x}_{\text{eq}}<0$, depending on the trajectory considered. This is shown in \ref{fig5}(c), where two different trajectories for $N=10^6$ are depicted.

It is remarkable that the simulations of the oscillator-spin system for very large $N$ resemble the \textit{separatrix} trajectory of the averaged system (\ref{b5a})-(\ref{b5b}), given by a $R(\tau)$ that solves (\ref{b5b}) with $C_n^{(0)}\equiv \delta_{n0}$. The averaged equations approximate the macroscopic equations (\ref{2.28})-(\ref{2.29}) obtained in the limit as $N\to\infty$ by ignoring fluctuations. However the separatrix of the averaged system that ends up at $R=0$ is not a trajectory of (\ref{2.28})-(\ref{2.29}), so it is surprising that the approximate average equations describe better the result of stochastic simulations for random initial conditions than the more exact macroscopic equations. It also comes as a surprise that stochastic simulations for a smaller spin system with random initial conditions resemble the solutions of the macroscopic equations (\ref{2.28})-(\ref{2.29}) for which the separatrix is not a trajectory.

We analyze the low temperature case ($\theta=0.37$) in figure \ref{fig6}. Parameter values are the same as in figure \ref{fig5}, except for a larger initial oscillator position, $x_0=2$. This choice enlarges the initial time window during which $R$ decreases linearly with time, cf (\ref{b3bb}). Again, the oscillator approaches the unstable ``trivial'' solution and not one of the stable states with $|\widetilde{x}_{\text{eq}}|\simeq 0.99$; see figure \ref{fig6}(a). The oscillation amplitude decays to zero: linearly with time according to (\ref{b3bb}) in a first stage where $2R\gg\theta$, whereas it decays exponentially according to (\ref{b5d}) once $2R/\theta\ll 1$, see figure \ref{fig6}(b). In figure \ref{fig6}(c), we show the behavior for a smaller system with $N=10^6$. In this case, the fluctuations are larger and they can drive the system out from the unstable state towards one of the two stable states, depending on the initial spin configuration. Again the simulations of the stochastic oscillator-spin system resembles the description given by the averaged equations for larger $N$ and the description given by the ``more exact'' macroscopic equations (\ref{2.28})-(\ref{2.29}) for a smaller $N$.

\section{Derivation of the results in section \ref{s3}}
\label{s5}
\subsection{Derivation of Result 1}
To derive an approximate solution of (\ref{2.28}) - (\ref{2.29}) as in Result 1 of section \ref{s3}, we make the multiple scales Ansatz
\numparts
\begin{eqnarray}
\widetilde{x}(t;\delta)=x^{(0)}(t,\tau)+\delta\, x^{(1)}(t,\tau) + \Or(\delta^2),
\label{b8a}\\
\widetilde{C_n}(t; \delta)=C_n^{(0)}(t,\tau)+ \delta\, C_n^{(1)}(t,\tau)+\Or(\delta^{2}),
\label{b8b}
\end{eqnarray}
\endnumparts
with $\tau=\delta t$. While $\widetilde{x}$ and $\widetilde{C_n}$ are functions of one variable $t$, we consider that $x^{(j)}$ and $C_n^{(j)}$ are functions of two \textit{independent} variables, $t$ and $\tau=\delta t$. This is not really so, and therefore we are free to impose an additional condition to the multiscale functions in order to determine the evolution in the slow time scale $\tau$. This additional condition is that the two-time functions $x^{(k)}$ and $C_n^{(k)}$ should be bounded for arbitrarily large values of $t$ and any fixed value of $\tau$. We will see later that this condition indeed determines the slow-time evolution in the scale $\tau$. Insertion of (\ref{b8a}) and (\ref{b8b}) in (\ref{2.28}) and (\ref{2.29}) produces the following hierarchy of equations
\numparts
\begin{eqnarray}
&&\frac{\partial^2 x^{(0)}}{\partial t^{2}}+ x^{(0)}=C_1^{(0)},   \label{b9a}\\
&&\frac{\partial C_n^{(0)}}{\partial t}= 0,\label{b9b}
\end{eqnarray}
\endnumparts
\numparts
\begin{eqnarray}
&&\frac{\partial^2 x^{(1)}}{\partial t^{2}}+ x^{(1)}-C_1^{(1)}=-2\frac{\partial^2
x^{(0)}}{\partial t\partial \tau},   \label{b10a}\\
&&\frac{\partial C_n^{(1)}}{\partial t}= \gamma(x^{(0)})(C_{n-1}^{(0)} +
C_{n+1}^{(0)})-2 C_n^{(0)}-\frac{\partial C_n^{(0)}}{\partial \tau},\label{b10b}
\end{eqnarray}
\endnumparts
and so on.

Eq.\ (\ref{b7}) implies that $C^{(0)}_n$ is independent of $t$. Then the
solution of (\ref{b9a}) is
\begin{eqnarray}
x^{(0)}(\chi,\tau)=C_1^{(0)}(\tau) + R(\tau)\,\sin\chi, \quad \chi\equiv t+\phi(\tau).
\label{b11}
\end{eqnarray}
The oscillator position is therefore slowly driven by the spin correlation $C_1^{(0)}(\tau)$ while it performs modulated oscillations with slowly-varying amplitude $R(\tau)$ and rapidly changing phase $\chi$.

To find the slowly-varying quantities $C_n^{(0)}(\tau)$, $R(\tau)$ and $\phi(\tau)$, we analyze the first order equations (\ref{b10a}) and (\ref{b10b}). The rhs of (\ref{b10b}) is $2\pi$-periodic in the fast time $t$ (or, equivalently, $\chi$). Thus it can be expanded in a Fourier series in $\chi$ whose zeroth harmonic should vanish for $C^{(1)}_n$ to be bounded as $t\to\infty$. The zeroth order harmonic of the rhs of (\ref{b11}) is its average over a period of $\chi$. This implies Eq.\ (\ref{b5a}) for $C^{(1)}_n(\tau)$ with the definition (\ref{b5}) for fast time averages. Eq.\ (\ref{b5a}) is similar to the Glauber dynamics with a corrected rate $\overline{\gamma(x^{(0)}(\chi,\tau))}$. Because of (\ref{b11}), $\overline{\gamma(x^{(0)}(\chi,\tau))}$ depends on $\tau$ through the functions $R(\tau)$, $\phi(\tau)$ and $C_1^{(0)}(\tau)$.

To find $R(\tau)$ and $\phi(\tau)$, we should first integrate (\ref{b10b}). To this end, we write the periodic function $\gamma(x^{(0)}(\chi,\tau))$ as a Fourier series
\numparts
\begin{eqnarray}
 && \gamma(x^{(0)}(\chi,\tau)) = \sum_{j=-\infty}^{+\infty} \gamma_j(\tau) e^{\rmi j \chi} \,, \label{3.10}\\
&&  \gamma_j(\tau) = \frac{1}{2\pi} \int_{-\pi}^{\pi} \rmd\chi e^{-\rmi j \chi} \gamma(x^{(0)}(\chi,\tau)) \equiv \overline{e^{-\rmi j\chi} \gamma(x^{(0)}(\chi,\tau))} \,. \label{3.11}
\end{eqnarray}
\endnumparts
Substituting (\ref{b5a}) in (\ref{b10b}), the latter equation can be written as
\begin{eqnarray}
    \frac{\partial C_n^{(1)}}{\partial t}&=&\left[ \gamma(x^{(0)})-\overline{\gamma(x^{(0)})}\right] \left[C_{n-1}^{(0)}+C_{n+1}^{(0)}\right] \nonumber \\
    &=&\sum_{j=-\infty,j\neq 0}^{+\infty} \gamma_j(\tau) e^{\rmi j \chi(t,\tau)} \left[C_{n-1}^{(0)}(\tau)+C_{n+1}^{(0)}(\tau)\right] \,.\label{3.12}
\end{eqnarray}
This equation can be immediately integrated with the result
\numparts
\begin{eqnarray}
&&C_n^{(1)}(t,\tau) = [C_{n-1}^{(0)}(\tau)+C_{n+1}^{(0)}(\tau)]
\sum_{j=-\infty,j\neq 0}^\infty \frac{\gamma_j(\tau)}{\rmi j} e^{\rmi j\chi}+D_n(\tau), \label{b14a}\\
&&\gamma_j=\overline{e^{-\rmi j\chi}\gamma(x^{(0)})}= \overline{\gamma(x^{(0)})\cos(j\chi)} -\rmi\, \overline{\gamma(x^{(0)})\sin(j\chi)}.   \label{b14b}
\end{eqnarray}
\endnumparts
Here the ``constant of integration'' $D_n(\tau)$ is a slowly-varying function that can be determined from higher order equations in the hierarchy. Inserting (\ref{b11}) and (\ref{b14a}) in (\ref{b10a}), we find
\begin{eqnarray}
\frac{\partial^2 x^{(1)}}{\partial t^{2}}+ x^{(1)}=(1+C_2^{(0)})
\sum_{j=-\infty,j\neq 0}^\infty \frac{\gamma_j}{\rmi j} e^{\rmi j \chi}+D_1 - 2
\frac{\partial^2 x^{(0)}}{\partial t\partial\tau}.   \label{b15}
\end{eqnarray}
For $x^{(1)}(\chi,\tau)$ to be bounded as $\chi\to\infty$, the right hand side of (\ref{b15}) cannot contain first harmonic terms, and therefore the following equations hold:
\numparts
\begin{eqnarray}
&&\frac{\rmd R}{\rmd\tau} = -(1+C_{2}^{(0)})\,\overline{\gamma(x^{(0)})\sin \chi},   \label{b16a}\\
&&R\,\frac{\rmd \phi}{\rmd\tau} =-(1+C_{2}^{(0)})\,\overline{\gamma(x^{(0)})\cos \chi}=0, \label{b16b}
\end{eqnarray}
\endnumparts
which are (\ref{b5b}). Equivalently, Eq.\ (\ref{b16a}) can be written as
\begin{eqnarray}
&&\frac{\rmd R}{\rmd\tau} = \frac{\theta}{2}(1+C_{2}^{(0)})\,\frac{\partial}{\partial R}
\overline{\ln\left(1+e^{-4(C_1^{(0)}+R\sin \chi)/\theta}\right)}.   \label{b17}
\end{eqnarray}

\subsection{Derivation of Result 2}

At the critical temperature $\theta=1$ we have a pitchfork bifurcation with nonzero stationary solutions that bifurcate from $R=C_1^{(0)}=0$ for $\theta<1$. We can obtain approximate solutions of the macroscopic system (\ref{2.28}) and (\ref{2.29}) directly by using bifurcation theory. See \ref{app_bif}. However to keep the unity of our theoretical description, it is instructive to study the dynamical behavior of the reduced system (\ref{b5a})-(\ref{b5b}) near the critical temperature. The stationary state $R=C_1^{(0)}=0$ is stable for $\theta>1$ and unstable for $\theta<1$ so that the pitchfork bifurcation is supercritical: nonzero stable stationary branches exist for $\theta<1$. From (\ref{2.25}), we anticipate that the stationary solutions bifurcating from $x^{(0)}=C_n^{(0)}=0$ are
\begin{equation}\label{4.1}
    x_{\text{eq}}^{(0)}= C_{1,\text{eq}}^{(0)}\sim \pm \sqrt{3 \left(1-\theta\right)} \,, \qquad \theta\to 1^-,
\end{equation}
whereas the other correlations are given by (\ref{3.21}). Therefore, by defining
\begin{equation}\label{4.2}
    \theta = 1 - \epsilon^2  \theta_2 \,,
\end{equation}
$ x_{\text{eq}}^{(0)}$ and  $C_{1,\text{eq}}^{(0)}$ are of order $\epsilon$, while $C_{n,\text{eq}}^{(0)}=\Or(\epsilon^n)$. The parameter $\epsilon$ therefore measures the amplitude of the bifurcating solution. We have introduced the parameter $\theta_2$ in (\ref{4.2}) to analyze simultaneously the cases $\theta<1$ ($\theta_2=1$) and $\theta>1$ ($\theta_2=-1$).

Let us now derive Result 2 for the pitchfork bifurcation. We expect that $ x^{(0)}$ and  $C_{1}^{(0)}$ are of order $\epsilon$ and $C_n^{(0)}=\Or(\epsilon^n)$ from the stationary solutions. Thus we introduce the following multiple scales Ansatz:
\numparts
\begin{eqnarray}
&&R=\epsilon\sum_{j=0}^2\epsilon^j R^{(j)}(\tau,s)+\Or(\epsilon^4),
\label{b20a}\\
&& C_1^{(0)}=\epsilon\sum_{j=0}^2 \epsilon^j \Gamma^{(j)}(\tau,s)+
\Or(\epsilon^4), \label{b20b}\\
&& C_n^{(0)}= \left(C_1^{(0)}\right)^{n}\left[\sum_{j=0}^2 \epsilon^j \varphi_n^{(j)}(\tau,s)+\Or(\epsilon^4)\right], \label{b20c}\\
&&s=\epsilon^2\tau. \label{b20d}
\end{eqnarray}
\endnumparts
All the unknown functions in these equations should be bounded in the limit $\tau\to\infty$. Note that $\varphi_0^{(j)}= \varphi_1^{(j)}=\delta_{j0}$. The Ansatz (\ref{b20c}) follows the form of the equilibrium solution and the slowly varying time scale $s$ is chosen to keep the functions $R^{(2)}$ and $\Gamma^{(2)}$ bounded as $\tau\to\infty$ (see below). Note that (\ref{b4b}) implies
\numparts
\begin{eqnarray}
&& \overline{x^{(0)}}= C_1^{(0)},\quad
\overline{x^{(0)}\sin \chi}=\frac{R}{2}, \label{b21a}\\
&& \overline{x^{(0)\, 3}}= C_1^{(0)\, 3}+C_1^{(0)}\frac{3R^2}{2},\quad
\overline{x^{(0)\, 3}\,\sin \chi}= \frac{3R}{8}\, (4C_1^{(0)\, 2}+R^2).\label{b21b}
\end{eqnarray}
\endnumparts
These equations are needed to calculate $\overline{\gamma}$ and $\overline{\gamma\,\sin\chi}$ up to terms of order $\epsilon^4$. Inserting (\ref{4.2}) and (\ref{b20a})-(\ref{b21b}) in (\ref{b5a}) and (\ref{b5b}) and equating like powers of $\epsilon$, we obtain a hierarchy of equations. The leading order equations are:
\numparts
\begin{eqnarray}
&&\partial_{\tau}R^{(0)}= -R^{(0)},\quad\partial_{\tau}\Gamma^{(0)}= 0,\label{b22a}\\
&&\partial_{\tau}\varphi_n^{(0)}= 2(\varphi_{n-1}^{(0)}-\varphi_n^{(0)}), \quad
\varphi_1^{(0)}\equiv 1.  \label{b22b}
\end{eqnarray}
\endnumparts
The solutions are $R^{(0)} =R^{(0)}(0) e^{-\tau}$, $\varphi_n^{(0)}= 1 +O(e^{-2\tau})$ (for $n\geq 2$) and $\Gamma ^{(0)}= \Gamma^{(0)}(s)$. For large enough values of the fast time $\tau$, these solutions become
\begin{eqnarray}
R^{(0)}= 0,\quad \Gamma^{(0)}= \Gamma^{(0)}(s),\quad\varphi_n^{(0)}=1,
\label{b23}
\end{eqnarray}
except for exponentially decreasing terms that we will ignore in what follows.
We now find an equation for $\Gamma^{(0)}(s)$. Equations (\ref{b21a})-(\ref{b21b}) become
\numparts
\begin{eqnarray}
&& \overline{x^{(0)}}=\sum_{j=0}^2 \epsilon^{j+1} \Gamma^{(j)}+\Or(\epsilon^4),\label{b24a}\\
&& \overline{x^{(0)}\sin \chi}\sim\frac{\epsilon^2}{2}
(R^{(1)}+\epsilon R^{(2)}), \quad\overline{x^{(0)\, 3}}\sim \epsilon^3 \Gamma^{(0)\, 3},  \quad \overline{x^{(0)\, 3}\,\sin \chi}= 0.\label{b24b}
\end{eqnarray}
\endnumparts
Then the equations for $R^{(j)}$ and $\Gamma^{(j)}$ with $j=1,2$ are
\numparts
\begin{eqnarray}
&& \partial_\tau R^{(1)}=- R^{(1)},\quad\partial_\tau \Gamma^{(1)}= 0,\label{b25a}\\
&&\partial_\tau R^{(2)}=- R^{(2)},\quad \partial_\tau \Gamma^{(2)}=-\partial_{s}\Gamma^{(0)}+2\theta_2 \Gamma^{(0)}-\frac{2 \Gamma^{(0)3}}{3}.
\label{b25b}
\end{eqnarray}
\endnumparts
Again $R^{(1)}=R^{(2)}=0$ up to terms that decrease exponentially in the $\tau$ scale. The solution $\Gamma^{(2)}$ of (\ref{b25b}) is bounded as $\tau\to\infty$ only if the right hand side of the equation vanishes, which produces the sought amplitude equation:
\begin{eqnarray}
\frac{d \Gamma^{(0)}}{ds}=2\theta_2 \Gamma^{(0)}-\frac{2}{3}\, \Gamma^{(0)3}.
\label{b26}
\end{eqnarray}
The solution of (\ref{b26}) for nonzero initial conditions is
\begin{eqnarray}
&&\Gamma^{(0)}(s) = \mbox{sign}[\Gamma^{(0)}(0)]\,\sqrt{\frac{3\,\theta_2}{1+e^{-4\theta_2 s}\left(\frac{3\,\theta_2}{\Gamma^{(0)}(0)^2}-1\right)}}.   \label{b27a}
\end{eqnarray}
As the slow time $s\to\infty$, this function tends to 0 if $\theta_2=-1$ ($\theta>1$) and to $\pm\sqrt 3$ if $\theta_2=1$ ($\theta<1$). In the original variables, (\ref{b27a}) becomes (\ref{b1a}). These solutions tend to the stationary values $C_1^{(0)}\sim \pm\sqrt{3(1-\theta)}$ corresponding to the stable equilibrium solutions if $\theta<1$. For $\Gamma^{(0)}(0)=0$, it is $\Gamma^{(0)}(s)=0$ for all $s$.

\subsection{Derivation of Result 3}
As $\theta\to 0$, the reduced equations (\ref{b5a}) and (\ref{b5b}) for the correlations $C_n^{(0)}$ and for the oscillation amplitude $R$ can be simplified.
Note that for $x^{(0)}\neq 0$ the argument in (\ref{2.30}) may become arbitrarily large, and therefore we have
\begin{eqnarray}
\gamma\sim [\mbox{sign } x^{(0)}]\left(1-2 e^{-4|x^{(0)}|/\theta}\right).\label{b31}
\end{eqnarray}
If $|C_1^{(0)}|<R$ we have that $x^{(0)}$ changes sign due to the oscillatory term and
\numparts
\begin{eqnarray}
&& \overline{\gamma}\sim [\mbox{sign } C_1^{(0)}] \left[\frac{2\chi_0}{\pi}\eta(R-|C_1^{(0)}|)+\eta(|C_1^{(0)}|-R)\right],\label{b32a}\\
&&\chi_0=\arcsin\left(\frac{|C_1^{(0)}|}{R}\right),\label{b32b}
\end{eqnarray}
\endnumparts
where $\eta(x)=1$ for $x>0$ and 0 otherwise. We also find (see \ref{app_lowtemp})
\numparts
\begin{eqnarray}
&& \overline{\gamma\left(\frac{2x^{(0)}}{\theta}\right)\sin\chi}\sim \frac{2\cos\chi_0}{\pi}=\frac{2}{\pi}\sqrt{1-\left(\frac{C_1^{(0)}}{R}\right)^2}, \quad\mbox{for } |C_1^{(0)}|<R, \label{b33a}\\
&&\overline{\gamma\left(\frac{2x^{(0)}}{\theta}\right)\sin\chi}\sim 2 e^{-4|C_1^{(0)}|/\theta} I_1\left(\frac{4R}{\theta}\right) , \quad\mbox{for } |C_1^{(0)}|>R,\label{b33b}
\end{eqnarray}
\endnumparts
Using Eq.\ (\ref{b11}) and (\ref{b31}), we get $\overline{\gamma}\sim S=$ sign\,$ C_1^{(0)}=\pm 1$ for $|C_1^{(0)}|>R$, which is (\ref{b2a}). The solution of the latter equation can be written as $C_n^{(0)}=S^n+\Delta_n(\tau)$, where $\Delta_n$ also satisfies the equations (\ref{b2a}) but with boundary condition $\Delta_0=0$. This system of equations can be solved by using a generating function whose moments are the $\Delta_n$. We show in \ref{app_lowtemp} that the result is
\begin{eqnarray}
C_n^{(0)}(\tau)= S^n-e^{-2\tau}  \sum_{j=1}^\infty [S^j-C_j^{(0)}(0)] [I_{|n-j|}(2\tau) - I_{n+j}(2\tau)] \,. \label{b35}
\end{eqnarray}
where $I_n(x)$ is the modified Bessel function of the first kind with integer index $n$. The correlations in (\ref{b35}) decay algebraically $C_n^{(0)}(\tau)\sim S^n-n /(4\pi\tau^3)^{1/2} \sum_{j=1}^\infty j [S^j -C_j^{(0)}(0)]$ as $\tau\to\infty$ (long time tails) \cite{Be99}. In fact, this is the particularization of Glauber's solution for the Ising model \cite{Gl63} to the zero temperature limit.

According to (\ref{b5b}) and (\ref{b33b}), $R$ satisfies (\ref{b2b}), whose right hand side is exponentially small. Thus $R(\tau)$ decays to zero only on an exponentially long time scale. Therefore $C_1^{(0)}$ approaches $S=\pm 1$ whereas $x^{(0)}$ oscillates on the time scale $\tau$ with almost constant amplitude $R(0)<1$ about it. It takes an exponentially long time for the oscillation amplitude to vanish. For equilibrium-like initial correlations $C_n^{(0)}(0)=r^n$, with $0<r<1$, we show in \ref{app_lowtemp} that $C_n^{(0)}(\tau)$ satisfies (\ref{5.8}).

For $R>|C_1^{(0)}|$, (\ref{b32a}) and (\ref{b33a}) yield (\ref{b3a}) and (\ref{b3b}). $R$ decays until $R\sim |C_1^{(0)}|$. Recapitulating, $C^{(0)}_n$ and $R$ are given by (\ref{b2a}) and (\ref{b2b}), respectively, if $R<|C^{(0)}_1|$ and by (\ref{b3a}) and (\ref{b3b}) if $R>|C_1^{(0)}|$. In any case, the correlations $C_n^{(0)}$ tend algebraically to $S^n=(\pm 1)^n$ as $C_n^{(0)}-S^n\propto\tau^{-3/2}$ ($\tau\to\infty$). According to (\ref{b4b}), the oscillator vibrates with its natural frequency about an equilibrium position that approaches slowly $x=\pm 1$, with an amplitude $R=\Or(1)$, which vanishes only after an exponentially long time.

\section{Conclusions}
\label{s6}

We have analysed the dynamics of a fast oscillator coupled to a one-dimensional chain of $N$ Ising spins $\sigma_i$, $i=1,\ldots,N$ in contact with a thermal bath at temperature $\theta$. The interaction energy between the oscillator and the spins is proportional to the oscillator position and to $\sum_i \sigma_i \sigma_{i+1}$. In the limit of infinitely many spins, there is a second order phase transition at the critical temperature $\theta=1$ and the fluctuations of the oscillator position and the spin correlations are negligible. Provided the oscillator natural period is much smaller than the relaxation time of the spins, the oscillator position performs modulated oscillations about its slowly varying equilibrium which is related to the correlation between nearest neighbor spins. The spin correlations decay to their equilibrium values over the slow spin relaxation time scale $\tau$. Using a multiple scale analysis that exploits the large separation between characteristic times, we derive modulation equations for the average oscillation envelope and phase of the oscillator and for the spin correlations.

The modulation equations  can be analyzed in two limits: near the critical temperature $\theta=1$ and in the limit of low temperatures. At the critical temperature, the ``trivial'' solution of the modulation equations (zero spin correlation and zero oscillator position) undergoes a supercritical pitchfork bifurcation: the trivial solution is stable for $\theta>1$ and two non-trivial stable stationary solutions bifurcate from it for $\theta<1$. We have constructed these solutions using multiple scales analysis and shown the typical slowing down near the critical temperature. Below the critical temperature, a study of the modulation equations show that most initial conditions evolve towards one of the stable stationary solutions revealed by the bifurcation analysis. The spin correlation $\widetilde{C_1}=\langle\sigma_i\sigma_{i+1}\rangle$ and the average oscillator position corresponding to these stationary solutions both approach $\pm 1$ as $\theta\to 0$. However, the instantaneous spin correlation for a fixed small temperature tends to a stable stationary value algebraically as $\tau\to\infty$. The oscillator position carries out modulated oscillations about the instantaneous spin correlation with an amplitude that decays to zero over an exponentially long time.

In the thermodynamic limit $N\gg 1$, perhaps the most striking result is the dynamical stabilization of the unstable trivial equilibrium state $\widetilde{x}_{\text{eq}}=\widetilde{C_{1,\text{eq}}}=0$ at any temperature $\theta<1$ for random initial conditions of the Ising system. In the limit of a very fast oscillator we are considering, the analysis of the averaged modulation equations shows that the oscillator position averaged over one natural period, $\overline{x(t)}$, tends to approach its instantaneous rest position $\widetilde{C_1}(t)$ (which evolves much more slowly towards equilibrium). But, for random initial conditions of the spin system $\widetilde{C_1}(t)=0$, and the envelope of the oscillator position tends to zero with relaxation time $\theta/\delta$. We have verified our theoretical predictions by direct numerical simulation of the oscillator-spin system and by numerically solving the macroscopic equations which ignore fluctuations and are valid in the thermodynamic limit. In general, the agreement between simulation and theory is excellent. For large enough size $N$ and random initial conditions, direct simulations show that the fluctuations are so small that the system is not able to depart from the unstable state, in every single trajectory. For smaller systems, fluctuations are larger and drive both the oscillator and the Ising system to one of the two stable states.

These results are somewhat surprising given that there is a Lyapunov functional for the oscillator-spin system guaranteeing that any initial condition evolves on the time scale $\tau$ towards a stable stationary state corresponding to a canonical equilibrium probability distribution. However, the corresponding H-theorem is proved for finitely many spins and therefore it does not necessarily applies to fluctuation-free results obtained after the limit $N\to\infty$ has been taken. Stabilization of the unstable trivial stationary state for random initial conditions is reminiscent of weak ergodicity breaking in glasses \cite{Bo92,BCKyM98}. It is different from noise-induced stabilization and selection of unstable states as studied by Freidlin \cite{fre01} and Muratov et al \cite{mur07}.

\ack

This research has been supported by the Spanish Ministerio de Ciencia e Innovaci\'on (MICINN) through Grants FIS2008-04921-C02-01 (LLB), FIS2008-01339 (AP, partially financed by FEDER funds), and FIS2008-04921-C02-02 (AC). The authors would like also to thank the Spanish National Network Physics of Out-of-Equilibrium Systems financed through the MICINN grant FIS2008-04403-E.

\appendix
\section{H-theorem}
\label{appH}
Let us define the functional
\begin{equation}\label{h1}
    H(t)=\int \rmd x\, \int\rmd p\, \sum_{\bm{\sigma}} {\cal P}(x,p,\bm{\sigma},t) \ln \left( \frac{{\cal P}(x,p,\bm{\sigma},t)}{{\cal P}_{0}(x,p,\bm{\sigma},t)} \right) \, ,
\end{equation}
where ${\cal P}(x,p,\bm{\sigma},t)$ and ${\cal P}_{0}(x,p,\bm{\sigma},t)$ are two different solutions of the master equation (\ref{2.17}). We have $H(t)\geq 0$ because $x\ln x\geq x-1$ for all $x={\cal P}/{\cal P}_{0}\geq 0$ and the probability densities are normalized to 1. By using (\ref{2.17}) and integrating by parts, the time derivative of $H$,
\begin{equation}\label{h2}
    \dot{H}(t)=\int \rmd x\, \int\rmd p\, \sum_{\bm{\sigma}}\left[ \partial_t {\cal P}\, \ln \left( \frac{{\cal P}}{{\cal P}_{0}} \right)-\left( \frac{{\cal P}}{{\cal P}_{0}} \right) \partial_t {\cal P}_0 \right] ,
\end{equation}
can be shown to be
\begin{eqnarray}
\dot{H}(t)&=& \delta\int \rmd x\, \int\rmd p\, \sum_{\bm{\sigma}} \sum_{\bm{\sigma'}} \left\{\ln \left( \frac{{\cal P}(x,p,\bm{\sigma},t)}{{\cal P}_{0}(x,p,\bm{\sigma},t)} \right)\right. \nonumber \\
& \times & \left[ W(\sigma'\to\sigma|x,p) {\cal P}(x,p,\bm{\sigma'},t)-W(\sigma\to\sigma'|x,p) {\cal P}(x,p,\bm{\sigma},t) \right]\nonumber\\
& - & \left[ W(\sigma'\to\sigma|x,p) {\cal P}_0(x,p,\bm{\sigma'},t)-W(\sigma\to\sigma'|x,p) {\cal P}_0(x,p,\bm{\sigma},t) \right]\nonumber\\
&\times& \left.\frac{{\cal P}(x,p,\bm{\sigma},t)}{{\cal P}_{0}(x,p,\bm{\sigma},t)} \right\}. \label{h3}
\end{eqnarray}
Here we have used the notation
\begin{equation}\label{h4}
    W(\sigma\rightarrow\sigma'|x,p)=\left\{ \begin{array}{ll}
    W_i(\bm{\sigma}|x,p) & \mbox{if $\sigma'=R_i\sigma$, $i=1,\ldots,N$} \\
    0   & \mbox{otherwise,}
    \end{array}\right.
\end{equation}
where $R_i\bm{\sigma}$ is the configuration obtained from $\bm{\sigma}$ by flipping the $i$-th spin. By interchanging $\bm{\sigma}$ and $\bm{\sigma'}$ in the second term on the rhs of (\ref{h3}), and defining
\begin{equation}\label{h5}
    f(x,p,\bm{\sigma},t)=\frac{{\cal P}(x,p,\bm{\sigma},t)}{{\cal P}_{0}(x,p,\bm{\sigma},t)} \,,
\end{equation}
equation (\ref{h3}) can be rewritten as
\begin{eqnarray}
\dot{H}(t) & = & \delta\int \rmd x\, \int\rmd p\, \sum_{\bm{\sigma}} \sum_{\bm{\sigma'}} W(\sigma'\to\sigma|x,p)\left\{ {\cal P}(x,p,\bm{\sigma'},t)\right.\nonumber \\
& \times& [ \ln f(x,p,\bm{\sigma},t)-  \ln f(x,p,\bm{\sigma'},t)] \nonumber\\
&-& \left. {\cal P}_{0}(x,p,\bm{\sigma'},t) [f(x,p,\bm{\sigma},t) - f(x,p,\bm{\sigma'},t) ] \right\} . \label{h6}
\end{eqnarray}
Let $f= f(x,p,\bm{\sigma},t)$ and $f'=f(x,p,\bm{\sigma'},t)$. Then we can write (\ref{h6}) as
\begin{eqnarray}
  \dot{H}(t) & = & \delta\int \rmd x\, \int\rmd p\, \sum_{\bm{\sigma}} \sum_{\bm{\sigma'}} W(\sigma'\to\sigma|x,p) {\cal P}'_{0} \left[f' \ln\left(\frac{f}{f'}\right)-f+f'\right]\nonumber \\
   &=& - \delta\int \rmd x\, \int\rmd p\, \sum_{\bm{\sigma}} \sum_{\bm{\sigma'}} W(\sigma'\to\sigma|x,p) {\cal P}'_{0} f \nonumber \\
&\times& \left[\left(\frac{f'}{f}\right)\ln\left(\frac{f'}{f}\right)- \left(\frac{f'}{f}\right)+1 \right]\leq 0.  \label{h7}
\end{eqnarray}
Since $Wf{\cal P}'_0\geq 0$ and $x\ln x-x+1\geq 0$ for $x\geq 0$, $H(t)$ decreases with time. Moreover the expression in square brackets in (\ref{h7}) vanishes only if $f=f'$. Thus $H(t)\geq 0$ decreases monotonically as time increases and is bounded from below. Hence
$H(t)$ tends to a limit as $t\to\infty$, such that $\lim_{t\to\infty}\dot{H}(t)=0$. According to (\ref{h7}), this is possible only if $f=f'$ for all pairs of connected states having $W(\sigma'\to\sigma|x,p)\neq 0$. For fixed $x,p$, all the spin configurations $\sigma$ are connected through a chain of transitions with non-zero probability, and therefore
\begin{equation}\label{h8}
    \lim_{t\to\infty}\frac{{\cal P}(x,p,\bm{\sigma},t)}{{\cal P}_{0}(x,p,\bm{\sigma},t)} =\widehat{f}(x,p) \,,
\end{equation}
independently of the spin configuration $\sigma$. We have shown that in the long time limit, the ratio of any two solutions of the master equation is a function of $x$ and $p$. Since the equilibrium distribution ${\cal P}_{\text{eq}}(x,p,\bm{\sigma})$ is also a solution of the master equation, any time-dependent solution satisfies
\begin{equation}\label{h9}
{\cal P}(x,p,\bm{\sigma},\infty)={\cal P}_{\text{eq}}(x,p,\bm{\sigma})\widehat{f}(x,p).
\end{equation}
The distribution ${\cal P}(x,p,\bm{\sigma},\infty)$ must be a stationary solution of the master equation (\ref{2.17}). By inserting (\ref{h9}) in (\ref{2.17}) and using detailed balance, we find that $\widehat{f}(x,p)$ is constant, independent of $x$ and $p$. Clearly $\widehat{f}(x,p)=1$ due to the normalization condition for ${\cal P}(x,p,\bm{\sigma},\infty)$. Then
\begin{equation}\label{h10}
    {\cal P}(x,p,\bm{\sigma},\infty)={\cal P}_{\text{eq}}(x,p,\bm{\sigma}) \quad\mbox{and} \quad  \lim_{t\to\infty}\frac{{\cal P}(x,p,\bm{\sigma},t)}{{\cal P}_{0}(x,p,\bm{\sigma},t)} =1.
\end{equation}
Equation (\ref{h10}) establishes that all the solutions of the master equation corresponding to $N$ spins tend to the canonical equilibrium distribution in the long-time limit. As $N\to\infty$, there is a second order transition at the critical temperature $\theta=1$. For $\theta>1$, there is only one phase given by the limit of the canonical equilibrium distribution as $N\to\infty$ with order parameter $\widetilde{x}_{\text{eq}}=\widetilde{C_1}_{\text{eq}}=0$. For $\theta<1$, there are three phases which appear as limits of the canonical distribution with order parameters $\widetilde{x}_{\text{eq}}=0$ (unstable phase) and $\pm|\widetilde{x}_{\text{eq}}|\neq 0$ (stable phases). These phases coalesce  at $\theta=1$ in a pitchfork bifurcation, as described in \cite{PBC10}. Each sign attracts different initial conditions. The multiplicity of (macroscopic) equilibrium solutions corresponding to extrema of the equilibrium distribution is a direct consequence of the nonlinearity of the macroscopic equation, which is compatible with the linearity of the master equation and the multiplicity of its stationary solutions as $N\to\infty$ \cite{vk92}.

\section{Pitchfork bifurcation in the macroscopic equations}
\label{app_bif}
Let us assume that $\delta=\Or(1)$ and seek a solution of the macroscopic equations (\ref{2.28})-(\ref{2.30}) near the critical temperature $\theta=1$ by means of the multiple scales Ansatz:
\begin{eqnarray}
&&\widetilde{x}=\epsilon\sum_{j=0}^2\epsilon^j \xi^{(j)}(t,\zeta)+\Or(\epsilon^4),
\label{b.1a}\\
&& \widetilde{C_1}=\epsilon\sum_{j=0}^2 \epsilon^j \Gamma^{(j)}(t,\zeta)+
\Or(\epsilon^4), \label{b.1b}\\
&& \widetilde{C_n}= \left(\widetilde{C_1}\right)^{n}\left[\sum_{j=0}^2 \epsilon^j \varphi_n^{(j)}(t,\zeta)+\Or(\epsilon^4)\right], \label{b.1c}\\
&&\theta=1-\theta_2\epsilon^2,\label{b.1d}\\
&&\zeta=\epsilon^2 t. \label{b.1e}
\end{eqnarray}
As in section \ref{s5}, $\epsilon$ measures the amplitude of the bifurcating solutions and all unknowns in (\ref{b.1a}) - (\ref{b.1c}) should be bounded as $t\to\infty$. Inserting (\ref{b.1a}) - (\ref{b.1e}) in (\ref{2.28})-(\ref{2.30}) and equating like powers of $\epsilon$, we get the following hierarchy of equations
\begin{eqnarray}\label{b.2a}
 \left( \frac{\partial^2}{\partial t^2}+1\right) \xi^{(0)} -\Gamma^{(0)}=0 ,  \\
  \label{b.2b}
\left(\frac{\partial}{\partial t}+2\delta\right)\Gamma^{(0)}-2\delta\xi^{(0)}=  0,
\end{eqnarray}
\begin{eqnarray}\label{b.3a}
 \left( \frac{\partial ^2}{\partial t^2}+1\right) \xi^{(1)} -\Gamma^{(1)}=0 ,  \\
  \label{b.3b}
\left(\frac{\partial}{\partial t}+2\delta\right)\Gamma^{(1)}-2\delta\xi^{(1)}=  0,
\end{eqnarray}
\begin{eqnarray}\label{b.4a}
 \left( \frac{\partial ^2}{\partial t^2}+1\right) \xi^{(2)} -\Gamma^{(2)} = -2 \frac{\partial^2\xi^{(0)}}{\partial t \partial\zeta},  \\
  \label{b.4b}
\left(\frac{\partial}{\partial t}+2\delta\right)\Gamma^{(2)}-2\delta\xi^{(2)}=  2\delta\xi^{(0)}\left(\theta_2+\Gamma^{(0)2}\varphi_2^{(0)}-\frac{8}{3}\xi^{(0)2}\right)-\frac{\partial\Gamma^{(0)}}{\partial\zeta},\\
\left(\frac{\partial}{\partial t}+2\delta\right)(\Gamma^{(0)2}\varphi_2^{(0)})=2\delta \xi^{(0)}\Gamma^{(0)},\label{b.4c}
\end{eqnarray}
and so on.

The solution of (\ref{b.2a})-(\ref{b.2b}) is
\begin{eqnarray}\label{b.5a}
\xi^{(0)}= K(\zeta)+ \mbox{Re}\left[A(\zeta) e^{(i\Omega-\delta)t}\right], \quad
\Omega=\sqrt{1-\delta^2},\\
\Gamma^{(0)}=K(\zeta)-2\delta\,\mbox{Re}\left[A(\zeta) e^{(i\Omega-\delta)t} (i\Omega-\delta)\right], \label{b.5b}
\end{eqnarray}
for $0<\delta<1$ and similar formulas for the case $\delta>1$. In both cases, we can ignore terms that decrease exponentially rapidly in the fast scale $t$ and set
\begin{eqnarray}\label{b.6}
\xi^{(0)}= K(\zeta)+ \mbox{EDT},\quad
\Gamma^{(0)}=K(\zeta) + \mbox{EDT},
\end{eqnarray}
where EDT stand for exponentially decreasing terms in the fast time scale. To find $K(\zeta)$, we insert (\ref{b.6}) in (\ref{b.4a}) - (\ref{b.4c}), ignore EDT and obtain:
\begin{eqnarray}\label{b.7a}
 \left( \frac{\partial ^2}{\partial t^2}+1\right) \xi^{(2)} -\Gamma^{(2)} = 0,  \\
  \label{b.7b}
\left(\frac{\partial}{\partial t}+2\delta\right)\Gamma^{(2)}-2\delta\xi^{(2)}=  2\delta K\left(\theta_2+\Gamma^{(0)2}\varphi_2^{(0)}-\frac{8}{3}K^{2}\right)-\frac{d K}{d\zeta},\\
\left(\frac{\partial}{\partial t}+2\delta\right)(\Gamma^{(0)2}\varphi_2^{(0)})=2\delta K^2. \label{b.7c}
\end{eqnarray}
The solutions of (\ref{b.7a}) and (\ref{b.7c}) are $\xi^{(2)}=\Gamma^{(2)}$+ EDT and $\Gamma^{(0)2}\varphi_2^{(0)}=K^2$+ EDT, respectively. Then (\ref{b.7b}) becomes
\begin{eqnarray}\label{b.8}
\frac{\partial \Gamma^{(2)}}{\partial t}=  2\delta K\left(\theta_2-\frac{1}{3}K^{2}\right)-\frac{d K}{d\zeta},
\end{eqnarray}
plus EDT which have been ignored. The solution $\Gamma^{(2)}$ is bounded as $t\to\infty$ provided
\begin{eqnarray}\label{b.9}
\frac{d K}{d\zeta}=  2\delta K\left(\theta_2-\frac{1}{3}K^{2}\right).
\end{eqnarray}
This amplitude equation is the same as (\ref{b26}) and therefore it has the same solution (\ref{b27a}):
\begin{eqnarray}
K(\epsilon^2 t)=[\mbox{sign}\,K(0)]\,\sqrt{\frac{3\, (1-\theta)}{1+e^{-4 (1-\theta)\delta t}\left(\frac{3\, (1-\theta)}{K(0)^2}-1\right)}}+\Or(|1-\theta|).\label{b.10}
\end{eqnarray}
In the original variables, we recover $\widetilde{x}=\widetilde{C_1}$ with $\widetilde{C_1}$ given by (\ref{b1a}), i.e., Result 2 is valid both for the macroscopic equations and for their averaged version in Result 1. If we keep EDT in the leading order approximation, we obtain the following composite expansion instead of (\ref{b1a})-(\ref{b1c}):
\begin{eqnarray}
&&\widetilde{x}(t)\sim K(\epsilon^2 t)+a(0)\, e^{-\delta t}\cos(\Omega t+\varphi),    \label{b.11a}\\
&& \widetilde{C_1}(t)\sim K(\epsilon^2 t)+2\delta\, a(0)\, e^{-\delta t}[\delta\,\cos(\Omega t+\varphi)+\Omega\,\sin(\Omega t+\varphi)],  \label{b.11b}\\
&& K(0)=r+2\delta v_0,\quad \widetilde{C_1}(0)=r,\, \widetilde{x}(0)=x_0,\, \dot{\widetilde{x}}(0)=v_0,\label{b.11c}\\
&& a(0) =\sqrt{(x_0-r-2\delta v_0)^2+\frac{[v_0(1-2\delta^2)+\delta(x_0-r)]^2}{1-\delta^2} }, \label{b.11d}\\
&&Ê\cos\varphi = \frac{x_0 -r-2\delta v_0}{a(0)}.\label{b.11e}
\end{eqnarray}
Note that $K(0)\to r$ and that $a(0)$ and $\varphi$ become $R(0)$ and $\pi/2-\phi$ given by (\ref{b7}) as $\delta\to 0$. The approximation (\ref{b.11a})-(\ref{b.11e}) is better than (\ref{b1a})-(\ref{b1c}) for relatively large values of $\delta$ but both approximations become indistinguishable as $\delta\to 0$.

\section{Averages at low temperature}
\label{app_lowtemp}
Let us assume that $|C_1^{(0)}|<R$ and that $C_1^{(0)}<0$ for example. Let us consider the limit $\theta\to 0+$. Then $x^{(0)}= C_1^{(0)}+R\sin\chi>0$ and therefore $\gamma(2x^{(0)}/\theta)\sim 1$ if $\chi_0<\chi\equiv t+\phi<\pi-\chi_0$ ($\sin\chi_0=|C_1^{(0)}|/R$ with $0\leq\chi_0\leq\pi/2$ as in \ref{b32b}) and $x^{(0)}<0$, $\gamma(2x^{(0)}/\theta)\sim -1$ for $-\pi<\chi<\chi_0$ or $\pi-\chi_0<\chi<\pi$. Then
\begin{eqnarray}
\overline{\gamma\left(\frac{2x^{(0)}}{\theta}\right)}\sim \frac{1}{2\pi}\left(\int_{\chi_0}^{\pi-\chi_0}d\chi-\int_{-\pi}^{\chi_0}d\chi-\int_{\pi-\chi_0}^\pi d\chi\right)=- \frac{2\chi_0}{\pi}, \label{a1}
\end{eqnarray}
and the other possibilities in (\ref{b32a}) are obtained using similar calculations.

For $0<-C_1^{(0)}<R$, we find
\begin{eqnarray}
\overline{\gamma\left(\frac{2x^{(0)}}{\theta}\right)\sin\chi}&\sim& \frac{1}{2\pi}\left(\int_{\chi_0}^{\pi-\chi_0}\sin\chi d\chi-\int_{-\pi}^{\chi_0} \sin\chi d\chi-\int_{\pi-\chi_0}^\pi \sin\chi d\chi\right)\nonumber\\
&=& \frac{1}{\pi}\int_{\chi_0}^{\pi-\chi_0}\sin\chi d\chi=\frac{2\cos\chi_0}{\pi}, \label{a2}
\end{eqnarray}
which is (\ref{b33a}) (the case $0<C_1^{(0)}<R$ gives the same formula). Eq.\
(\ref{b33b}) follows from using $\ln(1+x)\sim x$ in (\ref{b17}) and the integral formulas for the modified Bessel function.

For $C_1^{(0)}>R$, the equations for $\Gamma_n=C_n^{(0)}-1$ are
\begin{equation}\label{a3}
    \frac{\rmd \Gamma_n}{\rmd \tau}= \Gamma_{n-1}+ \Gamma_{n+1} - 2 \Gamma_n, \quad \Gamma_0=0.
\end{equation}
If we define $\Gamma_{-n}=-\Gamma_n$, the generating functional
\begin{equation}\label{a4}
 \Gamma(q,\tau)=\sum_{n=-\infty}^\infty e^{inq} \Gamma_n(\tau)= 2i\sum_{n=1}^\infty \Gamma_n(\tau)\,\sin(nq)
\end{equation}
obeys the equation
\begin{equation}\label{a5}
    \frac{\partial \Gamma}{\partial\tau}= -2(1-\cos q)\Gamma, \quad \Gamma(q,0)=2i\sum_{n=1}^\infty  [C_n^{(0)}(0)-1]\,\sin(nq).
\end{equation}
The solution of (\ref{a5}) is $\Gamma(q,\tau)=\Gamma(q,0)\, e^{-2(1-\cos q)\tau}$, from which we get
\begin{equation}\label{a6}
C_n^{(0)}(\tau)=1 -\frac{i}{\pi}\int_0^\pi \Gamma(q,0)\, e^{-2(1-\cos q)\tau}\sin(nq)\, dq.
\end{equation}
Inserting the initial condition (\ref{a5}) in (\ref{a6}) and using the integral representation formula for the modified Bessel functions \cite{Be99}, we obtain (\ref{b35}) with $S=1$:
\begin{equation}\label{a6a}
C_n^{(0)}(\tau)=1 - e^{-2\tau}\sum_{j=1}^\infty [1-C_n^{(0)}(0)]\, [I_{|n-j|}(2\tau)
- I_{n+j}(2\tau)].
\end{equation}

On the other hand, for an equilibrium-like initial condition, $C_n^{(0)}(0)=r^n$, $0<r<1$, the initial condition in (\ref{a5}) becomes
\begin{equation}\label{a7}
\Gamma(q,0)=2i\left(\frac{r\sin q}{1+r^2-2r\cos q}-\frac{\sin q}{2(1-\cos q)}\right).
\end{equation}
Inserting (\ref{a7}) in (\ref{a6}), we obtain
\begin{eqnarray}\label{a8a}
&&C_n^{(0)}(\tau)=1+\frac{2}{\pi}\int_0^\pi \frac{r\sin q\sin (nq)}{1+r^2-2r\cos q}\, e^{-2(1-\cos q)\tau} dq + Q_n(\tau),\\
&&Q_n(\tau)=-\frac{1}{\pi}\int_0^\pi \frac{\sin q\sin (nq)}{1-\cos q}\, e^{-2(1-\cos q)\tau} dq. \label{a8b}
\end{eqnarray}
Differentiating (\ref{a8b}), we find
\begin{eqnarray}\nonumber
    \frac{d Q_n}{d\tau}&=& \frac{2}{\pi}\int_0^\pi \sin q\sin (nq)e^{-2(1-\cos q)\tau} dq\\
    &=& \frac{e^{-2\tau}}{\pi}\int_0^\pi \{\cos[(n-1)q]-\cos[(n+1)q]\} e^{2\tau\cos q} dq \nonumber\\
    &=& e^{-2\tau}[I_{n-1}(2\tau)-I_{n+1}(2\tau)], \label{a9a}
\end{eqnarray}
whereas
\begin{eqnarray}
Q_n(0)&=& -\frac{1}{\pi}\int_0^\pi \frac{\sin q\sin (nq)}{1-\cos q}\, dq = -1. \label{a9b}
\end{eqnarray}
The solution of (\ref{a9a}) with initial condition (\ref{a9b}) is
\begin{eqnarray}
Q_n(\tau)= \int_0^\tau e^{-2t}[I_{n-1}(2t)-I_{n+1}(2t)]\, dt -1. \label{a10}
\end{eqnarray}
Equations (\ref{a8a}) and (\ref{a10}) yield (\ref{5.8}).

\end{document}